\documentclass[
superscriptaddress,
twocolumn,
amsmath,amssymb,
prapplied, aps,
nofootinbib
]{revtex4-2}

\usepackage{graphicx}
\usepackage{dcolumn}
\usepackage{bm}
\usepackage{mathtools}
\usepackage{comment}
\usepackage{hyperref}
\usepackage{placeins}

\DeclarePairedDelimiter{\ket}{\lvert}{\rangle}
\newcommand{\bb}[1]{\mathbf{#1}}

\begin{document}

\title{Simulation of integrated nonlinear quantum optics: from nonlinear interferometer to temporal walk-off compensator}

\author{Seonghun Kim}
\altaffiliation{These authors contributed equally.}
\author{Youngbin Kim}
\altaffiliation{These authors contributed equally.}
\affiliation{
School of Electrical Engineering, Korea Advanced Institute of Science and Technology (KAIST), Daejeon 34141, Republic of Korea}
\author{Young-Do Yoon}
\affiliation{Department of Physics, Korea Advanced Institute of Science and Technology (KAIST), Daejeon 34141, Republic of Korea}
\author{Seongjin Jeon}
\author{Woo-Joo Kim}
\affiliation{
School of Electrical Engineering, Korea Advanced Institute of Science and Technology (KAIST), Daejeon 34141, Republic of Korea}
\author{Young-Ik Sohn}
\email{youngik.sohn@kaist.ac.kr}
\affiliation{
School of Electrical Engineering, Korea Advanced Institute of Science and Technology (KAIST), Daejeon 34141, Republic of Korea}

\date{\today}

\begin{abstract}
Nonlinear quantum photonics serves as a cornerstone in photonic quantum technologies, such as universal quantum computing and quantum communications. The emergence of integrated photonics platform not only offers the advantage of large-scale manufacturing but also provides a variety of engineering methods. Given the complexity of integrated photonics engineering, a comprehensive simulation framework is essential to fully harness the potential of the platform. In this context, we introduce a nonlinear quantum photonics simulation framework which can accurately model a variety of features such as adiabatic waveguide, material anisotropy, linear optics components, photon losses, and detectors. Furthermore, utilizing the framework, we have developed a device scheme, chip-scale temporal walk-off compensation, that is useful for various quantum information processing tasks. We further show that the proposed device scheme can enhance the squeezing parameter of photon-pair sources and the conversion efficiency of quantum frequency converters without relying on higher pump power.
\end{abstract}

\maketitle

\section{Introduction}
Quantum optics has played a pivotal role in testing fundamental principles of quantum physics throughout the historical development of quantum information science. More recently, however, it has emerged also as a powerful tool for various useful applications, including quantum computing, quantum simulation, quantum communication, and quantum sensing. Due to its bosonic nature, photons rarely interact with their surroundings, and it makes them nearly free from decoherence even at room temperature. Therefore, when combined with mature fiber technologies, infrared photons are considered the best quantum information carrier for long-distance quantum communications. Furthermore, photons have recently emerged as one of the leading universal quantum computing platforms. Based on the very original one-way quantum computing architecture \cite{raussendorf2001} and its subsequent developments \cite{gimeno-segovia2015b, bartolucci2023}, utility-scale, error-corrected universal quantum computing may be possible in the foreseeable future. One of the main challenges toward this goal is to realize hardwares that can meet a formidable number of optical components, reliable operations, and very high performance metrics all at the same time. At present, there is little doubt that integrated photonics platform is one of the most promising approaches to fulfill such requirements.

Meanwhile, optical switches are crucial building blocks for photonic quantum technology, primarily due to the non-deterministic nature of key photonic quantum processes \cite{silverstone2016}. For instance, techniques like spontaneous downconversion, which is used to generate single photon states \cite{christ2013a}, and fusion operations that grow the size of a cluster state, are inherently non-deterministic \cite{browne2005,gimeno-segovia2015}. However, these processes can be multiplexed to create nearly deterministic sources of single photons or entangled states. Such multiplexing requires the use of reliable optical switches as its key components \cite{meyer-scott2020}. Likewise, an active feed-forward, which is the core functionality for photonic quantum computing, also requires the high-speed optical switching \cite{silverstone2016}. Among the various candidate technologies for switches, electro-optic (EO) technology stands out thanks to its ultra-high speed, very low insertion loss, and minimal power consumption \cite{wang2018a}. As for integrated quantum photonics, where minimizing loss and heat is crucial, the adoption of EO switch is necessary for the most of high-speed applications.
 
Given the demanding requirements of fault-tolerant quantum computing, especially the tremendous number of optical components, taking advantage of integrated photonics is an inevitable choice. Therefore, the various useful components of it such as tapers, curves, and optical anisotropy should be considered. Curves are crucial for routing different optical components and reducing device footprints. Tapers are used not only to connect different optical components with different waveguide geometry, but also to control optical modes in the waveguide in a desired fashion. In addition, to effectively utilize the Pockels effect, which is the primary principle behind EO switches, it is crucial to consider anisotropy; Pockels effect can be found in materials without centrosymmetry, such as piezoelectric lead zirconate titanate, BaTiO$_3$, and LiNbO$_3$ \cite{newnham2005}. When these materials are utilized in integrated photonics for optical switches, the c-axis is very often lie in the plane of the wafer due to the ease of fabrication and the better device performance \cite{alexander2018, abel2019, zhang2021}. When the c-axis is in the wafer plane, however, the effective index of waveguide modes depend on its relative direction as to wafer flat, because the angle between propagation direction and c-axis changes. It poses a significant challenge in the design process of optical components since it increases design complexity. Despite such difficulties, the monolithic fabrication of various optical components, including switches, on a single chip is still advantageous. It offers benefits in terms of low-loss operation and cost-effective manufacturing, making its adoption highly desirable. 

Another key aspect of integrated nonlinear photonics is the strong nonlinear interactions. The utilization of integrated waveguides for nonlinear interactions facilitates the concentration of optical power within a remarkably small mode area \cite{jankowski2021, wang2018}. Additionally, these waveguides confine optical modes in a way that extends the interaction time, surpassing the limitations set by the Rayleigh range in free space optics. Consequently, operating in the high-gain regime becomes more feasible and therefore important. However, transitioning to this regime also presents challenges: conventional theoretical models used in the low-gain regime lose their validity due to the significant contribution of higher order expansion terms in the time evolution operator \cite{quesada2014}. Additionally, third-order nonlinear effects, such as self-phase modulation (SPM) and cross-phase modulation (XPM), become crucial factors to consider \cite{triginer2020}. Therefore, accurate simulation of high-gain effects is a critical task in the design of efficient and bright nonlinear optical devices on an integrated photonics platform.

In summary, the capability to model and design nonlinear quantum optics processes on integrated photonics is pivotal for the scalability. The integrated photonics platform have many commonly used components, such as curves and tapers, which are essential building blocks for advanced optical circuits. Additionally, considering optical anisotropy and high-gain regime is essential for accurate prediction of the nonlinear processes on platforms equipped with EO switches.

In this work, we present a framework capable of simulating nonlinear quantum processes in integrated photonics platform. Based on the theoretical groundwork by Quesada \textit{et al.} \cite{quesada2018, quesada2020}, we have extended it further as a simulation framework tailored for nonlinear integrated photonics. We begin by providing the theoretical background, which is generalized for structures that vary in an adiabatic limit, such as curves and tapers. Subsequently, we demonstrate the framework through diverse case studies, including inhomogeneous spatial nonlinearity, waveguide tapers, anisotropic waveguides, and strongly pumped nonlinear processes. Additionally, our framework includes a model for the propagation loss, various linear optical components and detectors, which have been verified against experimental data. Based on these basic functionalities, we further introduce a more advanced chip-scale nonlinear quantum photonic circuit called temporal walk-off compensation (TWOC), or quasi-group velocity matching \cite{smith1998, huang2004}. Through an analysis of a two-mode vacuum squeezer and quantum pulse gate (QPG) enhanced by TWOC, we illustrate the versatility and applicability of our simulation framework.

\section{Theory}
\label{sec:theory}
In this section, we establish the theoretical groundwork for our simulation, which is primarily based on the previous studies \cite{quesada2018, vidrighin2016}. These works provide a theoretical model to calculate nonlinear quantum processes in waveguides, including the time-ordering effect and the third-order nonlinearity. Building upon the foundations, we extend the formulation further to include nonlinear propagation within slowly changing waveguides under the adiabatic limit, where the coupling between spatial eigenmodes of the waveguide is negligible. This approach enables us to accurately model the curvature, tapering, and anisotropy of the waveguide commonly used in integrated quantum photonics.

We derive the equation of motion (EOM) from Maxwell's equations, following the approach outlined by Lægsgaard \cite{laegsgaard2012}. Subsequently, by assuming the adiabatic limit, we simplify the model to exclude cross-coupling between spatial modes. We then carry out the quantization of the field by substituting bosonic operators for classical fields. Additionally, we incorporate linear optics and detectors into our framework using the Gaussian optics formalism, \cite{thomas2021}, further enhancing its capability.

\subsection{Equation of motion under adiabatic evolution}

We begin by defining electric and magnetic fields as the superposition of eigenmodes. The set of spatial eigenmodes at a certain frequency is determined by the waveguide geometry and the optical properties of the material. Each eigenmode is characterized by its field profile, propagation constant, group velocity \cite{bures2009}. From this set, we select a few modes that would engage in nonlinear interactions. The modal properties we consider include the central frequency $\bar{\omega}_m$, group velocity $v_m(z)$, normalized modal electric field $\vec{e}_m(\vec{r},t)$, normalized modal magnetic field $\vec{h}_m(\vec{r},t)$, and field amplitude $\psi_m(z,\omega)$ \cite{laegsgaard2012}:
\begin{subequations}
\begin{align}
\label{eq:field_decomp}
\vec{E}(\vec{r}, t) & =\frac{1}{\sqrt{4 \pi}} \sum_m \int d \omega \sqrt{\frac{\hbar \bar{\omega}_m}{v_m(z)}} \psi_m(z, \omega) \vec{e}_m(\vec{r},  t)\nonumber\\& \qquad\qquad\qquad\qquad\qquad\qquad+\mathrm{c . c}, \\
\vec{H}(\vec{r}, t) & =\frac{1}{\sqrt{4 \pi}} \sum_m \int d \omega \sqrt{\frac{\hbar \bar{\omega}_m}{v_m(z)}} \psi_m(z, \omega) \vec{h}_m(\vec{r},  t)\nonumber\\& \qquad\qquad\qquad\qquad\qquad\qquad+\mathrm{c . c},
\end{align}
\end{subequations}
where $z$ is the longitudinal coordinate along the waveguide. We neglect the frequency dependence of the modal field $\vec{e}_m(\vec{r},t)$, assuming that the spectral bandwidth of the optical modes is sufficiently narrow. Subsequently, the time-dependent modal fields $\vec{e}_m(\vec{r},  t), \ \vec{h}_m(\vec{r},  t)$ are expressed as:
\begin{subequations}
\begin{align}
    \vec{e}_m(\vec{r},  t)&=\vec{e}_m\left(\vec{r}_{\perp}, z\right)\exp \left\{ i\left[\int_0^z d z^{\prime} k_m\left(z^{\prime}, \omega\right)-\omega t\right]\right\},\\
    \vec{h}_m(\vec{r},  t)&=\vec{h}_m\left(\vec{r}_{\perp}, z\right)\exp \left\{i\left[\int_0^z d z^{\prime} k_m\left(z^{\prime}, \omega\right)-\omega t\right]\right\},
\end{align}
\end{subequations}
where $k_m(z,\omega)$ and $\vec e_m(\vec r_\bot, z)$ denote the propagation constant and normalized modal field of the spatial mode $m$ at local point $z$. The modal fields are normalized to conform the following normalization condition,
\begin{align}
2v_m(z)=\int d\vec{r}_\perp \ [&\vec e_m\left(\vec{r}_{\perp}, z\right)\times \vec h^*_m\left(\vec{r}_{\perp}, z\right) \nonumber \\ &-\vec h_m\left(\vec{r}_{\perp}, z\right)\times \vec e_m^*\left(\vec{r}_{\perp}, z\right)]\cdot \hat z,
\end{align}
where the integration is performed over the waveguide cross section.
This normalization allows the optical energy passing through the waveguide cross-section at position $z$ to be represented as \cite{jesipe2016}
\begin{equation}\label{eq:energy}
    E_m(z) = \int \hbar \omega d\omega \psi_m ^*(z,\omega)\psi_m(z,\omega).
\end{equation}
From the Maxwell's equations, the EOM governing the evolution of the field amplitudes is derived and takes the following form \cite{laegsgaard2012}:

\begin{align}\label{eq:eom_adiabatic}
\frac{\partial \psi_m}{\partial z}&=\frac{i}{2} \sqrt{\frac{\bar{\omega}_m}{\pi \hbar v_m(z)}} \int d a d t \vec{e}_m^*(\vec{r},  t) \cdot \delta \vec{P}(\vec{r}, t)\\
&+\sum_{n \neq m} \frac{1}{2} \frac{C_{m, n}}{\sqrt{v_n(z) v_m(z)}}\nonumber \\ 
&\times\exp{\left\{i\int_0^z dz'[k_n(z', \omega)-k_m(z', \omega)]\right\}} \psi_n(z, \omega),\nonumber
\end{align}
where $\delta \vec{P}(\vec{r}, t)$ represents the nonlinear polarization, and $C_{m,n}$ is cross-coupling coefficient between different mode $n$. The EOM captures two dynamics: the nonlinear interaction between optical modes and the linear coupling due to the variation in waveguide geometry, where each dynamics is expressed in the first and second terms, respectively. Note that the nonlinear polarization $\delta \vec{P}(\vec{r},t)$ can be further divided into the second order nonlinear polarization $\delta \vec{P}_{(2)}(\vec{r},t)$ and the third order nonlinear polarization $\delta \vec{P}_{(3)}(\vec{r},t)$:
\begin{subequations}
\begin{align}
\label{eq:nonlin_pol}
        &\delta P_{(2)}^{j}(\vec r,t) 
= \epsilon_0 \chi^{jkl}_{(2)}E^k(\vec r, t)E^l(\vec r, t),\\
        &\delta P_{(3)}^{j}(\vec r,t) = \epsilon_0 \chi_{(3)}^{jklm}E^k(\vec r, t)E^l(\vec r, t)E^m(\vec r, t),
\end{align}
\end{subequations}
where $\chi^{jkl}_{(2)}$ and $\chi^{jklm}_{(3)}$ are second-order and third-order nonlinearity tensor components \cite{boyd2019}. Among the combinations of mode mixing in nonlinear polarization, effective terms are considered where phase matching and energy conservation are satisfied simultaneously. 

The second term on the right-hand side of Eq. (\ref{eq:eom_adiabatic}) describes the linear coupling caused by waveguide geometry variation along its propagation. Under such conditions, the continuous translational symmetry is broken, and the set of eigenmodes depends on $z$. Therefore, the electric field after a small propagation should be expanded on another eigenmode basis. Such a process continuously happens as the wave propagates through the waveguide, and therefore the amplitude of each mode needs to be updated accordingly. In this work, however, we limit our study to the adiabatic regime, where propagation direction and waveguide geometry change slowly, so the cross-coupling can be ignored. Hence, we consider the first term solely while ignoring the second term of Eq. (\ref{eq:eom_adiabatic}).

As an illustration, we consider the spontaneous parametric down-conversion (SPDC) process, which occurs alongside SPM and XPM. In the SPDC process, we consider pump, signal, and idler modes with central frequencies $\bar{\omega}_p$, $\bar{\omega}_s$, and $\bar{\omega}_i$, respectively. A pump mode is excited by coupling a laser into a nonlinear waveguide, leading to the generation of photon pairs in the signal and idler modes, in accordance with the energy conservation ($\bar{\omega}_p = \bar{\omega}_s + \bar{\omega}_i$) and the phase matching conditions. In this process, third-order nonlinearity induces parasitic phenomena, particularly noticeable in a high-gain regime. The pump beam undergoes SPM, while the signal and idler modes are influenced by the pump through XPM. We ignore the XPM between the signal and idler since it is negligible compared to the XPM induced by the pump. It is straightforward to extend the EOM of SPDC to other similar nonlinear processes such as spontaneous four-wave mixing (SFWM) and quantum frequency conversion (QFC) as well. In what follows, we present a detailed description of the EOM targeting SPDC and QFC.

From the assumption of narrow spectral bandwidth, we first approximate the propagation constant to the first order:
\begin{equation}\label{eq:linearize}
    k_m(z,\omega) \approx \bar{k}_m(z) + \frac{1}{v_m(z)}(\omega- \bar{\omega}_m),
\end{equation}
where $\bar{k}_m(z)$ is the propagation constant at center frequency of mode m, $\bar{\omega}_m$. To avoid fast evolution of the phase of the pump amplitude, which requires time-consuming iterations in simulation, the nonlinear evolution is described in the reference frame where the pump envelope is stationary in time. On the frame, the pump envelope always arrives at $t=0$ everywhere and its amplitude is represented as
\begin{align}
    \beta_p(z,\omega) &= \sqrt{\hbar \bar{\omega}_p }\psi_p(z,\omega) \nonumber \\ &\times \exp\left({-i\int_0 ^ z dz^\prime \frac{\omega - \bar{\omega}_p}{v_p(z^\prime)}}\right).
\end{align} 
Applying the same reference frame, the field amplitudes of signal and idler modes are written as
\begin{align}
a_j(z,\omega)&= \psi_j(z,\omega)\nonumber \\ 
&\times\exp\left[ i\int_0^z dz' \left(\cfrac{1}{v_j(z')}-\cfrac{1}{v_p(z')}\right)(\omega - \bar{\omega}_j)\right].
\end{align}

Moving forward, we quantize the field amplitudes into operators as follows:
\begin{equation}
\begin{aligned}
& {\left[a_j\left(z, \omega\right), a_{j^{\prime}}^{\dagger}\left(z, \omega^{\prime}\right)\right]=\delta_{j, j^{\prime}} \delta\left(\omega-\omega^{\prime}\right),} \\
& {\left[a_j\left(z, \omega\right), a_{j^{\prime}}\left(z, \omega^{\prime}\right)\right]=0 ,}
\end{aligned}
\end{equation}
such that 
\begin{equation}
    a_j^{\dagger}\left(z, \omega\right)a_{j}(z,\omega)
\end{equation}
represents the spectral photon number density at position $z$ using the relation in Eq. (\ref{eq:energy}). As a result, we have the following equations of motion for the signal and idler, where both XPM and SPDC processes are included.
\begin{subequations}
\label{eq:eom_pair_PDC}
\begin{align}
\cfrac{\partial}{\partial z}a_s(z,\omega) &= i\Delta k_s(z,\omega)a_s(z,\omega)  \\
&+ i\cfrac{\gamma_{\mathrm{XPM},s}(z)}{2\pi}\int d\omega' \mathcal{E}_p(\omega-\omega')a_s(z,\omega') \nonumber\\
&+i\cfrac{\gamma_{\mathrm{PDC}}(z)\exp[{i\int dz'\Delta\bar{k}_{\mathrm{PDC}}(z')}]}{\sqrt{2\pi}}\nonumber \\ &\times \int d\omega' \beta_p(z,\omega+\omega')a_i^{\dagger}(z,\omega'),\nonumber\\
\cfrac{\partial}{\partial z}a_i(z,\omega) &= i\Delta k_i(z,\omega)a_i(z,\omega) \\
&+ i\cfrac{\gamma_{\mathrm{XPM},i}(z)}{2\pi}\int d\omega' \mathcal{E}_p(\omega-\omega')a_i(z,\omega') \nonumber \\
&+i\cfrac{\gamma_\mathrm{PDC}(z)\exp[{i\int dz'\Delta\bar{k}_{\mathrm{PDC}}(z')}]}{\sqrt{2\pi}} \nonumber \\ &\times \int d\omega' \beta_p(z,\omega+\omega')a_s^{\dagger}(z,\omega').\nonumber
\end{align}
\end{subequations}
The first term on the right-hand side causes a temporal walk-off between the pump and the mode $j(=s,i)$, where the rate of change in the spectral phase is defined as
\begin{equation}\label{eq:dkj}
    \Delta k _j (z,\omega) = \left(\cfrac{1}{v_j(z)}-\cfrac{1}{v_p(z)}\right)(\omega - \bar{\omega}_j).
\end{equation}
The central phase mismatch is
\begin{equation}\label{eq:c_mismatch}
    \Delta\bar{k}_{\mathrm{PDC}}(z) =\bar{k}_p(z)- \bar{k}_s(z)- \bar{k}_i(z),
\end{equation}
and the pump autocorrelation function is given by
\begin{equation}
	\mathcal{E}_p(\Delta\omega) = \int d\omega'\beta_p(z,\omega'-\Delta \omega)^*\beta_p(z, \omega').
\end{equation}
The dynamics of pump pulse within our simulation framework is also governed by Eq. (\ref{eq:eom_adiabatic}) under the assumption of a strong and undepleted pump. Accordingly, the effect of SPDC and XPM on the pump pulse is negligible. Consequently, the pump field dynamics is primarily influenced by SPM, leading to an EOM that is independent of other modes \cite{triginer2020}:
\begin{equation}
    \frac{\partial}{\partial z} \beta_{p}(z, \omega)=i \frac{\gamma_{\mathrm{SPM}}(z)}{2\pi} \int d \omega^{\prime} \mathcal{E}_{p}(\omega-\omega^{\prime}) \beta_{p}(z, \omega^{\prime}),
\end{equation}
where the parameters $\gamma_{\mathrm{PDC}}(z)$, ${\gamma_{\mathrm{SPM}}(z)}$, and ${\gamma_{\mathrm{XPM},j}(z)}$ represent the strength of nonlinear interactions. Their expressions are written in App. \ref{app:sec:nonlinear_coef}, as in the literature \cite{quesada2018}. However, in our EOM, parameters such as central phase mismatch, group velocity, and nonlinear coefficients are allowed to vary with $z$, making the EOM applicable to waveguides with slowly changing geometries.

A generalization to the QFC can be obtained with a small revision of Eq. (\ref{eq:eom_pair_PDC}). QFC typically involves either sum-frequency generation (SFG) for up-conversion or difference-frequency generation (DFG) for down-conversion of optical frequencies. Simply speaking, the sum-frequency generation combines two photons to produce a photon at a higher frequency, whereas the difference-frequency generation results in a photon at a lower frequency. The major difference of QFC from the squeezing is that we treat the signal (idler) mode as the non-vacuum state input for the upconversion (downconversion) process. Let's compare two QFC scenarios involving three optical modes: TE0 mode with 1550 nm wavelength as signal, TM0 mode with 775 nm wavelength as idler, and TM0 mode with 1550 nm wavelength as pump. When the input mode is signal, we expect an upconversion of the signal photon into an idler photon by SFG. Conversely, when the input mode is idler, we expect downconversion of idler photon into a signal photon by DFG. With the notation convention, the energy conservation expression becomes consistent as $\bar{\omega}_i = \bar{\omega}_s + \bar{\omega}_p$. In the following manner, the EOM of QFC for both SFG and DFG is
\begin{subequations}
\label{eq:eom_pair_QFC}
\begin{align}
\cfrac{\partial}{\partial z}a_s(z,\omega) & = i\Delta k_s(z,\omega)a_s(z,\omega) \\
&+ i\cfrac{\gamma_{\mathrm{XPM},s}(z)}{2\pi}\int d\omega' \mathcal{E}_p(\omega-\omega')a_s(z,\omega') \nonumber \\
&+i\cfrac{\gamma_{\mathrm{QFC}}^*(z)\exp[{-i\int dz'\bar{k}_\mathrm{QFC}(z')}]}{\sqrt{2\pi}} \nonumber \\  &\times \int d\omega' \beta_p^*(z,\omega'-\omega)a_i(z,\omega'),\nonumber\\
\cfrac{\partial}{\partial z}a_i(z,\omega) &= i\Delta k_i(z,\omega)a_i(z,\omega) \\
&+ i\cfrac{\gamma_{\mathrm{XPM},i}(z)}{2\pi}\int d\omega' \mathcal{E}_p(\omega-\omega')a_i(z,\omega') \nonumber \\
&+i\cfrac{\gamma_\mathrm{QFC}(z)\exp[{i\int dz'\Delta\bar{k}_{\mathrm{QFC}}(z')}]}{\sqrt{2\pi}} \nonumber \\ &\times \int d\omega' \beta_p(z,\omega - \omega')a_s(z,\omega'),\nonumber
\end{align}
\end{subequations}
where $\gamma_{\mathrm{QFC}}$ is a nonlinear coefficient for QFC written in App. \ref{app:sec:nonlinear_coef}, and the central phase mismatch is given by 
\begin{equation}
   \Delta\bar{k}_\mathrm{QFC}(z) = \bar{k}_p(z) + \bar{k}_s(z) - \bar{k}_i(z).
\end{equation}
By comparing equations (\ref{eq:eom_pair_PDC}) and (\ref{eq:eom_pair_QFC}), another major difference between the squeezing and the QFC can be found: QFC mixes the annihilation operators of the idler and the signal with each other, while the squeezing mixes the annihilation operator of the idler (signal) mode with the creation operator of the signal (idler) mode.

\subsection{Solving equations of motion}
\label{subsec:solve}
Here, we briefly summarize the procedure to find the solution of Eq. (\ref{eq:eom_pair_PDC}) following \cite{quesada2018}, where SPDC is modeled in detail. For numerical evaluation, the operators $a_j(z,\omega)$ are discretized into frequency mode operators. In the frequency range of interest, the frequency of each mode is $\omega_n = \omega_1 + (n-1)\Delta\omega|_{n=1}^{N_f}$, and annihilation operators at a single frequency is represented as $a_j(z,\omega_n)$. To simplify notations, let's group these operators into a vector as $\bb{a}_j(z) = (a_s(z, \omega_1),\dots, a_s(z, \omega_{N_f}))^\mathrm{T}$. Starting from Eq. (\ref{eq:eom_pair_PDC}), discretized EOM takes the following form:
\begin{align}
\label{eq:discretized_EOM}
\frac{\partial}{\partial z}\left(\begin{array}{c}
\bb{a}_s(z)\\
\bb{a}_i^{\dagger}(z)
\end{array}\right)=i\underbrace{\left[\begin{array}{c|c}
\mathbf{G}(z) & \mathbf{F}(z)\\
\hline -\mathbf{F}^{\dagger}(z) & -\mathbf{H}^{\dagger}(z)
\end{array}\right]}_{:=\mathbf{Q}(z)}\left(\begin{array}{c}
\bb{a}_s(z)\\
\bb{a}_i^{\dagger}(z)
\end{array}\right),
\end{align}
where each block is defined as
\begin{subequations}
\begin{align}
\mathbf{F}_{n,m}(z) & =\frac{\gamma_{\mathrm{PDC}}(z)}{\sqrt{2\pi}}\beta_{p}(z,\omega_{n}+\omega_{m}) \nonumber \\ 
                       &\qquad\times\exp\left[{i\int dz'\Delta\bar{k}_{\mathrm{PDC}}(z')}\right]\Delta\omega,\\
\mathbf{G}_{n,m}(z) & =\Delta k_{s}(z, \omega_{n})\delta_{m,n}\nonumber\\&\qquad+\frac{\gamma_{\text{XPM},s}(z)}{2\pi}\mathcal{E}_p(\omega_{n}-\omega_{m})\Delta\omega,\\
\mathbf{H}_{n,m}(z) & =\Delta k_{i}(z, \omega_{n})\delta_{m,n}\nonumber\\&\qquad+\frac{\gamma_{\text{XPM},i}(z)}{2\pi}\mathcal{E}_p^{*}(\omega_{n}-\omega_{m})\Delta\omega.
\end{align}
\end{subequations}
The solution can be represented using the propagator $\bb{U}(z,z_0)$ as follows:
\begin{align}
&\left(\begin{array}{c}
\bb{a}_s(z)\\
\bb{a}_i^{\dagger}(z)
\end{array}\right)  =\mathbf{U}(z,z_{0})\left(\begin{array}{c}
\bb{a}_s(z_0)\\
\bb{a}_i^{\dagger}(z_0)
\end{array}\right)\label{solK} \nonumber \\
 & =\left[\begin{array}{c|c}
\mathbf{U}^{s,s}(z,z_{0}) & \mathbf{U}^{s,i}(z,z_{0})\\
\hline (\mathbf{U}^{i,s}(z,z_{0}))^{*} & (\mathbf{U}^{i,i}(z,z_{0}))^{*}
\end{array}\right]\left(\begin{array}{c}
\bb{a}_s(z_0)\\
\bb{a}_i^{\dagger}(z_0)
\end{array}\right).
\end{align}
The propagator $\mathbf{U}(z,z_{0})$ can be obtained by Trotterization with arbitrary precision by choosing an arbitrary small $\Delta z$:
\begin{align}\label{eq:tf}
\mathbf{U}(z,z_{0})=\prod_{p=1}^{n}\exp\left(i\Delta z\mathbf{Q}(z_{p})\right) + \mathcal{O}(\Delta z ^2),
\end{align}
where $\mathcal{O}(\Delta z ^2)$ represents Trotterization error. The propagator $\bb{U}(z,z_0)$ is also called transfer matrix in some literature \cite{vidrighin2016}. If the SPDC is phase matched along a waveguide with continuous translational symmetry, where $\Delta\bar{k}(z)=0$ is satisfied for all $z$, the matrix $\bb{Q}$ is position independent. Therefore, along a straight and uniform waveguide, which starts at $z_0$ and ends at $z_1$, the propagator is obtained as
\begin{equation}
     \mathbf{U}(z_1,z_{0}) = \exp(i(z_1-z_0)\bb{Q}),
\end{equation}
without Trotterization error. The operators at different positions $z$ and $z_0$ are related by the transfer function as
\begin{subequations}\label{eq:inout}
\begin{align}
a_{s}(z, \omega)= & \int d \omega^{\prime} U^{s,s}\left(\omega, \omega^{\prime}; z, z_0\right) a_{s}\left(z_0,\omega^{\prime}\right) \nonumber \\
& +\int d \omega^{\prime} U^{s,i}\left(\omega, \omega^{\prime};z, z_0\right) a_{i}^{\dagger}\left(z_0, \omega^{\prime}\right), \\
a_{i}(z, \omega)= & \int d \omega^{\prime} U^{i,i}\left(\omega, \omega^{\prime}; z, z_0\right) a_{i}\left(z_0,\omega^{\prime}\right) \nonumber \\
& +\int d \omega^{\prime} U^{i,s}\left(\omega, \omega^{\prime}; z, z_0\right) a_{s}^\dagger\left(z_0,\omega^{\prime}\right),
\end{align}
\end{subequations}
where the continuous transfer function is related to the discrete transfer function as
\begin{equation}
    U^{i, j}\left(\omega_m, \omega_n ;z, z_0\right) = \left[ \mathbf{U}^{i,j}(z,z_{0}) \right]_{mn}/\Delta \omega.
\end{equation}
In the case of QFC, to obtain the input-output relation, we make a substitution in the operators within Eq. (\ref{eq:inout}) as $a_{s}^{\dagger}\left(\omega^{\prime}\right) \rightarrow a_{s}\left(\omega^{\prime}\right)$ and $a_{i}^{\dagger}\left(\omega^{\prime}\right) \rightarrow a_{i}\left(\omega^{\prime}\right)$.

Our numerical computation method works very efficiently in simulating nonlinear quantum processes in waveguides. For instance, in a homogeneous waveguide where SPM is disregarded, the $\bb{Q}(z)$ matrix needs to be computed only once, allowing the simulation to complete in just a few seconds on a personal computer for 300 frequency modes. For example, periodic poling is a commonly used technique to implement quasi-phase matching for many kinds of nonlinear crystals. In those cases, the $\bb{Q}(z)$ matrix is strongly position-dependent due to the fast-oscillating part in Eqs. (\ref{eq:eom_pair_PDC}, \ref{eq:eom_pair_QFC}), specifically the term $\exp[{i\int dz'\Delta\bar{k}_\mathrm{PDC(QFC)}(z')}]$. This results in the matrix exponential in Eq. (\ref{eq:tf}) being calculated multiple times along the coherence length, determined by the center phase mismatch. To enhance computational speed under these scenarios, we eliminate the fast oscillation by changing the frame of reference:
\begin{subequations}
\begin{align}
	a_s(z,\omega) = a_s'(z,\omega)\exp{\left[ i\int _0^z dz'\Delta \bar{k}_s(z')\right]}, \\
	a_i(z,\omega) = a_i'(z,\omega)\exp{\left[ i\int _0^z dz'\Delta \bar{k}_i(z')\right]}.
\end{align}
\end{subequations}
Here, $\Delta\bar{k}_\mathrm{PDC}(z) = \Delta\bar{k}_s(z)+ \Delta\bar{k}_i(z)$.
After substituting operators with these expressions, Eq. (\ref{eq:eom_pair_PDC}) no longer contains the oscillating term:
\begin{subequations}\label{eq:eom_adiabatic_frame}
\begin{align}
\cfrac{\partial}{\partial z}a_s(z,\omega) &= i\left[ \Delta k_s(z,\omega) -\Delta \bar{k}_s (z)\right] a_s(z,\omega) \nonumber\\  &+ i\cfrac{\gamma_{\mathrm{XPM},s}(z)}{2\pi}\int d\omega' \mathcal{E}_p(\omega-\omega')a_s(z,\omega')\nonumber \\
&+i\cfrac{\gamma_{\mathrm{PDC}}(z)}{\sqrt{2\pi}}\int d\omega' \beta_p(z,\omega+\omega')a_i^{\dagger}(z,\omega'),\\
\cfrac{\partial}{\partial z}a_i(z,\omega) &= i\left[ \Delta k_i(z,\omega)-\Delta \bar{k}_i(z)
\right] a_i(z,\omega)\nonumber \\&+ i\cfrac{\gamma_{\mathrm{XPM},i}(z)}{2\pi}\int d\omega' \mathcal{E}_p(\omega-\omega')a_i(z,\omega') \nonumber \\
&+i\cfrac{\gamma_\mathrm{PDC}(z)}{\sqrt{2\pi}}\int d\omega' \beta_p(z,\omega+\omega')a_s^{\dagger}(z,\omega').
\end{align}
\end{subequations}
Such modification yields a constant $\bb{Q}(z)$ matrix along the waveguide where parameters like $\gamma_{\mathrm{PDC}}$, $\gamma_{\mathrm{XPM}_j}$, $\Delta k_j$, and $\Delta \bar{k}_j$ are also independent on $z$. Implementing such change in the reference frame can accelerate the simulation of periodically poled waveguides by approximately 10 times, significantly enhancing the computational efficiency.

\subsection{Connection to Gaussian quantum optics}

We outline the connection between our simulation framework and Gaussian quantum optics, building upon the foundational work \cite{thomas2021}. Although the typical Hamiltonian for nonlinear interactions involves three and four bosonic operators, we can reduce the complexity to two bosonic operators by treating the pump classically. Consequently, the Hamiltonian for all nonlinear interactions of our interest are composed of quadratic bosonic operators, hence categorized as Gaussian processes. It means that all of our discussions so far stay within the Gaussian quantum optics framework \cite{eisert2007}. Let's take a look at an example of parametric downconversion from the Gaussian quantum optics viewpoint. Signal and idler input modes begin in a vacuum state, which are inherently Gaussian. Since Gaussian processes transform Gaussian states into other Gaussian states, the output modes also remain Gaussian. Furthermore, adopting a Gaussian optics formalism significantly simplifies the analysis when optical losses and photodetection need to be handled because they are relatively convenient to be implemented for Gaussian states.

To begin, we represent the collection of operators involved in the interaction as a vector of operators:
\begin{equation}
    \bb{A} = (\bb{a}_1, \dots, \bb{a}_{N_s}, \bb{a}^\dagger_1, \dots, \bb{a}^\dagger_{N_s})^\mathrm{T},
\end{equation}
where $N_s$ is the number of spatial modes. Each component of $\bb{A}$ represents a spatial mode, which is again composed of frequency modes within that spatial mode. The operators for frequency mode coincide with those introduced in the discretization of the EOM, as shown in Eq. (\ref{eq:discretized_EOM}). In such settings, a general quadratic Hamiltonian is formulated as
\begin{equation}\label{eq:generalquadratic}
    \bb{H} = \frac{1}{2}\bb{A}^\dagger \mathbb{H}\bb{A},
\end{equation}
with $\mathbb{H}$ being a matrix with scalar elements. The unitary evolution $\mathcal{U} = \exp{(-i\bb{H}})$ constructed from the Hamiltonian dictates the evolution of the mode operators $\bb{A}$ in the Heisenberg picture as $\mathcal{U}^\dagger \bb{A} \mathcal{U}$. By applying the Baker-Hausdorff lemma, it can be demonstrated that
\begin{equation}
    \mathcal{U}^\dagger \bb{A} \mathcal{U} = \bb{M}\bb{A},
\end{equation}
where $\bb{M}$ is a symplectic matrix defined as $\exp\left[-iK\mathbb{H}\right]$. The matrix $K$ is given by
\begin{equation}
    K = \begin{bmatrix}
        1 &  0 \\ 0 & -1 
    \end{bmatrix}\otimes \bb{1}_{N_f}.
\end{equation}
Note that the symplectic matrix $\bb{M}$ satisfies the symplectic condition, $\bb{M}K\bb{M}^\dagger = K$, to preserve the bosonic commutation relation. Importantly, without considering a displacement in the phase space, any given Gaussian state can be uniquely characterized by its covariance matrix $\sigma$. Consequently, the evolution of the system can be effectively described as a transformation of a covariance matrix:
\begin{equation}\label{eq:cov_evol}
    \sigma \rightarrow \bb{M}\sigma \bb{M} ^\dagger. 
\end{equation}
Our simulation framework specifically computes the transfer matrix $\bb{U}$, which acts on a subset of operators, namely $\bb{a}_s$ and $\bb{a}_i^\dagger$, from the full collection of operators $\bb{A}$. Therefore, this transfer matrix can be converted into the symplectic matrix $\bb{M}$ that propagates the covariance matrix. For the signal and idler modes, where the collection of operators is denoted as $\bb{A} = (\bb{a}_s, \bb{a}_i,\bb{a}^\dagger_s,\bb{a}^\dagger_i)^\mathrm{T}$, the corresponding symplectic matrix is found as follows:
\begin{equation}
    \bb{M} = \begin{bmatrix}
\bb{U}^{s,s} & \bb{0} & \bb{0} & \bb{U}^{s,i} \\
\bb{0}  & \bb{U}^{i,i} & \bb{U}^{i,s} & \bb{0} \\
\bb{0}  & {\bb{U}^{s,i}}^* & {\bb{U}^{s,s}}^* & \bb{0} \\
{\bb{U}^{i,s}}^* & \bb{0} & \bb{0} & {\bb{U}^{i,i}}^*
\end{bmatrix}.
\end{equation}

The covariance matrix approach in our simulation framework allows us to effectively model optical loss. To implement this, we prepare virtual radiation modes, representing channels into which the signal and idler photons can be lost, initially in a vacuum state. These signal and idler modes then interact with the radiation modes through a series of virtual beam splitters along a given waveguide. The process of optical loss is simulated by coupling these modes and subsequently tracing out the radiation modes. The symplectic matrix representing this coupling, or the ``beam splitter'' interaction, is given by: 
\begin{equation}
    \bb{M}_\mathrm{coupler} = \begin{bmatrix}
        \bb{M}_{SS} & \bb{M}_{SL} \\ 
        \bb{M}_{LS} & \bb{M}_{LL}
    \end{bmatrix},
\end{equation}
where the label $S$ denotes the set of spatial modes of interest (such as signal and idler), and $L$ represents the radiation modes into which the photon is lost. For instance, when applying loss to the set of spatial modes $S = \{\mathrm{signal},\mathrm{idler}\}$, $\bb{M}_{SS}$ is a diagonal matrix containing the transmittance of each spatial and frequency mode. Similarly, $\bb{M}_{SL}$ is a diagonal matrix representing the reflectivity. To construct the covariance matrix for modes $S$ and $L$, we utilize the fact that the covariance matrix of a vacuum is an identity matrix. Then, the combined covariance matrix before the coupling is a block-diagonal matrix. After the coupling, as per Eq.(\ref{eq:cov_evol}), the evolved covariance matrix becomes:
\begin{align}
\bb{M}_\mathrm{coupler}&\begin{bmatrix}
        \sigma_S & \bb{0} \\ 
        \bb{0} & \bb{1}_{2|L|}
    \end{bmatrix} \bb{M}_\mathrm{coupler}^\dagger \nonumber\\&
    =
    \begin{bmatrix}
        \bb{M}_{SS}\sigma_S\bb{M}_{SS}^\dagger + \bb{M}_{SL}\bb{1}_{2|L|}\bb{M}_{SL}^\dagger & \cdots \\
        \vdots & \ddots
    \end{bmatrix}.
\end{align}
Here, $|L|$ represents the number of modes labeled $L$, which is equal to $|S|$. By tracing out the radiation modes or simply discarding block matrices which correspond to the mode being traced out, the resulting covariance matrix after the loss is computed as:
\begin{equation}
    \sigma_S \rightarrow \bb{M}_{SS}\sigma_S\bb{M}_{SS}^\dagger + \bb{M}_{SL}\bb{1}_{2|L|}\bb{M}_{SL}^\dagger.
\end{equation}

After applying all the transformations, the resulting modes can be probed using a threshold detector, which is capable of distinguishing between the vacuum state and other Fock states. The probability of projection onto the vacuum state is given by \cite{takeoka2015, thomas2021}:
\begin{align}\label{eq:poff}
P_\mathrm{off}(S) &= \mathrm{Tr}\left[ \rho|\mathrm{vac}\rangle \langle\mathrm{vac}|_S\right]\nonumber \\&= \left( \det{\left[ (\bb{1}_{2|S|} +\sigma_S)/2 \right]} \right)^{-1/2}.
\end{align}
This equation calculates the likelihood of detecting a vacuum state in mode $S$. Consequently, the probability of the threshold detector registering a non-vacuum state, or ``detection event,'' is:
\begin{equation}
    P_\mathrm{on}(S) = 1-P_\mathrm{off}(S).
\end{equation}
Similarly, the probability of simultaneous detection, or coincidence, in modes $S$ and $S^\prime$ is
\begin{equation}
    P_\mathrm{coin}(S,S^\prime) = 1-P_\mathrm{off}(S)-P_\mathrm{off}(S^\prime)+ P_\mathrm{off}(S,S^\prime).
\end{equation}
Although this formulation pertains to ideal threshold detection, it can be generalized to include threshold detectors with dark counts and photon-number-resolving detectors. Such generalization allows our framework to accurately model a variety of experimental detection scenarios, enhancing its practical utility in quantum optics experiments.

\section{Simulation framework}
\label{sec:framework}
In this section, we begin by presenting a conceptual overview of our simulation framework, explaining the workflow of the simulator. Subsequently, we showcase the functionality of our simulator in a low-gain regime. The validation is conducted through comparisons with established methods in various configurations, such as conventional periodically poled lithium niobate (PPLN) nonlinear waveguides and more complex structures like apodized PPLN, tapered PPLN, angular phase matching (APM) waveguide, and nonlinear interferometers, as reported in prior studies \cite{xin2022, uren2005, poveda-hospital2023}. These examples show the adaptability of our framework to a wide range of configurations, including nonuniform nonlinearity profiles, adiabatically tapered waveguides, and material anisotropy. 

To further demonstrate the reliability of our framework in the high-gain regime, we compare its performance with existing, publicly accessible simulation tools known to be accurate in such settings \cite{christ2013}. Lastly, our models for optical loss and detection are verified against the empirical data from the recent experimental result \cite{shin2023}. Such comprehensive approach for validating our simulation not only confirms its reliability but also shows its potential for useful applications in integrated nonlinear quantum photonics.

\subsection{Overview of simulator}
\begin{figure}
\includegraphics[width=\linewidth]{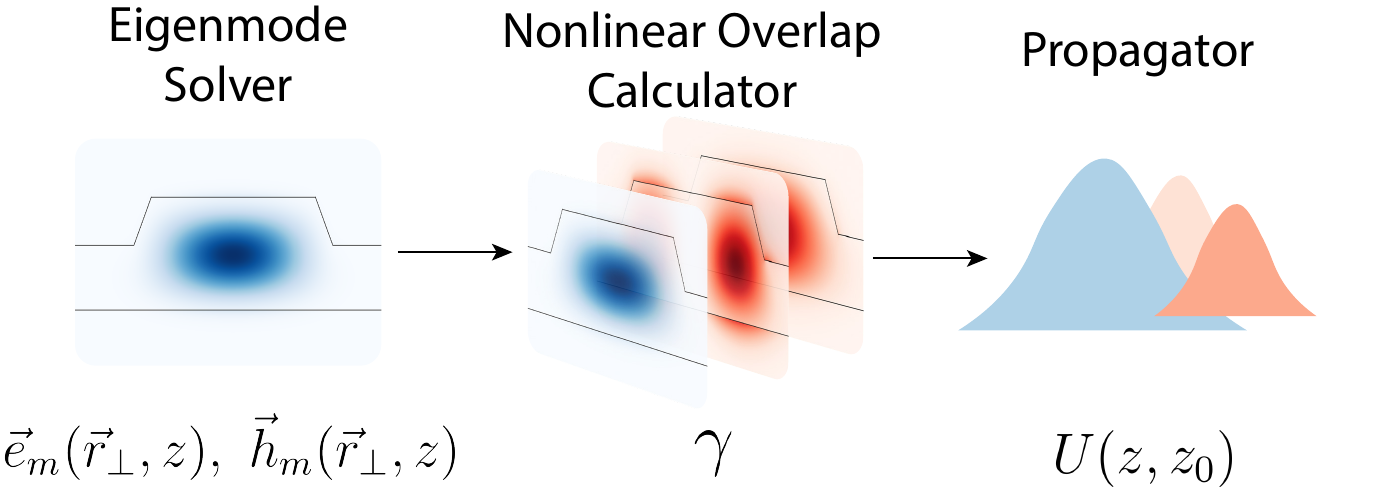}
    \caption{Workflow of the simulation framework.
} \label{fig:structure}
\end{figure}

The workflow of the simuator is outlined in Fig. \ref{fig:structure}. To model nonlinear interactions within nanophotonic structures, it is crucial to first determine the specific eigenmodes that participate in these processes. It begins with calculating the eigenmodes of the waveguide in its two-dimensional cross-section, a task accomplished using finite difference eigenmode (FDE) solvers. Commercial simulation software, such as Lumerical, is typically employed for this purpose. Once the relevant eigenmodes are identified, we obtain their characteristics, including the effective propagation constant, group velocity, and field profiles $\vec{e}(r_\perp,z),\  \vec{h}(r_\perp,z)$ at different longitudinal position $z$. These characteristics are then fed into a nonlinear overlap calculator, which computes the nonlinear coupling coefficient  $\gamma(z)$. The coefficient encompasses both material nonlinearity and mode overlap, essentially determining the strength of interaction among the eigenmodes. 

Equipped with nonlinear coefficients extracted from the previous step, along with the group velocity and propagation constant, we proceed to compute the quantum dynamics of the fields. This is done by solving the EOM to obtain the transfer function $U(z,z_0)$, as elaborated in Sec. \ref{sec:theory}.

\subsection{Simulation of integrated nonlinear waveguides}
In this subsection, we aim to showcase our simulation framework at various integrated waveguide settings by comparing with those produced by first-order perturbation, which is widely used in a low-gain regime. To begin, we provide a brief overview of first-order perturbation theory, which serves as the benchmark for the validation process.

To analyze the spectral profile of the SPDC process, we employ the joint spectral amplitude (JSA), a crucial characteristic of the photon pair that acts as a two-photon wavefunction in the spectral domain. We calculate the low-gain JSA, denoted as $f(\omega_s, \omega_i)$, using the conventional first-order approximation of the interaction Hamiltonian \cite{christ2013a}. The photon pair state generated by SPDC is expressed using JSA:
\begin{align}
    \ket{\psi}
    &\approx \ket{0} + \int d\omega_s d\omega_i f(\omega_s, \omega_i)\hat{a}_s^\dagger(\omega_s)\hat{a}_i^\dagger(\omega_i)\ket{0},
\end{align}
where the creation operators $\hat{a}_s^\dagger(\omega_s)$ and $\hat{a}_i^\dagger(\omega_i)$ satisfy the commutation relation:
\begin{equation}
    \left[ \hat{a}_j(\omega) , \hat{a}_k^\dagger (\omega')\right] = \delta(\omega-\omega') \delta_{jk}.
\end{equation}
In the low-gain regime, the JSA is formulated as a product of the pump spectral amplitude $\beta_p$ and the phase matching function $\Phi$: 
\begin{equation}\label{eq:jsa}
    f(\omega_s, \omega_i) = \beta_p(\omega_s + \omega_i)\Phi(\omega_s, \omega_i).
\end{equation}
The pump amplitude and the phase matching function constrain the wavefunction to conform to energy conservation and momentum conservation, respectively. The phase matching function, derived from the spatial integration of the nonlinear coupling coefficient $\gamma(z)$ and the local phase mismatch along the waveguide, is described by:
\begin{align}
\label{eq:phase_matching}
\Phi(L,\omega_s, \omega_i) =& \frac{1}{\sqrt{2\pi}}\int_0 ^L dz\gamma(z) \nonumber\\ &\times \exp\left[ {i\int_0^z dz' \Delta k(z',\omega_s, \omega_i)}\right],    
\end{align}
where $L$ is the length of the nonlinear waveguide, and the phase mismatch is defined as
\begin{equation}
    \Delta k(z,\omega_s, \omega_i) = k_s(z, \omega_s)+k_i(z, \omega_i)-k_p(z, \omega_s+\omega_i).
\end{equation}
Assuming each optical mode is sufficiently narrow in the frequency domain, the phase mismatch can be linearized as we did in Eq. (\ref{eq:linearize}):
\begin{align}
\label{eq:dk_expansion}
\Delta k (z, \omega_s , \omega_i) &= \Delta \bar{k}_{\mathrm{PDC}} (z)\nonumber \\&+ \left(\frac{1}{v_s(z)} - \frac{1}{v_p(z)} \right)(\omega_s - \bar{\omega}_s)\nonumber \\ &+ \left(\frac{1}{v_i(z)} - \frac{1}{v_p(z)} \right)(\omega_i - \bar{\omega}_i),
\end{align}
where the center phase mismatch $\Delta \bar{k}_\mathrm{PDC}(z)$ is defined in Eq. (\ref{eq:c_mismatch}).

\subsubsection{Periodically poled lithium niobate waveguide}
\label{sec:PPLN}
\begin{figure}
\includegraphics[width=\linewidth]{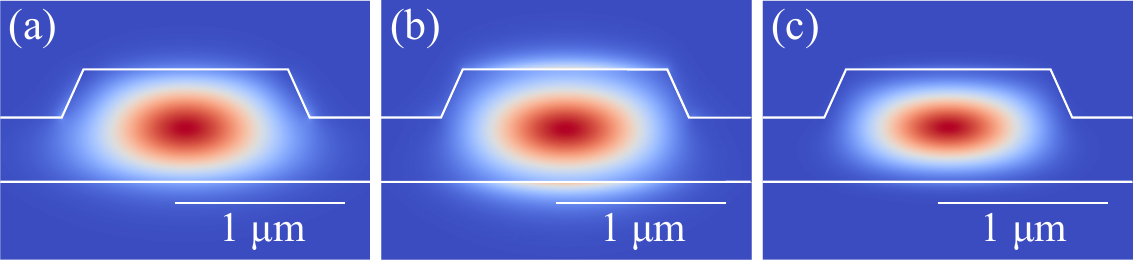}
\caption{The waveguide cross section and electric field intensity $|\vec{e}(\vec{r}_\perp)|^2$. (a) The signal mode, TE0 mode at 1550 nm. (b) The idler mode, TM0 mode at 1550 nm. (c) The pump mode, TM0 mode at 775 nm.} 
\label{fig:eigenmodes}
\end{figure}

\begin{figure}
\includegraphics[width=\linewidth]{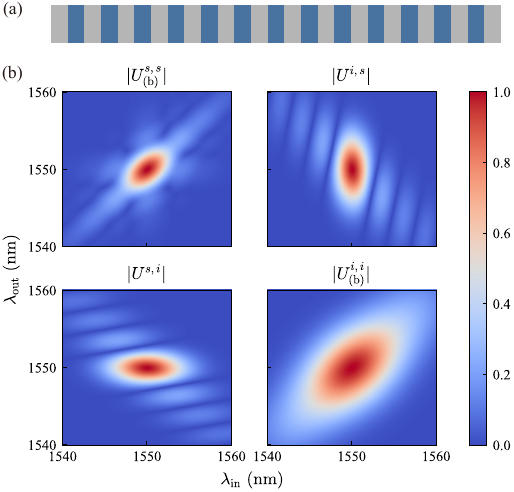}
\caption{(a) Schematic of the PPLN waveguide. (b) Normalized transfer functions of nonlinear interaction in the PPLN waveguide. Each transfer function is normalized to its maximum amplitude. The spectral purity of the output photon pair is 86.7\%. We define the broadband same mode transfer function as $\bb{U}^{{s,s(i,i)}}_{\mathrm{(b)}}(\omega, \omega')=\bb{U}^{{s,s(i,i)}}(\omega, \omega')-\delta(\omega-\omega').$ The color scale for each density plot is normalized independently.
    } \label{fig:JSAPPLN}
\end{figure}

As an initial demonstration of our simulation, we analyze the low-gain JSA of a PPLN nano-waveguide as illustrated in Fig. \ref{fig:JSAPPLN}(a). This waveguide, characterized by a uniform poling period and a uniform corss-section, is taken as a test case to validate our model against well-established results. Specifically, we simulate a rib-waveguide structure fabricated by semi-vertical etching of a thin-film lithium niobate (TFLN) platform. The waveguide is cladded by air on top and silica underneath. Top width, film thickness, etch depth, and sidewall angle are 1200 nm, 700 nm, 300 nm, and 62$^\circ$, respectively. The mode profiles for the given geometry are illustrated in Fig. \ref{fig:eigenmodes}. We use TE0 mode at 1550 nm as a signal, TM0 mode at 1550 nm as an idler, and TM0 mode at 775 nm as a pump. 

Utilizing an FDE solver, we calculate the mode profiles, group velocities, and propagation constants for each mode at these specific frequencies. To address the central phase mismatch between modes, a poling period of 3.22 $\mu \mathrm{m}$ is used, calculated by following the procedure used in first-order quasi-phase matching \cite{fejer1992}. For a given combination of modes, the nonlinear coupling coefficient for SPDC, $\gamma_\mathrm{PDC}$, is $-153.5\ \mathrm{W^{-1/2}m^{-1}}$.

In this example, we positioned the modes in the symmetric group velocity matching (sGVM) regime, where the pump mode's group velocity ($v_p$) is in between the group velocities of the signal ($v_s$) and idler ($v_i$) modes. This particular GVM regime is commonly chosen for generating spectrally pure single photons, as it allows for spectrally uncorrelated signal and idler modes in the frequency domain \cite{uren2005}. In our specific waveguide geometry, the group velocities follow the relationship $v_s > v_p > v_i$. To induce SPDC interaction, we propagate a Gaussian pump pulse with an intensity FWHM of 1.39 nm and an energy of 0.1 pJ through a 5 mm waveguide.

\begin{figure}
\includegraphics[width=\linewidth]{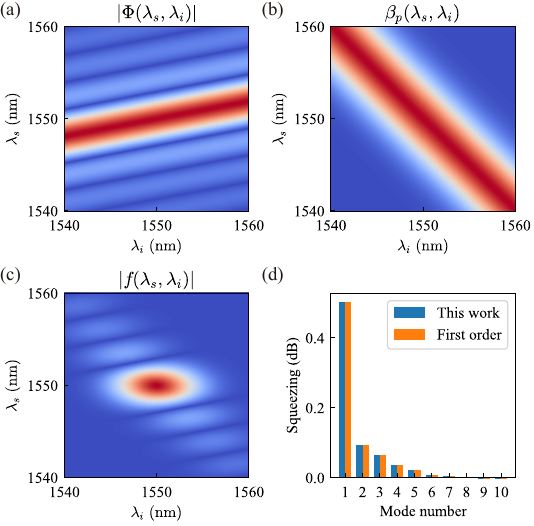}
    \caption{(a) Phase matching function of the PPLN waveguide. (b) Pump spectral amplitude. (c) JSA obtained by multiplying the phase matching function and pump spectral amplitude (d) Comparison of squeezing parameters between our framework and the first-order approximation method. The color scale for each density plot is normalized independently and follows the same color scale as used in Fig. \ref{fig:JSAPPLN}.
    } \label{fig:pmpump_plain}
\end{figure}

Using our simulation framework, we derived the transfer function representing the nonlinear interaction, as visualized in Fig. \ref{fig:JSAPPLN}(b). In the low-gain regime, the JSA is essentially equivalent to the cross-mode transfer function, either $U^{i,s}$ or $U^{s,i}$ \cite{quesada2020}. Specifically, we compare the cross mode transfer function $U^{s,i}$ with the JSA derived from the product of the pump’s spectral envelope function $\beta_p(\omega)$ and the phase matching function $\Phi(\omega_s, \omega_i)$ in Fig. \ref{fig:pmpump_plain}.

To confirm the match between our framework and the conventional JSA, we compare the squeezing parameter distribution of Schmidt modes. The result in Fig. \ref{fig:pmpump_plain}(d) shows excellent agreement between the two methods, completing the basic validation of our simulator.

\subsubsection{Apodized poling}
\label{sec:apdPol}

To demonstrate our framework's capability in handling inhomogeneous nonlinearity profiles, we report a simulation result of the nonlinear waveguide with apodized poling. Unlike the uniformly poled waveguide in the previous example, this waveguide exhibits a change in the sign of the nonlinear coefficient in an aperiodic manner. Despite the increased complexity, our simulation accurately reproduces previously known results, showcasing the robustness of our model.

Apodized poling has been employed in nonlinear quantum optics to enhance the spectral purity of single photons produced by detecting partner photons in SPDC \cite{branczyk2011,bendixon2013, dosseva2016, tambasco2016, graffitti2018, xin2022}. When operating in the sGVM regime as in the previous section, residual correlations arising from the sinc-shaped phase matching function impede achieving optimal purity. To mitigate detrimental effects from the sidelobes and boost spectral purity, the apodized poling technique can be utilized.

\begin{figure}
\includegraphics[width=\linewidth]{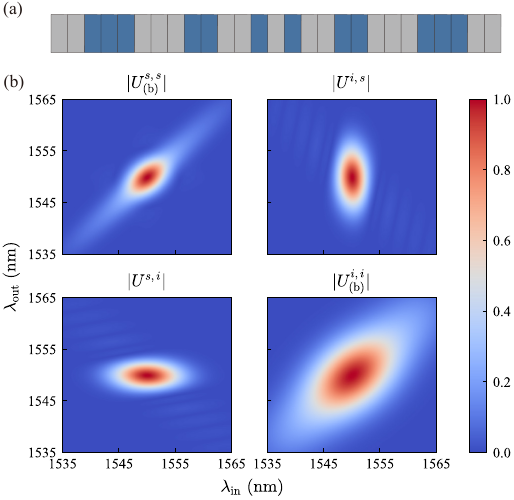}
    \caption{(a) Schematic of the apoLN waveguide. (b) Normalized transfer functions of the nonlinear interaction in apoLN. Spectral purity of the output two-photon state is 99.2\%. The color scale for each density plot is normalized independently.
    } \label{fig:poling_opt_tf}
\end{figure}

In this example, we simulated the nonlinear interaction in an apodized poling lithium niobate (apoLN) waveguide with the same geometry and optical modes as the previous example, but the poling pattern is aperiodic (see Fig. \ref{fig:poling_opt_tf}(a)). The poling pattern is optimized for the suppression of sidelobes in the phase matching function, following the approach in \cite{tambasco2016, graffitti2017}. The details of the strategy we utilized are outlined in App. \ref{app:poling}. The Gaussian pump pulse with the intensity FWHM of 1.85 nm and energy of 0.1 pJ was propagated through a 5 mm apoLN, where the pump bandwidth is optimized for spectral purity. Using our simulation framework, we obtained the transfer function, as shown in Fig. \ref{fig:poling_opt_tf}(b). Compared to the transfer function of the conventional PPLN waveguide, the sidelobes are significantly suppressed, giving a heralded single photon purity of 99.2\%.

\begin{figure}
\includegraphics[width=\linewidth]{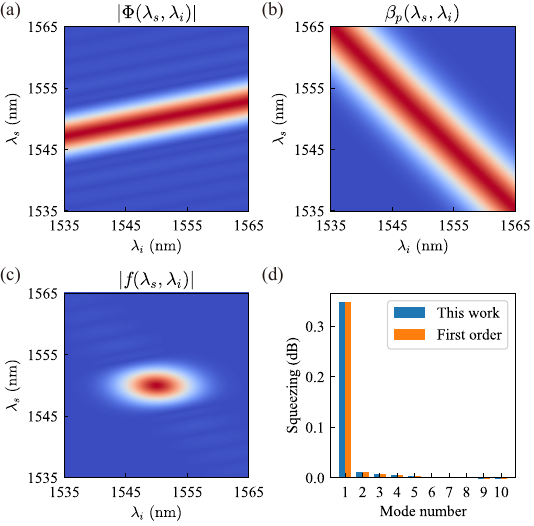}
    \caption{ (a) Phase matching function of the apoLN waveguide. (b) Pump spectral amplitude. (c) JSA obtained by multiplying the phase matching function and pump spectral amplitude (d) Comparison of squeezing parameters between our framework and the first-order approximation method. The color scale for each density plot is normalized independently and follows the same color scale as used in Fig. \ref{fig:poling_opt_tf}.
    } \label{fig:pmpump_opt}
\end{figure}

The transfer function is compared with the JSA using established conventional methods, as depicted in Fig. \ref{fig:pmpump_opt}. Again, we compared the squeezing parameters derived from both the simulation framework and the first-order perturbation, and confirming the validity of our approach.

\subsubsection{Periodically poled waveguide taper}
Adiabatically varied geometries are useful tools in nanophotonic circuit designs. Euler bends, for instance, are essential for routing between separate optical components without causing undesirable losses. Tapered waveguides are another example, often applied as a part of adiabatic directional couplers and polarization rotators. In the following three examples, we demonstrate our simulation framework's functionality to accommodate such adiabatic waveguide designs, including curves and tapers. 

In the first example, we simulate a tapered periodically poled lithium niobate (taperLN) waveguide. The geometry of the waveguide is similar to those of previous sections, except that we linearly tapered the width from 1175 nm to 1225 nm over a length of 5 mm. We employ the same spatial modes for the signal, idler, and pump as well. The entire procedure for the calculation stays the same; however, the amplitude of the local nonlinear coefficients $\gamma_\mathrm{PDC}$ and the phase mismatch $\Delta\bar{k}_\mathrm{PDC}$ are $z$-dependent due to the adiabatic change.

\begin{figure}
\includegraphics[width=\linewidth]{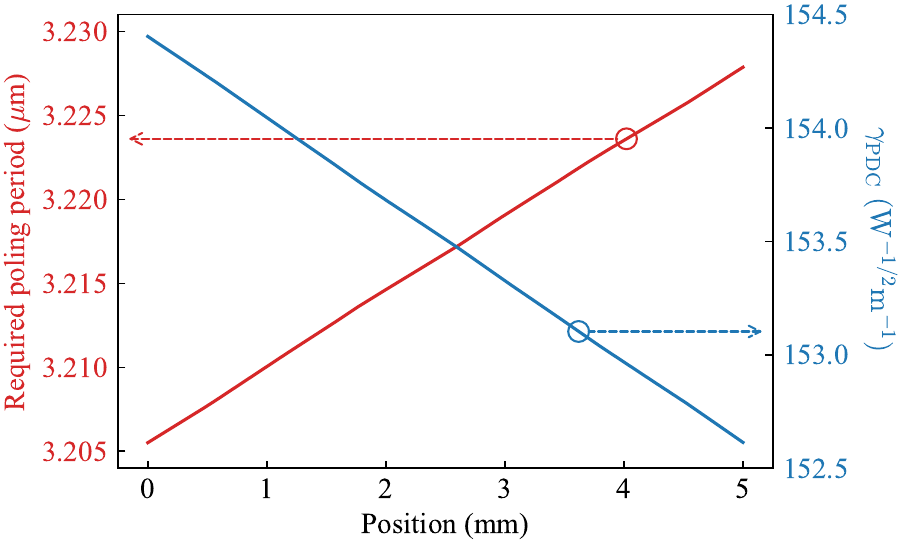}
    \caption[Required poling period and PDC coupling coeffiient of tapered PPLN]{Required poling period for the first-order phase matching and PDC coupling coefficient ($\gamma_\mathrm{PDC}$) as a function of longitudinal position $z$ in the taperLN.
    } \label{fig:pp_pdc}
\end{figure}

\begin{figure}
\includegraphics[width=\linewidth]{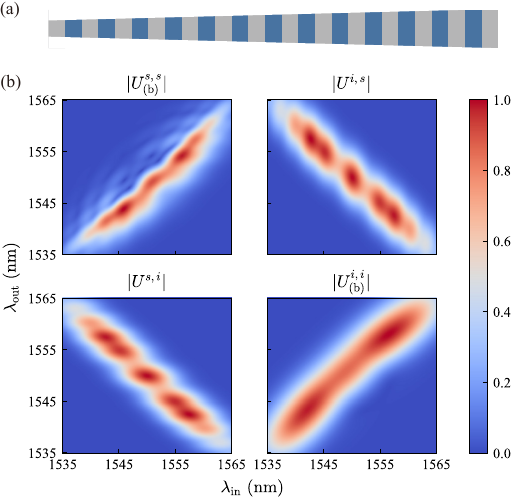}
    \caption{(a) Schematic of the taperLN waveguide. (b) Normalized transfer functions of the nonlinear process that happens in the taperLN nonlinear waveguide. The color scale for each density plot is normalized independently.
    } \label{fig:taper_tf}
\end{figure}

The test structure is a linearly tapered waveguide with a fixed poling period as shown in Fig. \ref{fig:taper_tf}(a). Although the required poling period is different for each waveguide width, we applied uniform periodic poling along the waveguide. To fully understand the working principle of such devices, we have shown the required poling periods for the quasi-phase matching and $\gamma_\mathrm{PDC}$. For a fixed set of wavelengths, both quantities vary linearly corresponding to the width change, as shown in Fig. \ref{fig:pp_pdc}. As the waveguide width increases, the required poling period at the specific combination of wavelengths increases in response to changes in the propagation constants. Conversely, the SPDC coupling coefficient decreases as the mode area increases, mostly due to decreased field intensities. In other words, at each local position along the waveguide, the perfect phase matching is met for different frequency combinations. Such an effect is evident in the broad phase matching bandwidth as shown in Fig. \ref{fig:taper_tf}(b).

\begin{figure}
\includegraphics[width=\linewidth]{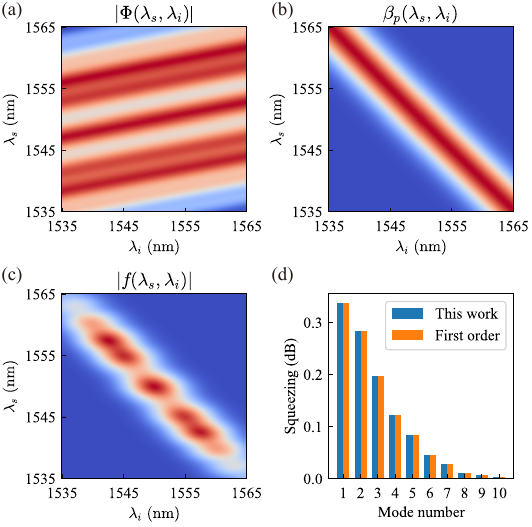}
    \caption{ (a) Phase matching function of taperLN waveguide. (b) Pump spectral amplitude. (c) JSA obtained by multiplying phase matching function and pump spectral amplitude (d) Comparison of squeezing parameters between our framework and the first-order approximation method. The color scale for each density plot is normalized independently and follows the same color scale as used in Fig. \ref{fig:taper_tf}.
    } \label{fig:pmpump_taper}
\end{figure}

We applied a Gaussian pump with an intensity FWHM of 1.41 nm and energy of 0.1 pJ, and the obtained output transfer function is shown in Fig. \ref{fig:taper_tf}(b). Following the same procedure as in previous examples, we compare our simulation results against the first-order perturbation method, as shown in Fig. \ref{fig:pmpump_taper}. Again, we have confirmed that two methods match in all aspects.

\subsubsection{Angular phase matching}
A phase matching scheme called APM, which utilizes anisotropy, was first introduced for nonlinear processes in microring resonators \cite{yang2007, yang2007a} and further adapted to interactions in waveguides \cite{poveda-hospital2023}. We introduce a similar device that was reported in \cite{poveda-hospital2023} and calculate important performance metrics, reproducing key features of APM. The goal of the work is to show that our simulator can accurately predict nonlinear processes with continuously changing nonlinearity profiles. In APM, the angular dependence of the nonlinear coefficient is exploited to achieve quasi-phase matching as well as a tailored profile of $\gamma_\mathrm{PDC}$ along the propagation. Here, we choose a design in which the nonlinear coefficient $\gamma(z)$ varies as follows:

\begin{align}
\label{eq:angle_modulation}
         \gamma(z) &= \gamma_0\exp\left[ -\frac{1}{2} \left( \frac{z}{L_\mathrm{eff}} \right)^2 \right] \left\vert \sin\left(\frac{2\pi z}{\Lambda}\right)\right\vert \nonumber\\ 
         &\approx  \gamma_0\exp\left[ -\frac{1}{2} \left( \frac{z}{L_\mathrm{eff}} \right)^2 \right] \left[ -\frac{4}{3\pi} \cos\left( \frac{4\pi z}{\Lambda} \right) \right],
\end{align}
where $L_\mathrm{eff}$ is the effective length of the waveguide, $\Lambda$ is the modulation period, and $\gamma_0$ is the maximum nonlinear coefficient. In particular, a Gaussian envelope is adopted to suppress sidelobes in the phase matching function. For the approximation in the second row, the Fourier expansion was applied, and only the dominant term was kept. The period $\Lambda$ is determined to satisfy the following equation:
\begin{equation}
    \Delta\bar{k}_\mathrm{PDC} - \frac{4\pi}{\Lambda} =0.
\end{equation}

\begin{figure}
\includegraphics[width=\linewidth]{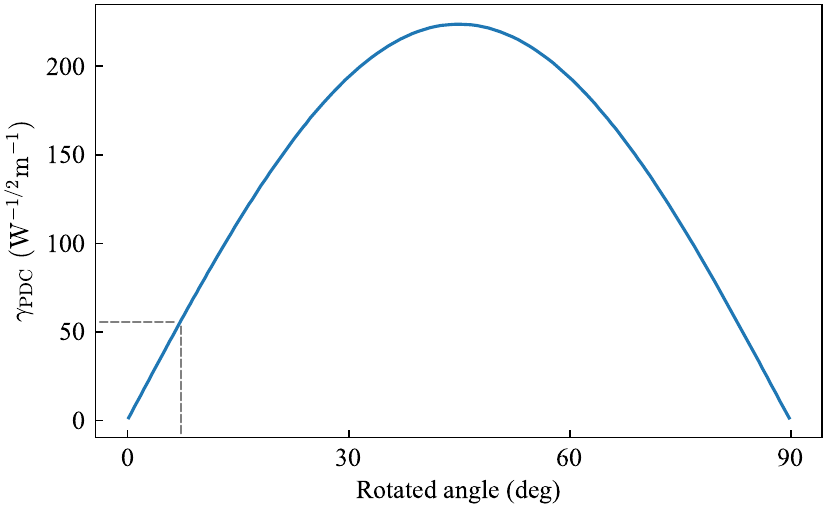}
    \caption{Nonlinear coupling coefficient for the SPDC process $\gamma_\mathrm{PDC}$ as a function of propagation angle. The maximum nonlinear coefficient is $223.0 \mathrm{\ W^{-1/2} m^{-1}}$ is achieved at a propagation angle $45^\circ$. The nonlinear coupling coefficient at a propagation angle of 7.25$^\circ$ is marked.
} \label{fig:apm_nonlinearity}
\end{figure}

Using the spatial integration formula in Eq. (\ref{eq:phase_matching}), the phase matching function takes the following form:
\begin{align}
\label{eq:pm_apm}
    \Phi(\omega_s, \omega_i) &\approx -\sqrt{2\pi}\frac{2\gamma_0 L_\mathrm{eff}}{3\pi} \nonumber\\ &\times \exp\left[ -\frac{L_\mathrm{eff}^2}{2}\left( \Delta k_s(\omega_s) + \Delta k_i(\omega_i) \right)^2 \right],
\end{align}
where $\Delta k_j $ is defined in Eq. (\ref{eq:dkj}). Here, we assume that the interval of integration extends to infinity for the Gaussian integration. This assumption is valid when the length of the nonlinear waveguide is considerably larger than $2L_\mathrm{eff}$. In practice, such periodic modulation can be achieved in a waveguide on an x-cut GaP wafer cladded by silica, as noted in \cite{poveda-hospital2023}. The optical modes adopted for the SPDC interaction are TE0 at 1550 nm for the pump, TE0 at 3100 nm for the signal, and TM0 at 3100 nm for the idler in a rectangular waveguide with a width of 1100 nm and a height of 2100 nm. Such a combination of modes requires a period $\Lambda$ of $12.03 \ \mu$m. We calculated the nonlinear coupling coefficient and its angular dependence from Eq. (\ref{eq:nonlin_coef}) using the material nonlinear coefficient \cite{shoji1997} and Miller's rule, as shown in Fig. \ref{fig:apm_nonlinearity}.

\begin{figure}
\includegraphics[width=\linewidth]{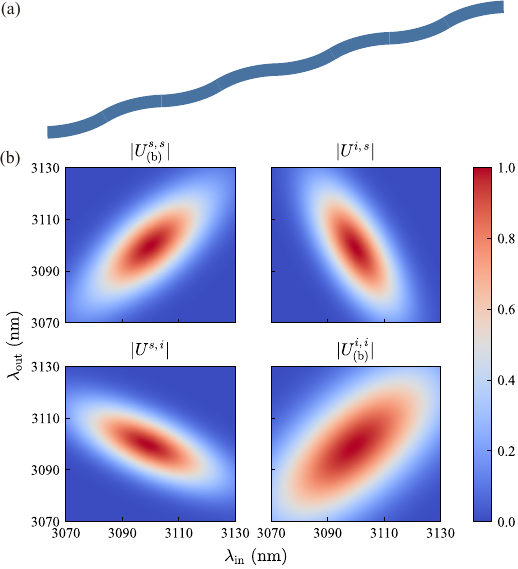}
    \caption{(a) The angular phase matching waveguide. (b) The transfer function of the nonlinear waveguide designed for APM. The color scale for each density plot is normalized independently.
} \label{fig:apm_tf}
\end{figure}

Although the maximum nonlinear coefficient is $223.0 \mathrm{\ W^{-1/2} m^{-1}}$ at $45^\circ$, increasing the propagation angle up to $45^\circ$ within a given period $\Lambda$ can lead to a very large curvature, resulting in potential radiation in the waveguide. Thus, choosing a smaller angle helps minimize losses due to the bends. Here, we consider a maximum angle of $7.25^\circ$, yielding a maximum nonlinearity $\gamma_0 = 55.75\mathrm{\ W^{-1/2} m^{-1}}$ as indicated in Fig. \ref{fig:apm_nonlinearity}. To achieve the desired nonlinearity profile as per Eq. (\ref{eq:angle_modulation}), we define the overall structure $z(\theta)$, where $z$ represents the propagation length and $\theta$ the propagation angle. The total length of the nonlinear waveguide is $8\mathrm{\ mm}$, and $L_\mathrm{eff}$ is $1 \mathrm{mm}$. The schematic of such a waveguide is illustrated in Fig. \ref{fig:apm_tf}(a).

\begin{figure}
\includegraphics[width=\linewidth]{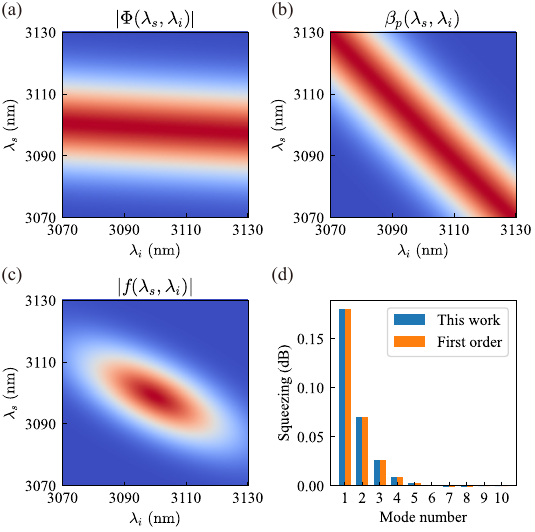}
    \caption{(a) Phase matching function of the APM waveguide. (b) Pump spectral amplitude. (c) JSA obtained by multiplying the phase matching function and pump spectral amplitude (d) Comparison of squeezing parameters between our framework and first-order approximation method. The color scale for each density plot is normalized independently and follows the same color scale as used in Fig. \ref{fig:apm_tf}.
} \label{fig:apm_pmpump}
\end{figure}

We used a Gaussian pump pulse with an intensity FWHM of 5 nm and a pulse energy of 1 pJ in the simulation, producing the output transfer function shown in Fig. \ref{fig:apm_tf}(b). The cross-mode transfer function $U^{s,i}$ corresponds to the JSA obtained from the first-order perturbation approach, as shown in Fig. \ref{fig:apm_pmpump}. The perfect match between the two models confirms that our framework accurately simulates a continuously varying nonlinearity profile along the propagation direction.

\subsubsection{Nonlinear interferometer}
\label{subsec:nonlinear_interferometry}
\begin{figure}
\includegraphics[width=\linewidth]{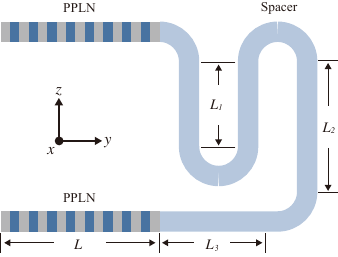}
    \caption{The nonlinear interferometer to demonstrate anisotropy. Two identical PPLN waveguides run in y-direction, but the spacer runs in z-direction. Due to anisotropy, group velocities in each direction is different. Therefore, the slopes of the phase matching function ($\theta_\mathrm{PM}$) and interference pattern ($\theta_\mathrm{Int}$) are different. $L = 3 \ \mathrm{mm}$, $L_1 = 1.5 \ \mathrm{mm}$, $L_2 = 1.7 \ \mathrm{mm}$, and $L_3 = 0.8\  \mathrm{mm}$ are used, and all of the curves are $\pi/2$-Euler bend with a radius of $200\ \mathrm{\mu m}$. The waveguide dimension is the same as the waveguide presented earlier in Fig. \ref{fig:eigenmodes}.
    } \label{fig:ccobul}
\end{figure}

In this example, we introduce the nonlinear interference of two squeezers to demonstrate the simulator's capability of handling rather complex structures. The nonlinear interferometer we study here consists of two PPLNs connected by a spacer as shown in Fig. \ref{fig:ccobul}. In this setup, the joint spectrum of each squeezer interferes, resulting in an output joint spectrum of the entire device \cite{uren2005}. To illustrate the phenomenon, we introduce two important slopes in the remainder of this example. The first is the phase matching angle that appears in the phase matching function $\Phi(\omega_s,\omega_i)$ of the single PPLN waveguide:
\begin{equation}
    \Phi(\omega_s,\omega_i) = e^{-i\Delta k L}\mathrm{sinc}({\Delta k L /2}),
\end{equation}
where $L$ represents the length of the PPLN waveguide.
Using Eq. (\ref{eq:dk_expansion}), the argument of the sinc function is expressed as:
\begin{align}
    \Delta k &(z, \omega_s , \omega_i)L \nonumber
\\&= \left(\frac{L}{v_s} - \frac{L}{v_p} \right)\Delta \omega_s + \left(\frac{L}{v_i} - \frac{L}{v_p} \right)\Delta \omega_i \nonumber
\\&=   \tau_s\Delta\omega_s+\tau_i\Delta\omega_i.
\end{align}
Here, $\tau_j = L/v_j - L/v_p$, the temporal walk-off between the pump and the mode $j$ at the end of the PPLN waveguide, represents the offset of the arrival time between two modes.
At $\Delta k L = 0$, the phase matching function is at its brightest condition; thus, the slope of the bright peak of the phase matching function on the $\omega_i - \omega_s$ plane is determined by the ratio of the temporal walk-offs:
\begin{equation}
    \tan(\theta_{PM}) = - \frac{\tau_s}{\tau_i}.
\end{equation}

After the second PPLN, the phase matching function of the entire process $\Phi_\mathrm{tot}$ can be obtained by multiplying the phase matching function of a single PPLN with a sinusoidal function:
\begin{subequations}\label{eq:nonlinear_interference}
\begin{gather}
    \Phi_\mathrm{tot}(\omega_s, \omega_i) = 2\Phi(\omega_s, \omega_i) \cos\left(\frac{\Delta\phi+T_s\Delta\omega_s+T_i\Delta\omega_i}{2}\right);\\
    \Delta \phi = \int_L ^{L+h} \Delta \bar{k}(z)dz,
\end{gather}
\end{subequations}
where $h$ denotes the length of the spacer, and $T_j$ represents the temporal walk-off between mode $j$ and the pump accumulated from the start of the first PPLN to the end of the spacer. Notably, for an inhomogeneous waveguide where group velocities are continuously changing, the temporal walk-off between mode $j$ and the pump can be calculated by integrating the group velocity offset along the waveguide:
\begin{equation}\label{eq:two_integration}
    \int \left(\frac{1}{v_j(z)} - \frac{1}{v_p(z)} \right)dz.
\end{equation}
Due to the cosine term in Eq. (\ref{eq:nonlinear_interference}), the slope of the interference pattern on the $\omega_i - \omega_s$ plane is determined by the ratio of temporal walk-offs evaluated at the end of the spacer:
\begin{equation}
    \tan(\theta_\mathrm{Int}) = - \frac{T_s}{T_i}.
\end{equation}

\begin{figure}
\includegraphics[width=\linewidth]{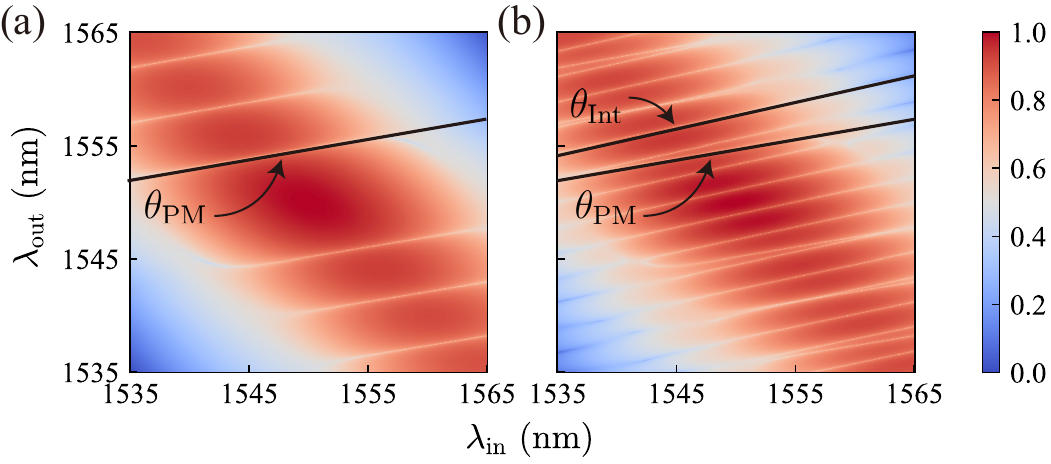}
    \caption{(a) Logarithmic plot illustrating the cross-mode transfer function $U^{s,i}$ for (a) the single PPLN waveguide and (b) the entire nonlinear interferometer. The color scale for each density plot is normalized independently.}\label{fig:ccobul_JSA}
\end{figure}

Whether the interference is constructive or destructive depends on the optical path length between the end of the first PPLN and the start of the second PPLN. The interference pattern occurs due to the different optical path lengths at various wavelengths, which is the manifestation of dispersion. The group velocity differences of the involved modes, representing the relative strength of dispersion, determine the slope of the interference pattern on the $\omega_i - \omega_s$ plane as they govern the temporal walk-offs. If the ratio of walk-offs in the spacer differs from that in the PPLN waveguide, two distinct slopes will appear in the joint spectrum—one from the phase matching function and the other from the interference pattern. We devised a device with two PPLNs and a spacer oriented such that the spacer and PPLN run in perpendicular directions, causing the two slopes to be distinct as shown in Fig. \ref{fig:ccobul_JSA}. Note that the difference in the degree of walk-off between the PPLN and spacer regions originates from the material anisotropy, which can be well managed by the simulation.

We calculated both angles $\theta_\mathrm{PM}$ and $\theta_\mathrm{Int}$ under the given conditions: $\theta_\mathrm{PM} = 10.1^\circ$ and $\theta_\mathrm{Int} = 12.3^\circ$. These slopes are found in the cross-mode transfer function, as in Fig. \ref{fig:ccobul_JSA}, and correspond to the slopes obtained through direct integration using Eq. (\ref{eq:two_integration}). From the result, we confirm that our framework is capable of managing complex nonlinear quantum photonic circuits. 

\subsection{Verification in high-gain regime}
In a high-gain regime, the first-order perturbation approach is no longer valid. It is because of the non-commuting time-dependent Hamiltonian at different times, which is called the time-ordering effect \cite{christ2013, quesada2014}. When considering the time-ordering effect, we find that the spectral mode and its distribution are distorted in an uncontrollable way, making the engineering of bright nonlinear devices challenging \cite{reddy2014, reddy2015, ansari2018b, houde2023}. Therefore, simulating the time-ordering effect is of great importance in designing nonlinear devices such as squeezers and quantum frequency converters. To simulate this effect, Christ \textit{et al.} focused on the form of Hamiltonian under a non-depleted classical pump assumption. In doing so, the Hamiltonian is quadratic in bosonic field operators, and the input-output relation becomes linear \cite{christ2013}. Based on the observation, Christ \textit{et al.} established an ansatz for numerical evaluation, and obtained an accurate result in the high-gain regime.

In the subsequent work, Quesada \textit{et al.} employed Trotterization of the propagator, enabling the calculation of dynamics with arbitrary precision. Similarly, our simulation framework utilizes the Trotter-Suzuki expansion to find the propagator in the high-gain regime. In this section, we present the Schmidt coefficients of the high-gain transfer function and compare these results with the ones calculated from the iterative method. The latter method is based on publicly available software offered by the original authors \cite{christ2013}.

\begin{figure}
\includegraphics[width=\linewidth]{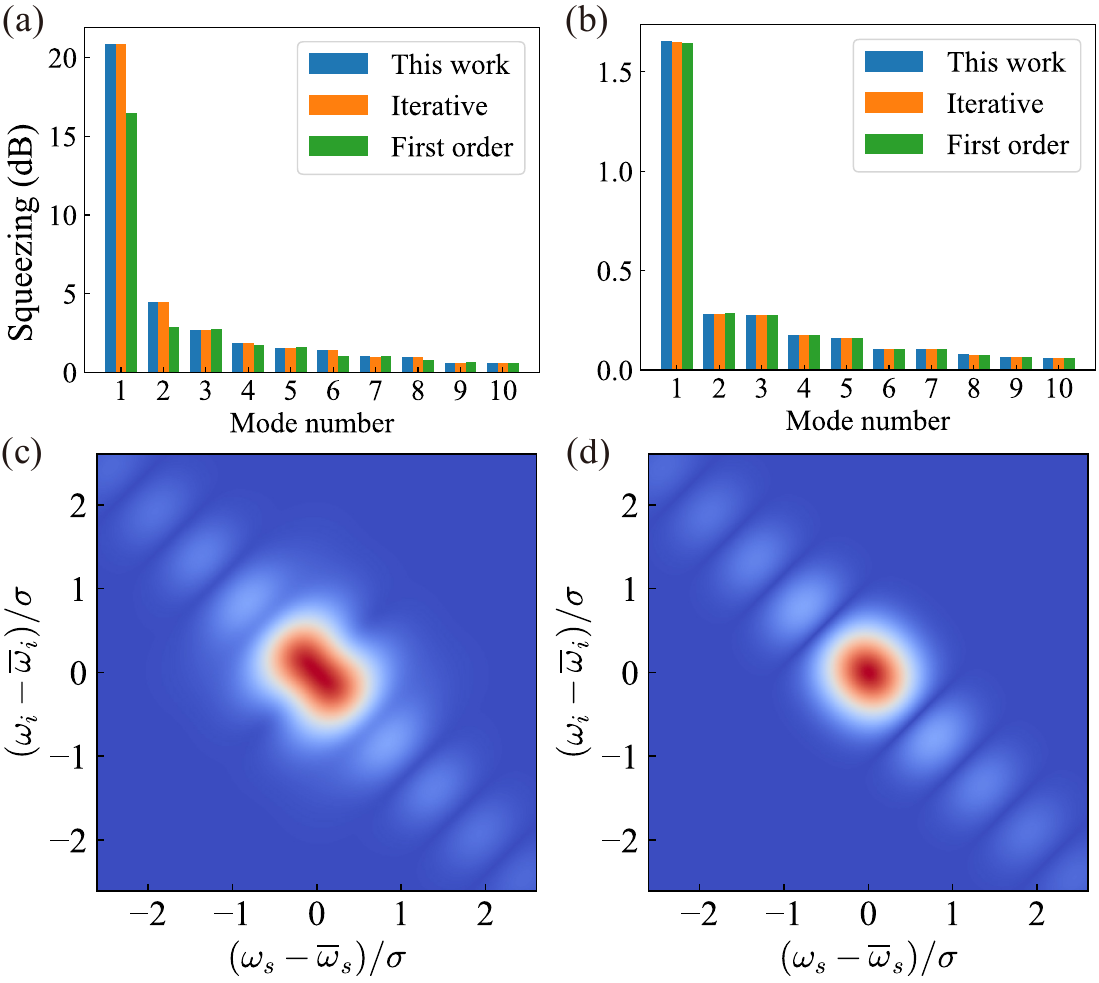}
\caption{Comparison of squeezing parameter distributions from different calculation methods: Our simulation framework, the iterative method introduced in \cite{christ2013}, and first-order perturbation. (a) low gain regime where the average photon number of signal mode $\langle n_s\rangle = 0.04.$ (b) high gain regime, where $\langle n_s \rangle = 30.96$. (c) JSA in the low-gain regime. (d) JSA in the high-gain regime. The color scale for each density plot is normalized independently and follows the same color scale as used in the previous figures.} \label{fig:christ_compare}
\end{figure}

In Fig. \ref{fig:christ_compare}, we compare three simulation methodologies. We used a phase-matched waveguide and adjusted pump energy to equalize the average photon number across the methods. In the low-gain regime, all simulations yield similar squeezing parameter distributions. On the contrary, in the high-gain regime, the analytic first-order solution does not match the other results, marking the inadequacy of first-order perturbation solutions. Meanwhile, our simulation method aligns with the result of the method introduced in \cite{christ2013}, showcasing its accuracy in the high-gain regime. 

To observe the time-ordering effect on the output photon spectrum, we use a broader definition of JSA $J(\omega,\omega')$ applicable in the high-gain regime \cite{quesada2014}:
\begin{align}
    &|\mathrm{TMSV}\rangle= \nonumber \\&\exp \left(\int d \omega d \omega' J(\omega, \omega') a_s^{(\mathrm{in}) \dagger}(\omega) a_i^{(\mathrm{in}) \dagger}(\omega')-\text {H.c.}\right)|\mathrm{vac}\rangle,
\end{align}
where TMSV denotes a two-mode squeezed vacuum, the output quantum state from parametric downconversion. This definition is equivalent to Eq. (\ref{eq:jsa}) in the low gain regime. We can observe distortion in the JSA of the high gain process in comparison to the JSA of the low-gain process, as shown in Figs. \ref{fig:christ_compare}(c) and \ref{fig:christ_compare}(d).

\subsection{Verification of loss model}
We have previously outlined the methodology for modeling loss in nonlinear quantum optics processes in Sec. \ref{sec:theory}. This procedure enables us to simulate propagation loss in nonlinear waveguides. In this section, we validate our simulation framework by matching results reported in the recent experiment \cite{shin2023}. Shin \textit{et al.} first developed an analytical model to describe waveguide propagation loss in a SFWM process with a degenerate pump mode. Subsequently, they fabricated waveguides of various lengths and measured key performance metrics of heralded single photon sources including heralding efficiency, coincidence-to-accidental ratio, and brightness. Here, we specifically focus on heralding efficiency to demonstrate the accuracy of our framework in modeling linear optical loss.

Shin \textit{et al.} provided an analytic expression for intrinsic heralding efficiency under propagation loss \cite{shin2023}:
\begin{align}\label{eq:he_analytic}
    &\mathrm{HE}\nonumber = \\&\frac{(\alpha L)^2 +(\Delta \bar{k}L)^2}{2\left(e^{\alpha L} - \frac{\alpha}{\Delta \bar{k}}\sin(\Delta \bar{k} L)-\cos(\Delta \bar{k} L)\right)}\mathrm{sinc}^2\left( \frac{\Delta\bar{k}L}{2} \right),
\end{align}
where the loss is assumed to be identical for all involved modes. In this expression, $L$ is the length of the nonlinear waveguide, $\Delta \bar{k} = \bar{k}_s+\bar{k}_i-2\bar{k}_p$ represents the central phase mismatch between modes, and $\alpha$ is the absorption coefficient. We simulated the SFWM interaction by applying the EOM for the SFWM detailed in App. \ref{app:sfwm}. For simulation purposes, the nonlinear waveguide is divided into $N$ sections. After each section, a virtual beam splitter couples the signal and idler modes with radiation modes. The transmission coefficient of the beam splitter that links the signal and idler modes to the radiation modes is $e^{-\alpha L/N}$. Additionally, the amplitude of the pump modes, which are treated classically, is updated after each section to account for the loss. At the end of the nonlinear waveguide, we determine the intrinsic heralding efficiency using the following formulas:
\begin{subequations}
\begin{align}
    \mathrm{HE}_s &= \frac{P_\mathrm{coin}(S,I)}{P_\mathrm{on}(S)} \\
    \mathrm{HE}_i &= \frac{P_\mathrm{coin}(S,I)}{P_\mathrm{on}(I)},
\end{align}
\end{subequations}
where $S$, $I$ is the label of signal and idler mode, respectively. Here, we consider an ideal threshold detector with unit efficiency and no dark counts. The simulation parameters, based on experimentally obtained values, are $\alpha = 2.22\ \mathrm{dB/cm}$ and $\Delta k = -4.01 \ \mathrm{m^{-1}}$. The waveguide is 6 cm long, and the number of sections is set to $N=6000$. A comparison between the experimental data of intrinsic heralding efficiency and our model demonstrates an excellent agreement, thereby verifying the validity of our model.
\begin{figure}
\includegraphics[width=\linewidth]{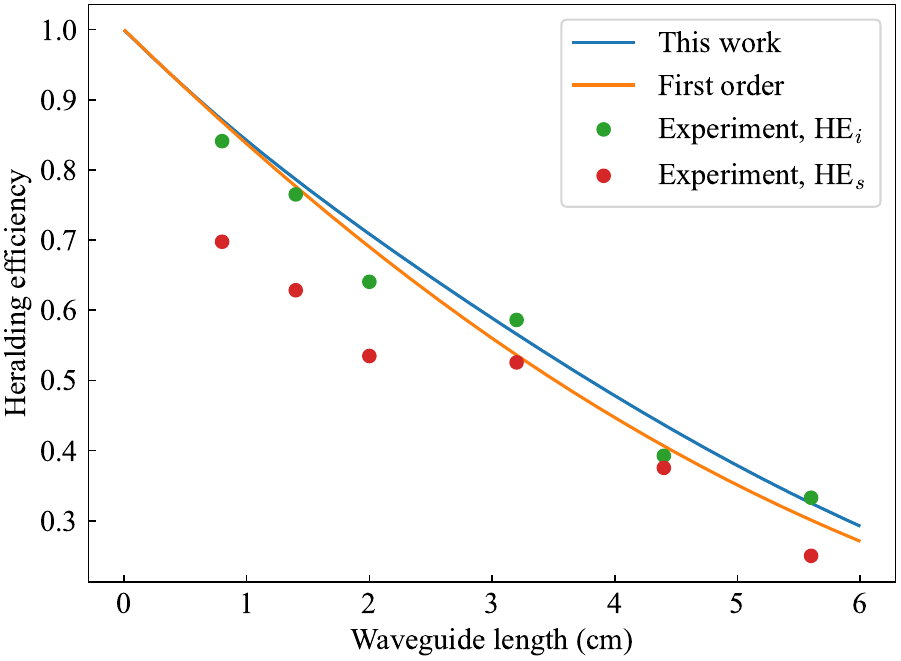}
    \caption{Comparison of results generated by our framework, analytic expression, and experimental data. $\mathrm{HE}_j$ is heralding efficiency when mode $j$ is used as herald. Analytic values are obtained from Eq. (\ref{eq:he_analytic}) using given $\alpha$ and $\Delta k$.
} \label{fig:he}
\end{figure}

\section{Temporal walk-off compensation}
\label{sec:twoc}

In this section, we present TWOC as an effective method for improving the performance of nonlinear quantum optical devices. We highlight its advantages through various simulations. Given TWOC's significant role in nonlinear interference, it has been extensively explored in free-space optics \cite{eisenberg2004,zhong2021,reddy2018}. We adapt TWOC for integrated platforms, enabling the design of scalable and efficient nonlinear devices. Implementing TWOC involves designing complex linear optical circuits, consisting of components such as polarizing beam splitters and phase shifters. Our analysis of the devices demonstrates the simulator's utility in designing and characterizing complex nonlinear circuits, accounting for anisotropy, adiabatically varying components, and optical losses.

TWOC can be effectively applied to the nonlinear interferometric setup outlined in Sec. \ref{subsec:nonlinear_interferometry}, which involves multiple nonlinear stages. TWOC addresses the issue of temporal walk-off between optical modes that accumulates during propagation. The analysis of TWOC yields two principal insights: (i) In a nonlinear interferometer comprising two nonlinear stages, temporal walk-off hinders constructive or destructive interference across the entire bandwidth between the stages. Through the application of TWOC, we can restore constructive or destructive interference across the full bandwidth; (ii) By implementing TWOC across multiple stages, it is possible to enhance nonlinear interactions without increasing pump pulse energy or sacrificing spectral bandwidth. This approach facilitates achieving significant nonlinear effects more easily and helps to avoid the time-ordering and third-order nonlinear effects which typically arise at high pump powers.

This section is organized into five subsections: First, we detail the effect of temporal walk-off in the frequency domain, where our simulator operates. Second, we examine a nonlinear interferometer using our simulator, analyzing the interference visibility and the shape of the transfer function with and without TWOC. Third, we transition to the temporal domain, highlighting how TWOC restores indistinguishability between the single photon amplitude from different squeezers. Fourth, we explore cascaded squeezers interconnected by TWOC, demonstrating a linear increase in the squeezing parameter proportional to the number of squeezers without affecting the bandwidth of the output photons. Furthermore, we show that TWOC offers advantages even when considering realistic losses.Finally, we address the limitations of QPG caused by the time-ordering effect and third-order nonlinearities, and how TWOC can be applied to overcome these limitations and achieve better performance.

\subsection{Temporal walk-off in frequency domain}
The temporal walk-off refers to a phenomenon where different optical modes gradually diverge in the time domain during propagation. For example, consider two optical modes, such as a pump and a signal, beginning their propagation simultaneously at the same position. The arrival times of the pump and signal envelopes start to differ from their initial synchronization as we move to the subsequent positions in the waveguide, and this temporal discrepancy grows as they propagate longer. Such an effect manifests in the $(z,\omega)$ domain, in which our simulation is working, as a linear spectral phase shift of the mode operators. For a mode with an index $j$, the linear phase accumulates according to the equation:
\begin{equation}
\frac{d}{dz} a_j(z,\omega) = i\frac{\omega - \bar{\omega}_j}{v_j}a_j(z,\omega),
\end{equation}
where $a_j$ is the annihilation operator, $\bar{\omega}_j$ is the center frequency, and $v_j$ is the group velocity of mode $j$. Since each mode propagates at its own group velocity, their linear phases evolve at different rates, representing temporal walk-off. When we deal with three-wave mixing interaction, it makes the problem simpler to consider the relative temporal walk-off between mode $j$ and the pump:
\begin{equation}
    \frac{d}{dz} a_j(z,\omega) = i\left(\frac{1}{v_j}-\frac{1}{v_p}\right)(\omega - \bar{\omega}_j)a_j(z,\omega).
\end{equation}

The effect of temporal walk-off can be observed in typical pulsed nonlinear interactions in a waveguide. Recall the phase matching function in Eq. (\ref{eq:phase_matching}) for homogeneous waveguide where $\gamma(z) = \gamma$ and $\Delta k (\omega_s, \omega_i,z) = \Delta k (\omega_s, \omega_i)$:
\begin{align}
    &\Phi(z, \omega_s, \omega_i)\nonumber \\&= \gamma\int ^z_0 dz^\prime\exp \Bigg\{ iz^\prime\bigg[ \left(\frac{1}{v_s} - \frac{1}{v_p} \right)(\omega_s - \bar{\omega}_s)\nonumber\\&\qquad\qquad\qquad\qquad  +\left(\frac{1}{v_i} - \frac{1}{v_p} \right)(\omega_i - \bar{\omega}_i)\bigg] \Bigg\}.
\end{align} 
In this expression, the interaction is phase matched at $(\bar{\omega}_s,\bar{\omega}_i)$, and $\Delta k (\omega_s, \omega_i)$ is linearized as Eq. (\ref{eq:linearize}). Under such a condition, there is a straight phase-matching line where the argument of the exponential term is zero in the $\omega_s - \omega_i$ plane. For the frequencies on the line, the amplitude of nonlinear coefficients constructively adds up since there is no relative phase difference. In contrast, for the frequencies away from the line, the added amplitudes at different positions have different phases. Hence, the addition of nonlinear amplitudes is performed with non-zero phase differences of those. When the frequency point on the $\omega_s - \omega_i$ plane is far away from the phase matching straight line, the relative phase difference becomes a fast-oscillating function of the waveguide position. Under such a condition, the addition of nonlinear amplitudes effectively becomes destructive interference as a total, yielding a negligible contribution to the phase matching function. More temporal walk-off induces faster oscillation for the points out of the straight line, leading to almost perfect destructive interference even at points close to the straight line. In other words, walk-off determines the bandwidth of the phase matching function in the frequency domain.

In the nonlinear interferometer that we studied in Sec. \ref{sec:framework}, temporal walk-off determines the period and direction of the interference pattern on $U^{s,i}(\omega, \omega')$, as given in Eq. (\ref{eq:nonlinear_interference}). Due to the temporal walk-off over the spacer, the constructive and destructive interference of the two PPLNs happen alternately over the bandwidth. With more temporal walk-off, the interference pattern oscillates faster in $\omega_s-\omega_i$ plane. In contrast, when the walk-off between each mode can be compensated, i.e., $T_s=T_i=0$, the resulting phase matching function becomes 
\begin{equation}
    \Phi_\mathrm{tot}(\omega_s,\omega_i) = 2\Phi(\omega_s, \omega_i)\cos \left( \Delta\phi/2 \right).
\end{equation}
Here, $\Phi(\omega_s, \omega_i)$ is the phase matching function of single waveguide, and $\Delta \phi$ is an integrated phase mismatch as illustrated in Eq. (\ref{eq:nonlinear_interference}).

With the perfect walk-off compensation, interference of phase matching function occurs consistently across the full bandwidth. In a low-gain approximation, the JSA of a single PPLN is a product of pump spectral amplitude and phase matching function. When $\Delta \phi=0$, the phase matching function of the total system $\Phi_\mathrm{tot}$ is simply doubled, and therefore the JSA is also doubled. In turn, the output photon number is quadrupled in a low-gain scenarios.

In the next subsection, we compare the interference between cascaded squeezers revealed in the visibility of the output average photon number with and without TWOC.

\subsection{Nonlinear interferometry with TWOC}
\label{subsec:nli_twoc}

The temporal walk-off compensation in nonlinear quantum optics has been realized previously by synchronizing the arrival time of signal, idler, and pump at each stage with the use of material dispersion \cite{uren2005}. In the integrated photonics platform, another type of architecture was introduced to achieve the same functionality for the second harmonic generation: in this architecture, the faster mode is coupled into the delay line while the slower mode stays in the original waveguide, and both of them are routed to the second interaction stage to be re-synchronized \cite{huang2004}. Such synchronization of interacting fields is the key ingredient of temporal walk-off compensation.

In general three-wave mixing processes, the delay of both the signal and idler with respect to the pump need to be compensated, thereby requiring two couplers and two auxiliary delay lines for each mode. As to chip-scale implementation, it is necessary to have two separate designs of mode selective couplers and delay lines, rendering the practical realization very challenging.

\begin{figure}
\includegraphics[width=\linewidth]{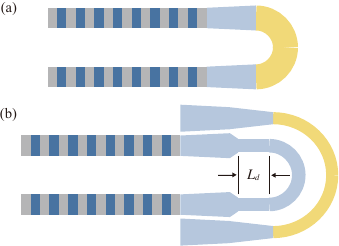}
    \caption{(a) Two-stage nonlinear interferometer without TWOC. Two PPLN waveguides are connected via linear tapers and two $\pi/2$-Euler bends with effective radius of 50 $\mathrm{\mu m }$. (b) Two-stage nonlinear interferometer with TWOC. After the first PPLN, the auxiliary waveguide is attached via APBS, which selectively couples out the signal mode. The temporal delay is adjusted by tweaking the length $L_d$.}
    \label{fig:TWOC}
\end{figure}

In contrast, our approach for walk-off compensation utilizes a specific group velocity scheme: asymmetric group velocity matching (aGVM), where the group velocity of the pump matches that of another mode. Throughout the TWOC discussion, we use the design that pump and idler group velocities are the same while that of the signal is different. Once the aGVM condition is met, the pump and idler modes do not experience temporal walk-off relative to each other. Consequently, the walk-off compensation is only necessary between the signal mode and the other two. In our design, we direct the signal mode, which propagates faster than the others, into an auxiliary delay waveguide via a polarization-dependent beam splitter; it selectively couples out the signal mode while allowing the other modes to continue in the primary waveguide. To demonstrate that such a device is implementable with current integrated photonics technology, we designed an adiabatic polarization-dependent beam splitter (APBS) that features low-loss and broadband operation. The construction of the device is illustrated in Fig. \ref{fig:TWOC}(b), and further details can be found in App. \ref{app:integrated_photonics}. In what follows, the term ``TWOC device'' is reserved to refer to the complete set of devices for walk-off compensation, including APBS, auxiliary delay line, and primary delay line in this work. We adjust the length $L_d$ of the straight part of the primary delay line to compensate the temporal walk-off between the pump/idler and signal modes.

We compare the nonlinear interferometer configurations with and without TWOC. Two cascaded PPLN waveguides connected with a spacer without TWOC is also employed as shown in Fig. \ref{fig:TWOC}(a) for the comparison to the device with TWOC. The spacer can be further divided into two identical linear tapers and two $\pi/2$-Euler bends; the bends are designed identically to the inner bend of the TWOC device. 

\begin{figure}
\includegraphics[width=\linewidth]{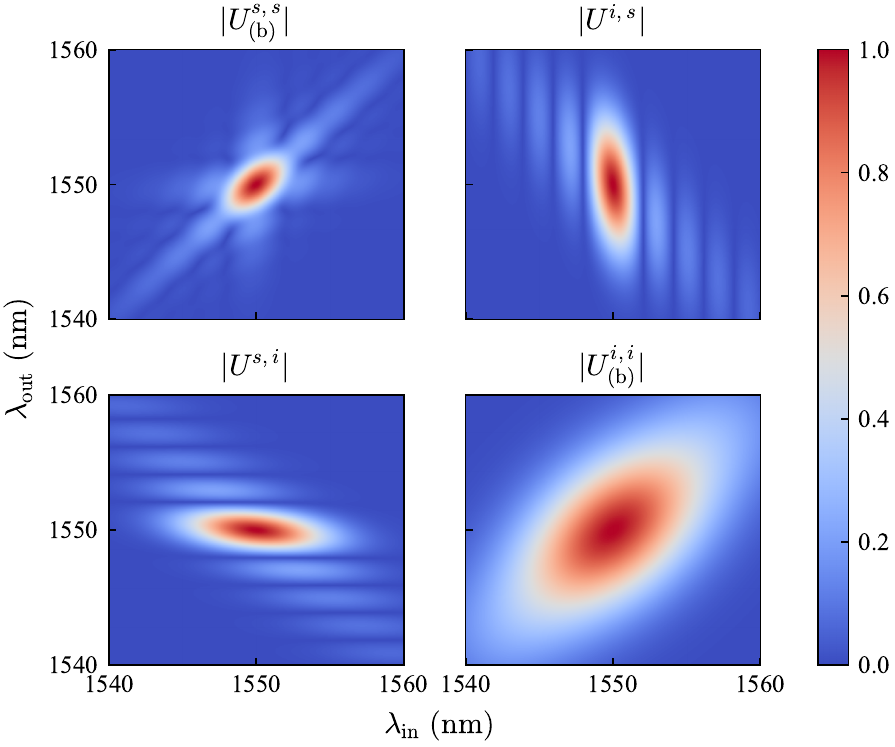}
    \caption{The transfer function of a single pass through the proposed PPLN squeezer. The color scale for each density plot is normalized independently.
    } \label{fig:single_pass_tf}
\end{figure}

As a starter, we describe the single PPLN device that is employed in the nonlinear interferometer. To achieve aGVM, we use a 6 mm long waveguide with cross-section that has a width of 890 nm, a film thickness of 500 nm, an etch depth of 300 nm, and a sidewall angle of 68$^\circ$ on an x-cut TFLN platform. Note that the design for aGVM is different from the PPLN used in Sec. \ref{sec:PPLN}. For the optical modes, we utilize the TM0 at 775 nm for the pump and the TM0 and TE0 at 1550 nm for the idler and signal, respectively. The geometry allows the group velocities of the pump and idler modes to be nearly identical. The required poling period for the first-order quasi-phase matching is 2.262 $\mu \mathrm{m}$, and the nonlinear coupling coefficient is $\gamma_\mathrm{PDC}=-182.9\ \mathrm{W^{-1/2}m^{-1}}$. We set the pump pulse energy and intensity FWHM at 2 pJ and 1.32 nm, respectively, and the corresponding transfer function of a single PPLN is given in Fig. \ref{fig:single_pass_tf}. The output mean photon numbers in both the signal mode and the idler mode are 0.270.

To manifest the effect of TWOC, we simulate the average output photon number from the cascaded squeezers. For the device without TWOC (Fig. \ref{fig:TWOC}(a)), the interference pattern in the transfer function is determined by the phase mismatch integrated over the spacer, which is defined as
\begin{equation}\label{eq:ipm_a}
    \Delta\phi = \int dz \left[k_s(z, \bar{\omega}_s)+k_i(z, \bar{\omega}_i)-k_p(z, \bar{\omega}_p)\right].
\end{equation}
In the structure where TWOC is added, as shown in Fig. \ref{fig:TWOC}(b), the integrated phase mismatch is modified as 

\begin{align}\label{eq:ipm_b}
    \Delta\phi = &\int_\mathrm{primary} dz \left[k_i(z, \bar{\omega}_i)-k_p(z, \bar{\omega}_p)\right] \nonumber\\&+\int_\mathrm{auxiliary} dz k_s(z, \bar{\omega}_s),
\end{align}
where the integration is performed along the primary and auxiliary routing waveguides and summed up. The integrated phase mismatch can be controlled using local phase shifter, such as thermo-optic or electro-optic devices, which is highlighted in yellow in Fig. \ref{fig:TWOC}.

\begin{figure*}
\includegraphics[width=\textwidth]{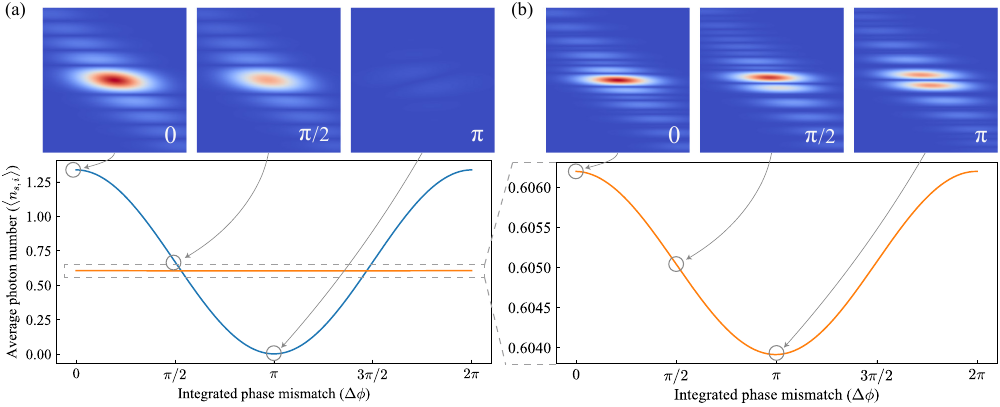}
    \caption{Interference of output photon numbers from a nonlinear interferometer (a) with and (b) without a walk-off compensator. The inset displays the transfer function $U^{s,i}$ for various values of integrated phase mismatch. The color scale is normalized based on the maximum value across the three figures in (a) and (b), respectively, following the same color scale as used in Fig. \ref{fig:single_pass_tf}.}\label{fig:interference}
\end{figure*}

We compare the two devices in Fig. \ref{fig:TWOC} by applying the same pump pulse energy and bandwidth that was previously used for the single PPLN. In Fig. \ref{fig:interference}(b), the interference of the device without TWOC shows a change in the mean photon number as a function of the integrated phase mismatch. The maximum and minimum output photon number of signal mode $\langle n_s \rangle$ are 0.606 and 0.604, respectively, yielding the interference visibility of output photons
\begin{equation}
V_1 = \frac{\langle n_s \rangle_\mathrm{max} - \langle n_s \rangle_\mathrm{min}}{\langle n_s \rangle_\mathrm{max} + \langle n_s \rangle_\mathrm{min}} = 0.002.
\end{equation}
In contrast to the previous case, the interferometer equipped with TWOC shows interference visibility $V_2=0.997$ (see Fig. \ref{fig:interference}(a)). The maximum output photon number is 1.339, which is more than four times that of a single squeezer, while the minimum photon number is 0.002.

The small visibility without TWOC can be explained by investigating the output transfer function. As we studied in Sec. \ref{subsec:nonlinear_interferometry}, the addition of the transfer function amplitude from the two squeezers is constructive or destructive depending on the wavelength (see Eq. (\ref{eq:nonlinear_interference})). Due to the fast alternation between destructive and constructive addition, the average intensity of the transfer function, which is the average output photon number as in Eq. (\ref{eq:output_photon_num}), almost remains the same even in the change of the integrated phase mismatch. Simply speaking, the change in the integrated phase mismatch hardly influences the average photon number. Mathematically, it is resulting from the fact that the cosine term in Eq. (\ref{eq:nonlinear_interference}) is strongly dependent on $(\omega_{s}, \omega_{i})$. On the other hand, with TWOC, whether the addition of amplitude is constructive or destructive does not depend on the $(\omega_{s}, \omega_{i})$, but depends only on the integrated phase mismatch $\Delta \phi$. Therefore, the average intensity of the output transfer function significantly varies with $\Delta \phi$, consequently yielding high visibility of the output photon number. 

The qualitative features regarding the transfer function $U^{s,i}$ we have discussed are observed. Without TWOC, the shape of $U^{s,i}$ strongly depends on $\Delta \phi$ as evident in Fig. \ref{fig:interference}(b), while the scale of its absolute value does not change much. On the contrary, with TWOC, the shape of $U^{s,i}$ does not change with $\Delta \phi$, while the scale of its absolute value changes dramatically for all the frequency range of interest. In other words, the high visibility can be attributed to the adequate TWOC as it recovers indistinguishability between the two squeezers such that the transfer function adds constructively or destructively over the entire bandwidth. For the discussion on the indistinguishability and interference, see Sec. \ref{subsec:twoc_timedomain}. Furthermore, the TWOC-equipped constructive interferometer generates photons more than four times the value produced from the single nonlinear waveguide. The scaling of photon numbers with respect to the number of squeezers is investigated in further detail in Sec. \ref{subsec:cascaded_squeezers}.

\subsection{Temporal walk-off compensation in time domain}
\label{subsec:twoc_timedomain}

We can understand the difference in interference visibility with and without TWOC using the availability of temporal which-path information. To begin with, we describe the condition for interference to occur when the optical intensity of the idler mode is measured.

Interference happens when idlers from two different regions are indistinguishable by all means. At first sight, idlers from two different regions seem already indistinguishable because their group velocity is matched to the group velocity of the pump. If we use the arrival time of pump pulse as a reference, it ensures their arrival time at the virtual detector after the nonlinear interaction is the same regardless of their origin. However, the interference visibility without TWOC was as low as 0.002.

This is because, \textit{in principle}, idler photons can be distinguishable due to the distinguishability of signal photons that are born in separate nonlinear regions. Without TWOC, we can distinguish signal photons born in different nonlinear regions by measuring arrival times with respect to the pump. The signal photons generated in the first nonlinear region arrive at the end of the second nonlinear region earlier than those generated in the second crystal when referenced to the pump. Because the signal and idler photons are entangled, from the which-path information encoded in the signal, we can distinguish in which region the idler photon was born \cite{kim2000, kim2000c}.

At this point, we can interpret the feature of TWOC in terms of induced coherence \cite{zou1991}. Induced coherence refers to the single-photon interference of the signal (idler) photon from two regions when the origin of the corresponding idler (signal) photon is fundamentally unknown by erasing which-path information \cite{chekhova2016}. We can interpret TWOC as it "induces" coherence on the idler photon by making signal modes in two regions common through delaying the signal born in the first crystal. This can be confirmed by the high interference visibility in the idler mode when TWOC applied.

\begin{figure}[h]
\includegraphics[width=\linewidth]{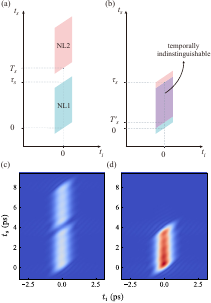}
    \caption{(a) Schematic of temporal transfer function for a double pass in a PPLN waveguide under aGVM conditions. (b) Same as (a), but with TWOC. (c) Temporal transfer function, $\tilde U^{s,i}(t_s,t_i)$, for the scenario in (a), which is 2D Fourier transform of the spectral transfer function in the first inset of Fig. \ref{fig:interference}(b). (d) Temporal transfer function, $\tilde U^{s,i}(t_s,t_i)$, for the scenario in (b), which is a 2D Fourier tranform of the spectral transfer function in the first inset of Fig. \ref{fig:interference}(a). The color scale for the transfer functions in (c) and (d) is normalized to the maximum value across each, employing the same color scale as in Fig. \ref{fig:single_pass_tf}.} \label{fig:temporal_tf}
\end{figure}

The effect is schematically expressed in Fig. \ref{fig:temporal_tf} using the temporal transfer function picture. The temporal transfer function $\tilde U^{s,i}(t_s,t_i)$ is a 2D Fourier transform of the spectral transfer function $U^{s,i}(\omega_s, \omega_i)$ which is calculated as follows:
\begin{subequations}
\begin{gather}
\label{eq:jta_ft}
    \text{SPDC:}\nonumber \\ \tilde U^{s,i}(t_s,t_i) = \cfrac{1}{2\pi} \int d\omega_s d\omega_i U^{s,i}(\omega_s, \omega_i) e^{i\omega_s t_s}e^{i\omega_i t_i}, \\
\text{QFC:}\nonumber \\
\tilde U^{s,i}(t_s,t_i) = \cfrac{1}{2\pi} \int d\omega_s d\omega_i U^{s,i}(\omega_s, \omega_i) e^{i\omega_s t_s}e^{-i\omega_i t_i}.
\end{gather}
\end{subequations}
The temporal transfer function provides the timing information of photon generation. Its amplitude at $(t_s, t_i)$ represents the signal (idler) photon was generated at $t_s$ ($t_i$) with respect to the reference timing $t=0$, which represents the peak of the pump pulse. In other words, we are using a moving timing reference that is fixed at the middle of the pump, and all the other timings are measured with respect to it. 

To study the effect of TWOC in the time domain, we examine the devices depicted in Fig. \ref{fig:TWOC}. We begin by performing a 2D Fourier transform on the spectral transfer function $U^{s,i}$ of these devices with an integrated phase mismatch $\Delta\phi=0$ as shown in Fig. \ref{fig:temporal_tf}. As the pump pulse propagates, the signal photon walks off and proceeds ahead. Such behavior manifests as an elongation of $\tilde U^{s,i}$ along the $t_s$-axis over time; for example, the non-zero $\tilde U^{s,i}$ amplitude at $t_s = t_0$ represents that the signal photon arrives at an observation point $t_0$ earlier than the peak of the pump. Therefore, a large value of $t_0$ is only possible when there is a significant walk-off. The elongation along $t_s$ axis stops at $\tau_s$, which is the amount of walk-off between the signal and pump modes over the entire first squeezer (see Fig. \ref{fig:temporal_tf}(a)). Meanwhile, there is no elongation along $t_i$ axis since there is no walk-off of the idler due to the aGVM condition. Therefore, it is almost synchronized with the pump, giving a relatively narrow width along $t_i$ axis. Still, the width along $t_i$ axis is non-zero due to the finite temporal width of the pump \cite{reddy2013}.

The signal photon's walk-off continues over the spacer between the two squeezers, giving $T_s$ walk-off at the start of the second squeezer. Over the length of the second squeezer, which is equivalent to the first one, the $\tau_s$ walk-off happens again, giving an elongation pattern from $T_s$ to $T_s + \tau_s$. Therefore, the dynamics without TWOC yields the temporal distinguishability between squeezers by revealing which-time information. Meanwhile, employment of TWOC could reverse the temporal walk-off ($T_s' < \tau_s)$, translating $\tilde U^{s,i}$ from the second squeezer downward along $t_s$ axis: when the amount of walk-off compensation $T_s - T_s'$ is more than $T_s - \tau_s$, two temporal transfer functions start overlapping. In general, when $T_s - T_s'$ is in between $T_s-\tau_s$ and $T_s+\tau_s$, the two islands of $\tilde U^{s,i}$s partially overlap and therefore indistinguishability is also partially recovered. When the amount of walk-off compensation is exactly $T_s$, two $\tilde U^{s,i}$s overlap completely, and hence the temporal indistinguishability is fully restored: the measurement of timing doesn't tell apart where the signal photon is generated between the two squeezers.

\subsection{Cascaded squeezers}
\label{subsec:cascaded_squeezers}
In the previous subsection, we showed that the temporal walk-off degrades the performance of cascaded squeezers and prohibits complete interference over the entire bandwidth. However, when the temporal walk-off is compensated, the interference revives and the average output photon number has more than quadrupled at constructive interference. In this subsection, we demonstrate that the squeezing parameter of the output field scales linearly with respect to the number of cascaded squeezers when temporal walk-off is properly compensated. In turn, as the photon number scales exponentially with respect to the squeezing parameter, we find that the output photon number scales exponentially with respect to the number of squeezing stages.

In theoretical research, Onodera \textit{et al.} showed a quadratic scaling of output photon number in the number of cascaded microresonators for SFWM in the low-gain regime \cite{onodera2016}. The quadratic scaling was achieved by keeping the constructive interference of microring resonators sharing a single bus waveguide. They found that the condition for the constructive interference over the photon bandwidth is that the pump pulse should be spectrally narrower than the resonance linewidth. When the pump pulse is spectrally broad, the phase of the pump pulse is significantly modified by the resonators, resulting in an destructive addition of the biphoton wavefunction. Also, note that the temporal walk-off was not a limiting factor in the study, because the signal, idler and pump modes are spectrally close in a typical SFWM.

\begin{figure}
\includegraphics[width=\linewidth]{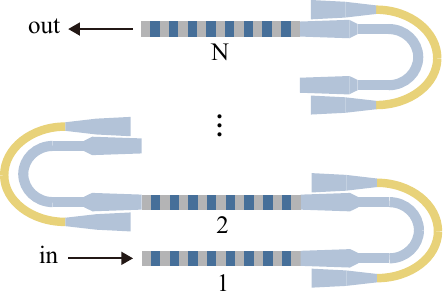}
    \caption{PPLN waveguides cascaded via TWOC.}
    \label{fig:nstages}
\end{figure}

In contrast, when it comes to three-wave mixing, where the group velocities of the participating modes are notably different, temporal walk-off plays a significant role. As we have demonstrated in the previous subsection, we could boost the coherence between multiple squeezers by compensating temporal walk-off. Furthermore, we achieved nearly superradiant photon number scaling with the two cascaded squeezers equipped with TWOC. Therefore, it is natural to question what would happen if we could cascade more than two squeezers where TWOC is attached to all spacers as shown in Fig. \ref{fig:nstages}.

To investigate the physics of such devices, we compare three distinct configurations: a fully walk-off compensating (FC) device; a partially walk-off compensating device; and a setup with no temporal walk-off compensation (NC). In the FC configuration, a delay line completely compensates for the temporal walk-off. In this case, the length of the adjustable delay line $L_d$ is 306.8 $\mathrm{\mu m}$. In the case of PC, $L_d$ is adjusted such that the signal mode arrives 2.00 ps earlier than the pump mode at the starting point of the second squeezer, yielding $L_d = 427.9\ \mathrm{\mu m}$. The configuration NC does not utilize TWOC; therefore, the time delay at the second squeezer is 4.01 $\mathrm{ps}$. The configurations of FC and PC are depicted in Fig. \ref{fig:TWOC}(b), and NC is illustrated in Fig. \ref{fig:TWOC}(a). The TWOC design employed here is the same as given in Sec. \ref{subsec:nli_twoc}.

\begin{figure*}
\includegraphics[width=\textwidth]{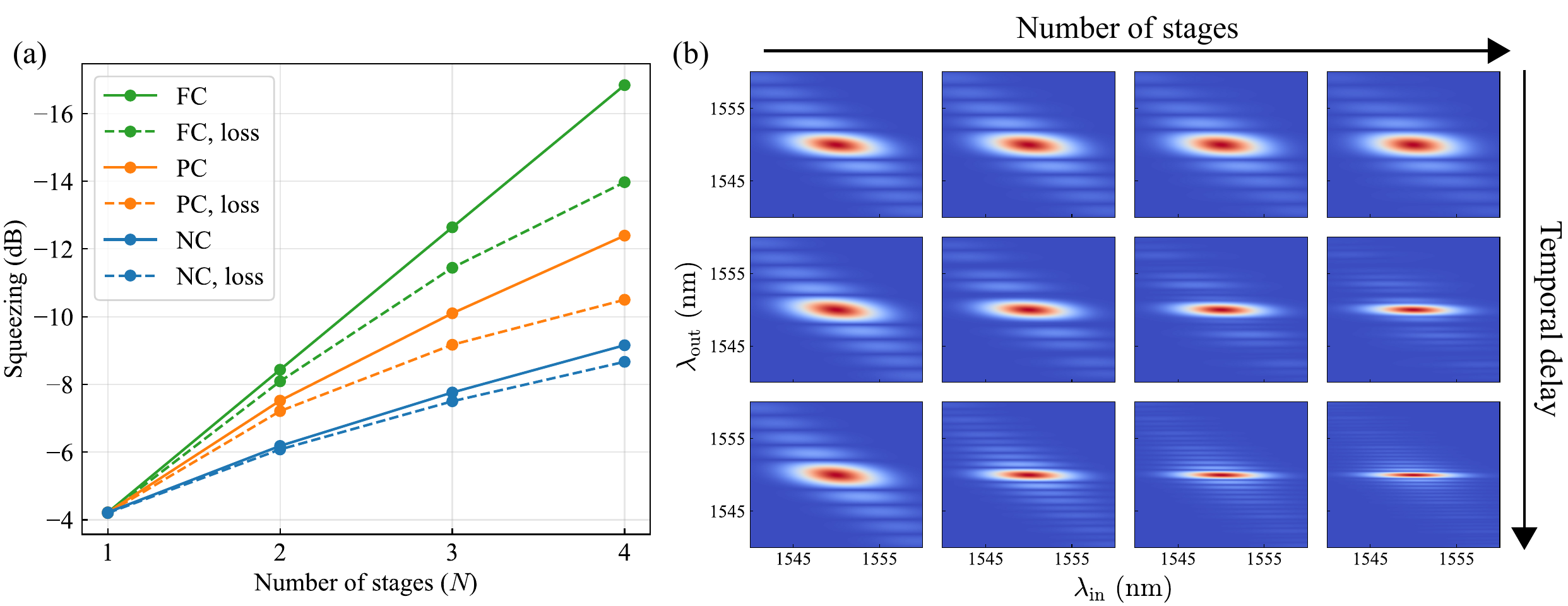}
    \caption{Scaling of squeezing parameters along multiple stages. (a) Scaling of squeezing parameters as the number of stages $N$ increases. The cascaded squeezers in the FC configuration shows linear scaling of squeezing parameters with the number of squeezers. In contrast, in the NC configuration, the growth of the squeezing parameter is much slower. The squeezing values are obtained under the consideration of realistic losses. (b) Cross-mode transfer function $U^{s,i}$ at each configuration with different stages in the lossless case. In the configuration FC, the spectral structure is retained, but the spectral bandwidth decreases with the number of stages in other configurations. The color scale for each density plot is normalized independently and follows the same color scale as used in Fig. \ref{fig:single_pass_tf}.
    } \label{fig:stage_scaling} 
\end{figure*}

Remarkably, in the configuration FC, the bandwidth of the output photons remained constant across all stages as we eliminated the temporal walk-off after each stage. Conversely, in the configurations of PC and NC, the output photon bandwidths shrank as the number of stages increased; this is a direct consequence of accumulated temporal walk-offs in each stage (see Fig. \ref{fig:stage_scaling}). In fact, the $N$-stage NC can be seen as the simple PPLN waveguide that is $N$ times longer than the PPLN of the single stage. For a single PPLN, the phase matching bandwidth is inversely proportional to the total length of the waveguide, which is the same as our argument for NC.

We further investigate the cascaded squeezers under the influence of loss. We assumed a propagation loss of 0.03 dB/cm, which is achieved with a monolithic lithium niobate waveguide \cite{zhang2017}. Also, expected losses from physical origins are illustrated in App. \ref{app:integrated_photonics} in detail. Along the squeezer, we iteratively applied propagation loss after a small nonlinear propagation and obtained the covariance matrix using Eq. (\ref{eq:cov_evol}). The loss in TWOC is applied at once after the linear evolution through TWOC because the phase evolution and optical loss can be applied in any order. The resulting covariance matrix is eigendecomposed to obtain the squeezing of the eigenmode with the largest squeezing \cite{roslund2014}. We find that the squeezing increases slower than in the case without loss, and the effect is more significant on the brighter device.

A key observation from our study is the linear increase in the squeezing parameter across different stages. This advancement significantly boosts nonlinear optical interactions, contributing to the development of squeezers of practical use such as heralded single-photon sources and bright squeezed vacuum generation. For example, a bright two-mode squeezed vacuum can be exploited as a source for several CV quantum information processing, e.g., Gaussian boson sampling \cite{zhong2020, arrazola2021}, cluster state generation \cite{larsen2019,asavanant2019}, and quantum teleportation \cite{takeda2013,he2022}.

\subsection{Quantum pulse gate}

In this subsection, we explore another unique application of TWOC for a practical purpose: the quantum pulse gate (QPG). We analyze the limitations of a QPG device in a high-gain regime where the time-ordering and third-order nonlinear effects are evident. After that, we apply TWOC to a QPG device to overcome such limitations and achieve high performance metrics. 

A QPG is a device designed to manipulate and detect quantum information encoded in spectro-temporal modes. The spectro-temporal modes are defined in the energy or frequency degree of freedom to represent quantum information. Spectro-temporal modes can be encoded by the SPDC process \cite{ansari2018a}, and take several advantages compared to a photon's other degrees of freedom: (i) Different temporal modes can be in the same spatial degree of freedom, thereby the quantum information can be transmitted through waveguides and existing single-mode fiber networks; (ii) The dimension of the Hilbert space where the temporal modes live is unbounded in principle, allowing the high-dimensional encoding; (iii) The quantum information in spectro-temporal modes is robust against linear dispersion and polarization rotation \cite{brecht2015, reddy2015, ansari2018b}. The QPG plays a crucial role in quantum information science utilizing the spectro-temporal mode, and improving its performance is an important task. 

The QPG basically utilizes the QFC process, which is characterized by the cross-mode transfer function $U^{s,i}(\omega, \omega')$. As expressed in Eq. (\ref{eq: qfc_sch_decom}), we perform Schmidt decomposition of the transfer function, which gives input Schmidt modes $\tau_i^{(l)}$, output Schmidt modes $\rho_s^{(l)}$, and Schmidt coefficients $\sin(r_l)$ for each mode labeled $l$. The Schmidt coefficient can be interpreted as follows: when a single photon with spectral profile $\tau_i^{(l)}$ enters the QPG, it is converted into a photon with spectral profile $\rho_s^{(l)}$ with a conversion efficiency $\eta_l = \sin^2(r_l)$. Note that we label the Schmidt modes in decreasing order of conversion efficiency.

An ideal QPG device converts only the target spectro-temporal mode with unit efficiency. The performance can be described using two important figures of merit: separability and selectivity, as defined in App. \ref{app:quantity}. The separability $\sigma_j$ characterizes the ability of a QPG device to discriminate a mode $j$ from other modes, where unity is obtained only when $\eta_k(k\ne j)$ is zero and $\eta_j$ is non-zero. The selectivity simultaneously quantifies the ability of mode discrimination and its conversion efficiency, expressed as the multiplication of the separability of dominant mode $\sigma_1$ and its conversion efficiency $\eta_1$. Achieving unity in selectivity is possible only when the conversion efficiency of the dominant Schmidt mode $\eta_1$ is one, whereas those of all other Schmidt modes are zero. 

Moreover, for efficient conversion, it is crucial that the spectrum of the target photon matches the first input Schmidt mode, $\tau_i^{(1)}$. Thus, the primary engineering objectives are twofold: (i) achieve near-unity selectivity; and (ii) align $\tau_i^{(1)}$ with the target photon's spectrum. The condition of aGVM is beneficial in this context because it provides high separability, and $\tau_i^{(1)}$ can be readily engineered by shaping the spectral shape of the pump pulse. Under aGVM condition where the group velocities of pump and input (idler) modes match, $\eta_1$ becomes dominant and $\tau_i^{(1)}$ resembles the pump spectral shape \cite{mejling2012a, reddy2017a}.

Given the high separability achieved using aGVM condition, the conversion efficiency of the first Schmidt mode needs to be increased to achieve high selectivity. In the low-gain regime, an increase in the pump power predominantly boosts the conversion efficiency of the first Schmidt mode, leaving the other modes largely unaffected. However, at some point, the conversion efficiencies of undesired modes $\eta_j(j\ne 1)$ start to increase faster than $\eta_1$, resulting in lowered selectivity in a high conversion regime. Such a phenomenon limits the single stage selectivity to around 0.8 \cite{reddy2017a}. Such constraint can be attributed to the distortion of the transfer function due to the time-ordering effect, which is the inherent feature of nonlinear dynamics.

\begin{figure*}
\includegraphics[width=\textwidth]{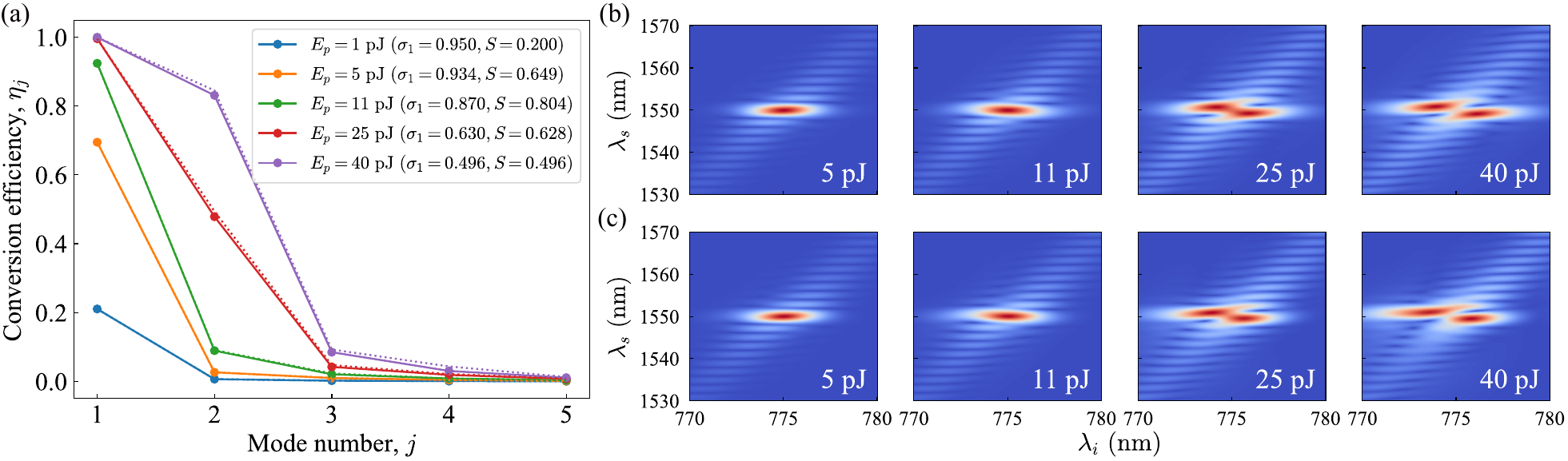}
    \caption{(a) Conversion efficiency of five dominant Schmidt modes on various pump energies. The dotted lines and solid lines represent conversion efficiency with and without a third-order nonlinear effect, respectively. The separability and selectivity corresponding to each pump energy is indicated in the legend. (b), (c) Cross-mode transfer function $U^{s,i}$ of QPG at different input pump energies with and without a third-order nonlinear effect. The color scale for each density plot is normalized independently and follows the same color scale as used in Fig. \ref{fig:single_pass_tf}.
    } \label{fig:qfc_timederdering}
\end{figure*}

We observe this effect from our simulation result as we increase the input pulse energy (see Fig. \ref{fig:qfc_timederdering}). We simulated a single PPLN waveguide in an aGVM condition. The pump and input (idler) modes propagate with the same group velocity while the output (signal) mode goes with a different group velocity. The PPLN waveguide has the same geometry as in the previous subsection \ref{subsec:nli_twoc}: a length of 6 mm, a width of 890 nm, a film thickness of 500 nm, an etch depth of 300 nm, and a sidewall angle of 68 degrees with air-cladding. In such a QPG device, we induce DFG interaction by pumping TM0 mode at 1550 nm wavelength with a Gaussian spectrum and an intensity FWHM of 8.33 nm, satisfying energy conservation $\omega_i - \omega_p = \omega_s$. It converts a photon in an idler mode, TM0 mode at 775 nm wavelength, into a photon in a signal mode, TE0 mode at 1550 nm wavelength with a poling period of 2.262 $\mu$m. The combination of modes yields the nonlinear coupling coefficient $\gamma_\mathrm{QFC}= -258.6 \ \mathrm{W^{-1/2}m^{-1}}$. As the pump energy is increased, distortion of the transfer function is observed (see Fig. \ref{fig:qfc_timederdering}(b)), causing the selectivity to peak at 0.80 with a pump energy of 11.2 pJ; this result matches the findings in \cite{reddy2013} very well. 

We further analyze the effect of the third-order nonlinearities, such as SPM and XPM, which may affect the spectral properties of the QPG device. To simulate such effects, we calculate nonlinear coupling coefficients from Eq. (\ref{eq:nonlin_coef}), resulting in $\gamma_\mathrm{SPM} = 0.97\  \mathrm{W^{-1}m^{-1}}$, $\gamma_{\mathrm{XPM}, s} = 0.93 \mathrm{\ W^{-1}m^{-1}}$, and $\gamma_{\mathrm{XPM}, i} = 5.22 \ \mathrm{W^{-1}m^{-1}}$. The components of the $\chi^{(3)}$ tensor used in the calculation is detailed in App. \ref{app:sec:nonlinear_susceptibility}. The third-order nonlinear effects result in asymmetry in the transfer function as shown in Fig. \ref{fig:qfc_timederdering}(c). At first glance, the third-order effects appear to significantly impact selectivity; however, both conversion efficiency and selectivity are not affected seriously as shown in Fig. \ref{fig:qfc_timederdering}(b). We found that the selectivities undergo changes less than 2\% with and without third-order effects.

\begin{figure}[hbt!]
\includegraphics[width=\linewidth]{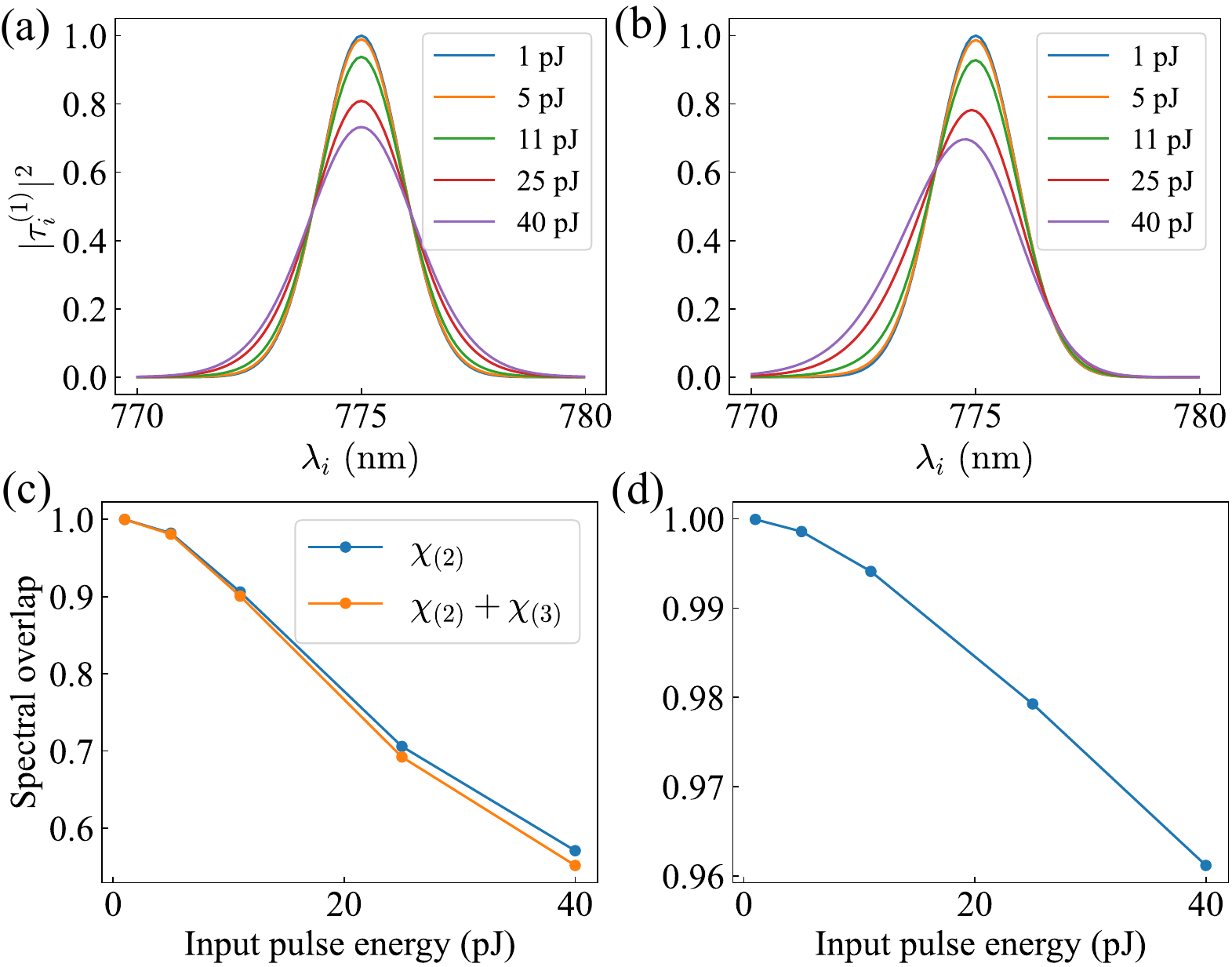}
    \caption{Spectral intensity of the first input Schmidt modes (a) without 3rd order nonlinear effects, and (b) with third-order nonlinear effects across different pulse energies. (c) Spectral mode overlap between the input Schmidt mode at a pulse energy of 1 pJ and at higher energies. In the data labeled $\chi_{(2)}$ ($\chi_{(2)}+\chi_{(3)}$), we consider DFG interaction alone (DFG interaction combined with third-order nonlinearities). (d) Spectral overlap between the input Schmidt modes in (a) with and in (b) without third-order nonlinearities, across different pulse energies.
    } \label{fig:qfc_skewing}
\end{figure}

To match the target photon spectrum and the first input Schmidt mode, precise calculation of the input Schmidt mode $\tau_i^{(1)}$ is the first step. We point out that the first-order perturbation fails to account for high-gain effects in the calculation of the input Schmidt mode spectrum, and therefore a rigorous simulation beyond the first-order approach is required. To quantify the matching, we define the spectral overlap between two normalized spectra $\phi(\omega)$ and $\psi(\omega)$ as   
\begin{equation}
O(\phi, \psi) \equiv \left| \int d\omega \phi^*(\omega)\psi(\omega) \right|^2.
\end{equation}
Then, the conversion efficiency of the input photon with spectral profile $\phi_\mathrm{in}$ into a first output Schmidt mode is a product of $\eta_1$ and spectral mode overlap between $\phi_\mathrm{in}$ and $\tau_i^{(1)}$ \cite{brecht2014}:
\begin{equation}\label{eq:qpg_actual_ce}
    O(\phi_\mathrm{in}, \tau_i^{(1)})\eta_1.
\end{equation}
Therefore, not only achieving high selectivity but also engineering the first input Schmidt mode is essential to an efficient QPG device. Based on this, we analyzed the time-ordering and third-order nonlinear effects in the first input Schmidt mode. We plot the spectrum of $\tau_{i}^{(1)}$ with and without third-order nonlinearities in Fig. \ref{fig:qfc_skewing}. For the large pump energy, from 25 pJ and above, there is a noticeable broadening and skewing of $\tau_{i}^{(1)}$, which is significantly different from that at the low powers.

To quantify the distortion of the first input Schmidt mode and assess its impact on the conversion efficiency as defined in Eq. \ref{eq:qpg_actual_ce}, we start with the assumption that the first input Schmidt mode in the low-gain regime is perfectly aligned with the input photon spectrum. This implies that the input photon spectrum coincides with $\tau_i^{(1)}$ at an input pulse energy of 1 pJ. We then calculate the overlap between $\tau_{i}^{(1)}$ at higher pump energies and the input photon spectrum, discovering a significant decrease in overlap as pulse energy is increased; it is attributable to the spectral broadening, as depicted in Fig. \ref{fig:qfc_skewing}. This observation underscores the importance of accounting for the time-ordering effect when designing an efficient QPG device aimed at a specific spectrum. Furthermore, we explore the impact of third-order nonlinear effects by calculating the overlap between the input photon spectrum and the first input Schmidt mode $\tau_i^{(1)}$ including third-order nonlinearity, noting a marked decrease in mode overlap primarily due to the time-ordering effect (see Fig. \ref{fig:qfc_skewing}(c)). 

We further investigate the sole contribution of third-order nonlinearities by comparing results with and without it. To achieve this goal, we calculated the overlap of the two first input Schmidt modes $\tau_i^{(1)}$s, each extracted from the simulation with $\chi^{(3)}$ turned on and off. As shown in Fig. \ref{fig:qfc_skewing}(d), the impact of third-order nonlinearities is relatively small compared to the time-ordering effect. However, note that careful consideration of the third-order nonlinearity is necessary if very high performance of the QPG is required.

In what follows, we further investigate QPG with and without TWOC in integrated photonics settings. Previously, Reddy \textit{et. al.} proposed the multi-stage QPG to mitigate the time-ordering effect and break the selectivity limit of a single-stage QPG \cite{reddy2015}. Afterwards, they experimentally showed that such a scheme is indeed helpful in achieving high-selectivity beyond the single-stage limit \cite{reddy2018}. In a multi-stage configuration, it is possible to attain near unit efficiency in the first Schmidt mode using lower pump power than that would be required for a single-stage scenario. Under these circumstances, the required pump power is less demanding, and the system is free from the detrimental time-ordering effect. Therefore, the multi-stage configuration is crucial for achieving high selectivity. Interestingly, the principles of \cite{reddy2015} are the same as those of TWOC studied in subsection \ref{subsec:nli_twoc}; in fact, our TWOC proposal was inspired by this previous work, applying it to the integrated photonics platform.

\begin{figure*}
\includegraphics[width=\textwidth]{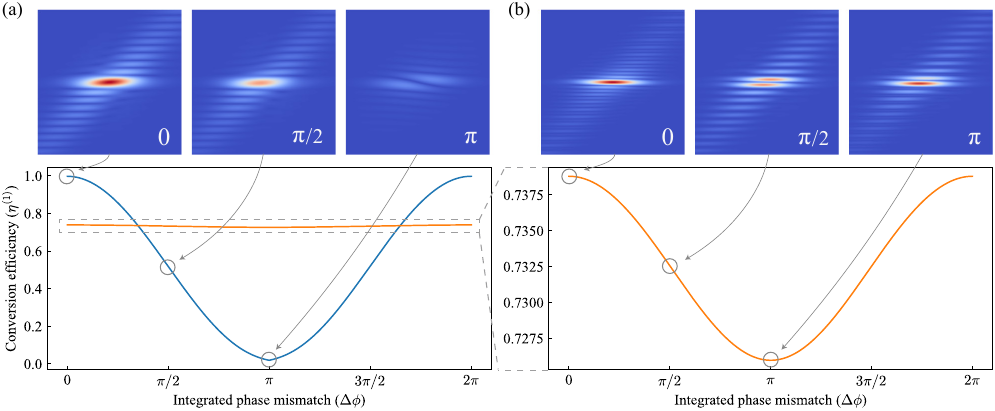}
    \caption{Interference of the first Schmidt mode conversion efficiency, $\eta^{(1)}$, in a two-stage QPG: (a) with TWOC and (b) without TWOC. The inset displays the transfer function $U^{s,i}$ for various values of integrated phase mismatch. The color scale is normalized based on the maximum value across the three figures in (a) and (b), respectively, following the same color scale as used in Fig. \ref{fig:single_pass_tf}.} \label{fig:qpg_interference}
\end{figure*}

To show the utility of TWOC in a two-stage QPG scenario, we connected two PPLN waveguides as in Fig. \ref{fig:TWOC}, and compared the results with and without TWOC. As a reference, we first pumped the single-stage QPG composed of the same 6-mm-long PPLN waveguide, where we used a pump of energy 2.8 $\mathrm{pJ}$ and intensity FWHM 8.33 nm as before, providing $\eta_1 = 0.49$. When we connect two PPLNs as in Fig. \ref{fig:TWOC}, an interference of conversion efficiency along with the integrated phase mismatch $\Delta \phi$ is observed. Without TWOC, the visibility of $\eta_1$ is observed to be $V_\mathrm{1} = 0.01$ as shown in Fig. \ref{fig:qpg_interference}(b). The maximum conversion efficiency was $\eta_1= 0.74$, and the maximum selectivity $S=0.707$ was achieved at a constructive interference. In contrast, with TWOC, we found $V_\mathrm{2}=0.963$ where the maximum conversion efficiency of $\eta_1=0.998$, and the maximum selectivity $S = 0.912$ beyond the single stage limit is obtained at the constructive interference as illustrated in Fig. \ref{fig:qpg_interference}(a). Similarly to the cascaded squeezers studied in Sec. \ref{subsec:cascaded_squeezers}, the parameter $r_1$ in the first Schmidt coefficient experiences an almost linear increase at constructive interference $\Delta \phi = 0$, thus the maximum conversion efficiency is approximately
\begin{equation}
\eta_1^\mathrm{2-stage}(\Delta \phi = 0) \simeq \sin^2(2r_1) = 4\eta_1(1-\eta_1).
\end{equation}
Note that the relation is not exact because there exists slight mismatch between the output Schmidt mode of the first stage and the input Schmidt mode of the second stage even under exact temporal walk-off compensation \cite{reddy2015}. Similarly to the cascaded squeezers, once again, the variation in interference visibility with and without TWOC can be attributed to the presence or absence of fast oscillation in the joint spectrum as illustrated in Fig. \ref{fig:qpg_interference}. Judging from the various simulation results obtained so far, we confirm that the scheme proposed in \cite{reddy2015} was well adapted to an integrated photonics platform, thereby overcoming the limitations of single-stage selectivity.

Furthermore, by utilizing a QPG equipped with TWOC, we can achieve a very high selectivity exceeding 0.99 through extending the device's length and applying the poling optimization technique described in App. \ref{app:poling}. To examine the impacts of increased length and apodized poling, we analyze four device configurations: (a) a single-stage QPG with a 6 mm PPLN waveguide, as previously introduced; (b) a two-stage QPG with a 6 mm PPLN and TWOC; (c) a two-stage QPG with an 18 mm PPLN and TWOC; (d) a two-stage QPG with 18 mm apoLN and TWOC. The pump energy involved in each configuration is adjusted to yield maximum selectivity. Transitioning from (b) to (c), the extended length ensures enhanced separability owing to a higher aspect ratio of the transfer function, as detailed in \cite{reddy2013}. Moving from (c) to (d), apodized poling mitigates the sidelobes in the transfer function, thereby increasing selectivity. The stepwise enhancements are documented in Table \ref{tab:selectivity}. Especially in the configuration (d), the selectivity reaches 0.997, which is comparable with the theoretical value of four-stage scheme with TWOC studied in \cite{reddy2015}. The apodized poling could effectively reduce the required number of nonlinear stages, thereby making it more realistically implementable.

\begin{table}[]
\caption{The selectivity of QPG with different configurations. (a) single-stage QPG with a 6 mm PPLN waveguide (b) two-stage QPG with two PPLNs of a length 6 mm (c) two-stage QPG with two PPLNs of a length 18 mm. (d) two-stage QPG with two apoLNs of a length 18 mm.
}
\begin{ruledtabular}
\begin{tabular}{cccc}
                            & $E_p \ \mathrm{(pJ)}$ & $\sigma_1$& $S$    \\
\hline
(a)                         &   11  &   0.870&   0.804 \\
(b)                         &  2.8 &   0.914&   0.912\\
(c)                         & 0.9  &   0.967&   0.965\\
(d)                         & 2.1  &   0.997&   0.997
\end{tabular}
\end{ruledtabular}
\label{tab:selectivity}
\end{table}

\section{Conclusion}
In this work, we introduced the simulation framework for integrated nonlinear quantum photonics. Building upon the previous foundational works \cite{triginer2020, quesada2020}, we extended the model to include the slowly changing waveguide in an adiabatic limit. Furthermore, we connect our simulation formalism to Gaussian optics, enabling photodetection in experimental scenarios. To verify the simulation, we compared our simulation results with the conventional calculation method both in low- and high-gain regimes. In addition, for the purpose of showing the utility of our simulation framework, we proposed TWOC, which can be applied to various scenarios in nonlinear quantum optics, including bright squeezing and QPG. Through accurate prediction of the performance of devices equipped with TWOC, we proved the functionality of our framework to simulate complex circuits, including typically used linear optical components in integrated photonics. For the complex devices covered in our work, it is not straightforward to calculate only with analytic methods.

In summary, our framework enables the design and simulation of nonlinear quantum optical devices in an integrated optics context. Its functionalities include the followings, but not exhaustively: (i) high-gain effects such as time-ordering, SPM, and XPM resulting from the large optical confinement; (ii) adiabatically changing structures including taper and curve that can be used for multiple purposes such as mode conversion or optical routing; and (iii) the implementation of detectors and linear losses by making connections with a Gaussian optics framework. With these key functionalities, our simulator can serve as an efficient toolbox for designing complex quantum optical circuits such as TWOC devices.

Utilizing our simulation framework, we demonstrated the significant potential of the chip-scale TWOC technique across various scenarios. We showcased its effectiveness in designing a bright squeezer and a highly selective QPG. In cascaded squeezers, we observed a linear increase in the squeezing parameter with the number of stages while maintaining the spectral shape of the output mode across the stages. Despite the impact of realistic losses, we established that incorporating TWOC still offers substantial benefits. For QPGs, we identified that accounting for high-gain effects is essential for high efficiency in photon conversion, with TWOC and poling apodization enabling near-unity selectivity. This indicates that TWOC allows for the enhancement of nonlinear interactions without suffering from detrimental effects such as time-ordering, SPM, and XPM.

In the future, we expect our simulation framework would include other physics models that affect the nonlinear optical process. Notable examples are two-photon absorption, free-carrier absorption in silicon, broadband Raman noise in silicon nitride, and photorefraction in lithium niobate. Inclusion of these features will allow us to estimate the limitations and potentials of each material platform. Furthermore, models for handling fabrication imperfections in integrated photonics can be added \cite{chen2024}, enhancing our simulator to become a powerful toolbox that can predict most of the conceivable experimental metrics in integrated quantum photonics.

\begin{acknowledgments}
The authors thank Prof. Heedeuk Shin and co-authors for sharing experimental data of \cite{shin2023}. This work was supported by the National Research  Foundation of Korea Grant funded by the Korean Government(NRF-2021R1C1C1006400, NRF-2022M3K4A1094782, NRF-2022M3E4A1077013, NRF-2022M3H3A1085772).
\end{acknowledgments}

\appendix
\section{Observable quantities}
\label{app:quantity}
\paragraph{Squeezing parameter}From the simulation, we routinely obtain transfer functions that represent the unitary evolution of input photon modes to the output mode after propagating in nonlinear waveguides. From the transfer function, various quantities can be deduced. The transfer function of squeezing process, including SPDC and SFWM, can be decomposed into Schmidt modes \cite{christ2013, quesada2018}:
\begin{subequations}
\begin{align}
U^{s, s}\left(\omega, \omega^{\prime} ; z, z_0\right) & =\sum_l \cosh \left(r_l\right)\left[\rho_s^{(l)}(\omega)\right]\left[\tau_s^{(l)}\left(\omega^{\prime}\right)\right]^*, \\
U^{s, i}\left(\omega, \omega^{\prime} ; z, z_0\right) & =\sum_l \sinh \left(r_l\right)\left[\rho_s^{(l)}(\omega)\right]\left[\tau_i^{(l)}\left(\omega^{\prime}\right)\right], \\
{\left[U^{i, i}\left(\omega, \omega^{\prime} ; z, z_0\right)\right]^* } & =\sum_l \cosh \left(r_l\right)\left[\rho_i^{(l)}(\omega)\right]^*\left[\tau_i^{(l)}\left(\omega^{\prime}\right)\right], \\
{\left[U^{i, s}\left(\omega, \omega^{\prime} ; z, z_0\right)\right]^* } & =\sum_l \sinh \left(r_l\right)\left[\rho_i^{(l)}(\omega)\right]^*\left[\tau_s^{(l)}\left(\omega^{\prime}\right)\right]^*,
\end{align}
\end{subequations}
where $r_l$ is the squeezing parameter, $\rho^{(l)}_j$ is the output Schmidt mode, and $\tau^{(l)}_j$ is the input Schmidt mode of the transfer function \cite{quesada2020}. The order of Schmidt modes is in decreasing order of the squeezing parameter $r_l$.
These Schmidt modes are orthogonal and complete:
\begin{subequations}
\begin{align}
\int d \omega \rho_s^{(l)}(\omega)\left[\rho_s^{\left(l^{\prime}\right)}(\omega)\right]^*  &=\delta_{l, l^{\prime}}, \\
\sum_l \rho_s^{(l)}(\omega)\left[\rho_s^{(l)}\left(\omega^{\prime}\right)\right]^*  &=\delta\left(\omega-\omega^{\prime}\right).
\end{align}
\end{subequations}

To derive the squeezing parameters from lossy waveguides, we employed eigendecomposition to examine the output covariance matrix. The eigenvalue represents the variance, while the associated eigenvector represents the spectral field of the mode. The minimum and maximum eigenvalues correspond to the variances of the squeezing and anti-squeezing quadratures, respectively. Given that the resultant state is a multi-mode two-mode squeezed vacuum, it features two degenerate modes characterized by identical eigenvalues \cite{roslund2014}.

\paragraph{Spectral purity}From the distribution of Schmidt coefficients, Schmidt number $K$, the number of temporal mode pairs, can be obtained as \cite{ansari2018b}:
\begin{equation}
    K = \frac{\left( \sum_l \sinh(r_l)^2 \right)^2}{\sum_l \sinh(r_l)^4},
\end{equation}
and from this, spectral purity $\mathcal{P}$ is obtained as the follows
\begin{equation}
    \mathcal{P} = {1/K}.
\end{equation}

\paragraph{Average photon number}The average photon number of the signal mode using the input-output relation in Eq. (\ref{eq:inout}) is
\begin{align}\label{eq:output_photon_num}
    \left\langle N_s\right\rangle &=\int d \omega\left\langle a_s^{(\mathrm{out}) \dagger}(\omega) a_s^{(\mathrm{out})}(\omega)\right\rangle \\ &=  \int d\omega d\omega' \left|U^{s,i}(\omega, \omega')\right|^2 =\sum_l \sinh ^2\left(r_l\right),\nonumber
\end{align}
where the third equality holds in the lossless case. 

\paragraph{Separability and selectivity}For the QFC process, the Schmidt decomposition yields the following \cite{christ2013}:
\begin{subequations}
\begin{align}
\label{eq: qfc_sch_decom}
U^{s, s}\left(\omega, \omega^{\prime} ; z, z_0\right) & =\sum_l \cos \left(r_l\right)\left[\rho_s^{(l)}(\omega)\right]\left[\tau_s^{(l)}\left(\omega^{\prime}\right)\right]^*, \\
U^{s, i}\left(\omega, \omega^{\prime} ; z, z_0\right) & = -\sum_l \sin \left(r_l\right)\left[\rho_s^{(l)}(\omega)\right]\left[\tau_i^{(l)}\left(\omega^{\prime}\right)\right]^*, \\
{U^{i, i}\left(\omega, \omega^{\prime} ; z, z_0\right) } & =\sum_l \cos \left(r_l\right)\left[\rho_i^{(l)}(\omega)\right]\left[\tau_i^{(l)}\left(\omega^{\prime}\right)\right]^*, \\
{U^{i, s}\left(\omega, \omega^{\prime} ; z, z_0\right) } & =\sum_l \sin \left(r_l\right)\left[\rho_i^{(l)}(\omega)\right]\left[\tau_s^{(l)}\left(\omega^{\prime}\right)\right]^*,
\end{align}
\end{subequations}

The index of Schmidt modes is given in the decreasing order of $r_l$. The conversion efficiency of $j$-th Schmidt mode is given by $\eta_j = \sin^2(r_j)$. In particular, when the QFC process is used in a QPG context, there are two important figures of merit: selectivity and separability. The selectivity of the QPG device is 
\begin{equation}
    S = \eta_1\cdot \frac{\eta_1}{\sum_{k=1}^\infty \eta_k},
\end{equation}
and the separability for the $j$-th Schmidt mode among the $N$ modes is defined as 
\begin{equation}
\sigma_j = \frac{\eta_j}{\sum_{k=1}^{N} \eta_k}.
\end{equation}

\section{Spontaneous four-wave mixing}
\label{app:sfwm}
We simulated the SFWM process in the silicon nonlinear waveguide to reproduce experimental results in lossy structures. Here, we introduce the EOM of non-degenerate dual pump SFWM, ignoring the effect of XPM. Although it is usual to choose a degenerate pump and a non-degenerate photon pair or vice versa, we first assumed that they are all different in its central frequencies. The EOM for the usual case can be obtained by applying an appropriate limit to the general equation. 
\paragraph{Pump dynamics}In the reference frame of the first pump mode denoted as $p1$, the EOM of the first pump can be written as
\begin{equation}
\begin{split}
\frac{\partial}{\partial z} \beta_{p1}(z,\omega) =i\frac{\gamma_\mathrm{SPM}(z)}{2\pi}\int d\omega' \mathcal{E}_{p1}(\omega - \omega') \beta_{p1} (z,\omega').
\end{split}
\end{equation}
Similarly, in the frame of second pump mode $p2$, the EOM of the second pump mode can be obtained simply by substituting $p1$ into $p2$.

\paragraph{Photon-pair dynamics}We choose to work in the reference frame of $p1$, and in this frame of reference, the EOM of output photon modes can be written as
\begin{equation}
\begin{split}
\frac{\partial}{\partial z} a_s (z,\omega)& =
i(\omega - \bar{\omega}_s) \left(\frac{1}{v_s} -\frac{1}{v_{p1}}\right)a_s(z,\omega)
\\
&+ 
i \frac{\gamma_\mathrm{SFWM}}{2\pi} \int d\omega' B_p (z, \omega + \omega') a_i ^\dagger (z,\omega' ),\\
\frac{\partial}{\partial z} a_i (z,\omega)& =
i(\omega - \bar{\omega}_i) \left(\frac{1}{v_i} -\frac{1}{v_{p1}}\right)a_s(z,\omega)
\\
&+ 
i \frac{\gamma_\mathrm{SFWM}}{2\pi} \int d\omega' B_p (z, \omega + \omega') a_s ^\dagger (z,\omega' ),
\end{split}
\end{equation}
where
\begin{equation}
\begin{split}
B_p(z,\omega&) = \int d\omega' \beta_{p1}(z, \omega-\omega')\beta_{p2}(z,\omega')\\ 
&\times\exp\bigg\{-i\int dz' \bigg[ (\frac{1}{v_{p1}(z')} - \frac{1}{v_{p2}(z')})(\omega'-\bar{\omega}_{p2}) \\ 
&\qquad\qquad+ \Delta \bar{k}_\mathrm{SFWM}(z') \bigg] \bigg\};
\end{split}
\end{equation}

\section{Nonlinear optical susceptibility}
\label{app:sec:nonlinear_susceptibility}
A nonlinear polarization $\vec{P}$ is expressed in a power series of an electric field $\vec{E}$ as 
\begin{equation}
\label{eq:pol_density}
P^i = \epsilon_0 \chi_{(2)}^{ijk}E^jE^k + \epsilon_0 \chi_{(3)}^{ijkl}E^jE^kE^l + \cdots,
\end{equation}
where $\chi_{(2)}^{ijk}$ and $\chi_{(3)}^{ijkl}$ are the nonlinear susceptibility tensors. To estimate the nonlinear coupling strength of various processes, it is crucial to find the nonlinear tensor components. We provide how we obtained these values and how we calculate nonlinear coupling coefficients from the information of optical modes and tensors.

\subsection{Second order susceptibility}
In this work, we simulated a nonlinear process in a 5\% MgO-doped lithium niobate waveguide. The elements of the effective second order nonlinearity tensor $d$, defined as $d = \frac{1}{2}\chi_{(2)}$, had been obtained from the second-harmonic generation (SHG) experiment with the fundamental mode at 1064 nm \cite{shoji1997}. The non-zero elements of contracted $d$ matrix are $d^{33} = -25 \ \mathrm{pm/V}$, $d^{31} = -4.4 \ \mathrm{pm/V}$, and $d^{22} = 2.2 \ \mathrm{pm/V}$. These values were used to obtain the nonlinearity matrix at fundamental mode 1550 nm using Miller's rule: 
\begin{equation}
\label{eq:miller_rule}
\cfrac{d^{ijk}(2\omega; \omega, \omega)}{(\epsilon_1^i(2\omega)-1)(\epsilon_1^j(\omega)-1)(\epsilon_1^k(\omega)-1)} = \text{constant}
\end{equation}
 The estimated $d$ matrix elements at 1550 nm wavelength are: $d^{33} = -22.6 \ \mathrm{pm/V}$, $d^{31} = -3.9 \ \mathrm{pm/V}$, and $d^{22} = 2.0 \ \mathrm{pm/V}$. 

\subsection{Third order susceptibility}
The LN is a 3m-class crystal, which its $\chi_{(3)}$ has 37 nonzero elements with 14 independent variables. Despite the numerous researches regarding Kerr nonlinearity, the entire $\chi_{(3)}$ elements are still ambiguous for LN, especially MgO:LN, which is yet rarely studied \cite{stegeman1997, kozub2023}. In this work, we simulated a third-order nonlinear process by estimating the values from known information. We assumed that the MgO doping does not significantly affect the $\chi_{(3)}$ values. 

Recently, the nonlinear refractive index $n_2$ of LN was measured through the x-axis and z-axis at 1550 nm wavelength (i.e., $n_2^{xxxx}$ and $n_2^{zzzz}$) \cite{shams-ansari2022}. Reported values are $n_2^{xxxx} = 1.61\times 10^{-19} \ \mathrm{m^2/W}$ and $n_2^{zzzz} = 1.74\times 10^{-19} \ \mathrm{m^2/W}$. It is well known that $n_2$ and $\chi_3$ follow the relation Eq. (\ref{eq:n2_chi3}) \cite{boyd2019}. Since the refractive indices along the $x$- and $z$- axis are $n^x(\text{1550 nm}) = 2.208$ and $n^z(\text{1550 nm}) = 2.13$, the estimated diagonal components are $\chi_{(3)}^{xxxx} = \chi_{(3)}^{yyyy} = 2779 \ \mathrm{pm^2/V^2}$ and $\chi_{(3)}^{zzzz} = 2795 \ \mathrm{pm^2/V^2}$. 
\begin{equation}
\label{eq:n2_chi3}
n_2 (\mathrm{m^2/W}) = \cfrac{282.55}{n^2}\chi_{(3)}(\mathrm{m^2/V^2})
\end{equation}

Throughout our entire works, even though it is not fully verified by experimentally measured values, we assume that $\chi_{(3)}$ is dispersion-free, nearly isotropic, and it satisfies the full-permutation symmetry. Diagonal components are set to be $\chi_{(3)}^{xxxx} = \chi_{(3)}^{yyyy} = \chi_{(3)}^{zzzz}$ and only non-diagonal components such as $\chi_{(3)}^{xxyy}=\chi_{(3)}^{xxzz} = \chi_{(3)}^{xyxy} = ...$ are considered. We set the diagonal components to be the average of the estimated value from the reported work, $2787 \ \mathrm{pm^2/V^2}$. If the material is considered to be near isotropic, non-diagonal components can be determined by the relation $\chi_{(3)}^\mathrm{non-diag} = \chi_{(3)}^\mathrm{diag}/3$. Consequently, the 18 non-diagonal components were all set to be $929 \ \mathrm{pm^2/V^2}$. 

\subsection{Nonlinear coefficients}
\label{app:sec:nonlinear_coef}
We calculated nonlinear coefficients using Eq. (\ref{eq:nonlin_pol}) and confirmed it matches well with the result given in the work we based on \cite{quesada2020, quesada2022}. The difference is that our expression can accommodate longitudinal dependence. These equations are expressed with electric fields, and the field data can be obtained directly from the eigenmode solver.

The nonlinear coefficients that appear in the main text can be classified into three-wave mixing coefficients and four-wave mixing coefficients. The three wave mixing coefficients are $\gamma_\mathrm{PDC}$ and $\gamma_\mathrm{QFC}$, and the four-wave mixing coefficients are $\gamma_\mathrm{SPM}$, $\gamma_{\mathrm{XPM},s}$, $\gamma_{\mathrm{XPM},i}$, and $\gamma_\mathrm{SFWM}$. These coefficients are obtained by substituting Eq. (\ref{eq:nonlin_pol}) into Eq. (\ref{eq:eom_adiabatic}) and taking field expansion as in Eq. (\ref{eq:field_decomp}). Their full expressions are given in Eq. (\ref{eq:nonlin_coef}). 
\begingroup
\allowdisplaybreaks
\begin{widetext}
\begin{subequations}
\begin{gather}
\label{eq:nonlin_coef}
\gamma_\mathrm{PDC}(z)  = \epsilon_0\sqrt{\cfrac{\bar{\omega}_s\bar{\omega}_i}{2v_p(z)v_s(z)v_i(z)}}\int d\vec{r}_\perp\chi_{(2)}^{lmn}(z)[e_s^l(\vec{r}_\perp, z)]^*[e_i^m(\vec{r}_\perp, z)]^*[e_p^n(\vec{r}_\perp, z)] \\
\gamma_\mathrm{QFC}(z)  = \epsilon_0\sqrt{\cfrac{\bar{\omega}_s\bar{\omega}_i}{2v_p(z)v_s(z)v_i(z)}}\int d\vec{r}_\perp\chi_{(2)}^{lmn}(z)[e_s^l(\vec{r}_\perp, z)][e_i^m(\vec{r}_\perp, z)]^*[e_p^n(\vec{r}_\perp, z)]^* \\
\gamma_\mathrm{SPM}(z) = \epsilon_0 \cfrac{3\bar{\omega}_p}{4v_p^2(z)} \int d\vec{r}_\perp\chi_{(3)}^{lmno}(z)[e_p^l(\vec{r}_\perp, z)]^*[e_p^m(\vec{r}_\perp, z)]^*[e_p^n(\vec{r}_\perp, z)][e_p^o(\vec{r}_\perp, z)] \\
\gamma_{\mathrm{XPM},s}(z) = \epsilon_0 \cfrac{3\bar{\omega}_s}{2v_p(z)v_s(z)} \int d\vec{r}_\perp\chi_{(3)}^{lmno}(z)[e_p^l(\vec{r}_\perp, z)]^*[e_s^m(\vec{r}_\perp, z)]^*[e_p^n(\vec{r}_\perp, z)][e_s^o(\vec{r}_\perp, z)] \\
\gamma_{\mathrm{XPM},i}(z)  = \epsilon_0 \cfrac{3\bar{\omega}_i}{2v_p(z)v_i(z)} \int d\vec{r}_\perp\chi_{(3)}^{lmno}(z)[e_p^l(\vec{r}_\perp, z)]^*[e_i^m(\vec{r}_\perp, z)]^*[e_p^n(\vec{r}_\perp, z)][e_i^o(\vec{r}_\perp, z)] \\
\gamma_{\mathrm{SFWM}}(z)  = \epsilon_0 \cfrac{3\sqrt{\bar{\omega}_s\bar{\omega}_i}}{2 \sqrt{v_{p1}(z)v_{p2}(z)v_{s}(z)v_{i}(z)}} \int d\vec{r}_\perp\chi_{(3)}^{lmno}(z)[e_{s}^l(\vec{r}_\perp, z)]^*[e_{i}^m(\vec{r}_\perp, z)]^*[e_{p1}^n(\vec{r}_\perp, z)][e_{p2}^o(\vec{r}_\perp, z)]
\end{gather}
\end{subequations}
\end{widetext}
\endgroup
When light propagates along a curve of anisotropoic crystal, the nonlinearity tensors $\chi_{(2)}$ and $\chi_{(3)}$ are rotated following the tensor rotation relation \cite{newnham2005}:
\begin{subequations}
\begin{align}
\label{eq:chi_rot}
 \chi_{(2)}^{ijk}(z)  &= R_x^{im}[\theta(z)]R_x^{jn}[\theta(z)]R_x^{ko}[\theta(z)]\chi_{(2)}^{mno},\\
 \chi_{(3)}^{ijkl}(z)  &= R_x^{im}[\theta(z)]R_x^{jn}[\theta(z)]R_x^{ko}[\theta(z)]R_x^{lp}[\theta(z)]\chi_{(3)}^{mnop},
\end{align}
\end{subequations}
where the rotation tensor is given as
\begin{equation}
R_x(\theta)  = \begin{bmatrix}
1 & 0 & 0 \\
0 & \cos(\theta) &  \sin(\theta)\\
0 & -\sin(\theta) & \cos(\theta)
\end{bmatrix}.
\end{equation}

An example of $\gamma_\mathrm{PDC}(z)$ along the curve is illustrated in Fig. (\ref{fig:apm_nonlinearity}). It reproduces the result of the reference \cite{poveda-hospital2023}, especially that the nonlinearity of GaP crystal vanishes at propagation angles of 0$^\circ$, 90$^\circ$ and maximized at 45$^\circ$. In this particular example, the eigenmode has no angular dependence, hence no logitudinal dependence; $e(\vec{r}_\perp, z)=e(\vec{r}_\perp)$; therefore, only the nonlinearitiy tensor has dependence on angles. In general, Eq. (\ref{eq:nonlin_coef}) combined with Eq. (\ref{eq:chi_rot}) can be applied, even when both eigenmode and nonlinearity tensor take angular dependence.

\section{Poling optimization}
\label{app:poling}
We briefly summarize the method detailed in \cite{tambasco2016, graffitti2017} and show that the resulting phase matching function mimics the optimization target. First, we obtained the amplitude of the phase matching function at an intermediate point $z$ in the waveguide from Eq. (\ref{eq:phase_matching}). At the central frequencies of the signal and idler, the amplitude of the phase matching function is only a function of $z$: $\Phi(z, \bar{\omega}_s, \bar{\omega}_i)$.

To mimic the Gaussian nonlinearity profile, we optimize the poling direction $\gamma(z)(=1,-1)$ of each domain to the target function, which gives a Gaussian phase matching function:
\begin{equation}
     \Phi(z, \bar{\omega}_s, \bar{\omega}_i) = -\mathrm{i} c\left(\operatorname{erf}\left(\frac{L-2 z}{2 \sqrt{2} \sigma}\right)-\operatorname{erf}\left(\frac{L}{2 \sqrt{2} \sigma}\right)\right),
\end{equation}
where $L$ is the length of the nonlinear waveguide and $\sigma$ is the bandwidth of the Gaussian nonlinearity profile. By properly choosing the poling direction, the amplitude of the optimized phase matching function can be increased or retained after a single poling domain. Using this behavior, the realized phase matching function can follow the target function. However, when the target function is too steep because the prefactor $c$ is too large, such that we are always required to choose increase, the realized pattern tends to periodic poling, thereby harming the spectral purity of the heralded single photon. In contrast, when the prefactor $c$ is too small, the purity is guaranteed, but brightness is severely compromised. In this work, we take $\sigma = L/4$ and $c = \sqrt{2/\pi}\sigma$, which provide sufficient purity without significantly compromising the brightness, as is chosen in \cite{tambasco2016}.

\begin{figure}[h]
\centering
\includegraphics[width=\linewidth]{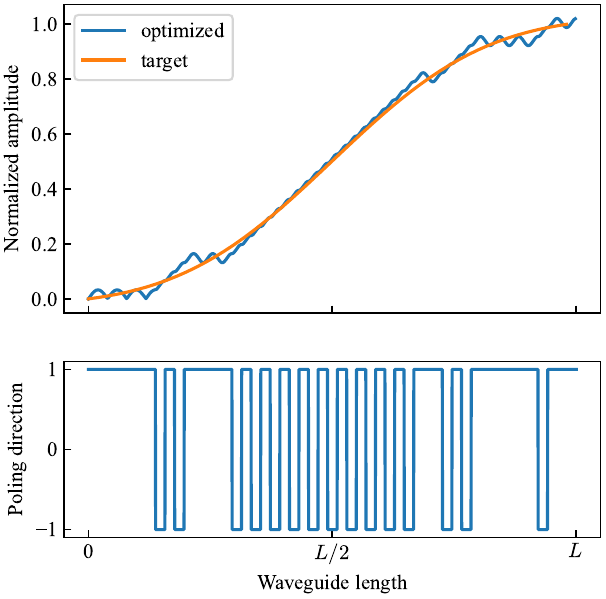}
    \caption{The amplitude of the target phase matching function and the amplitude of the optimized phase matching function. 
    } \label{fig:poling_opt}
\end{figure}

As an illustration, we have plotted both the target function and the amplitude of the phase matching function, along with the optimized poling pattern, in Fig.(\ref{fig:poling_opt}). Such visualization demonstrates how the poling pattern has been adjusted to achieve the desired phase matching characteristics. In our specific example used in Sec. \ref{sec:framework}, the length of the nonlinear waveguide, 5 mm, is set to contain 50 coherence lengths, and the number of poling domains is 3116. As a result, we found the optimized poling pattern follows the target function as shown in Fig. \ref{fig:poling_opt}.

\section{Integrated photonics design}
\label{app:integrated_photonics}

We propose the on-chip TWOC applicable to a specific group velocity relation, aGVM. Our device consists of three parts: APBS, adiabatic taper, and a partial Euler bend. In this section, we label the modes in line with \ref{subsec:nli_twoc} and accordingly denote 1550 nm, TE0 mode as signal; 1550 nm, TM0 mode as idler; and 775 nm, TM0 mode as pump. The labeling between the pump and idler should be swapped for QPG operation. The adiabatic polarization beam splitter picks up the signal mode from the primary waveguide, converts TE0 mode into TE1 mode, and leaves the other modes unaffected. The adiabatic taper at the auxiliary waveguide converts TE1 mode into TM0 mode, such that the light can enter the curve and proceed with negligible radiation loss. The Euler bend is designed for minimum optical loss, and it routes the beams to the next nonlinear interaction stage, placed in parallel with the previous one, reducing the footprint of the entire device. 

To compensate the temporal delay between modes, the group delays of optical modes through TWOC are calculated using two different methods: direct integration of time laps calculated from the local group velocities and the use of numerical electromagnetic simulation software such as the eigenmode expansion (EME) solver or FDTD solver. From the computation result, the length of the delay line is adjusted. The length is 306.8 $\mu$m for the compensation of the signal delay after propagation along 6 mm PPLN that is engineered for aGVM in Sec. \ref{sec:twoc}. Moreover, the wavelength-dependent insertion loss through the TWOC is estimated, and evaluated to be less than 0.3 dB for all three modes over the entire wavelength range. 

\subsection{Adiabatic polarization beam splitter}
\label{app:sec:apbs}

APBS is made of a series of adiabatic couplers carefully designed to meet the required functionality. Adiabatic couplers enjoy low loss, high fabrication error tolerance, and large operation bandwidth, and therefore they are suitable for integrated quantum photonics applications \cite{chung2017, chen2021}. Here, we introduce a novel APBS design on x-cut lithium niobate platform. To the best of our knowledge, this is the first proposal of APBS on TFLN that selectively couples TE mode while passing TM mode.

\begin{figure}
\centering
\includegraphics[width=\linewidth]{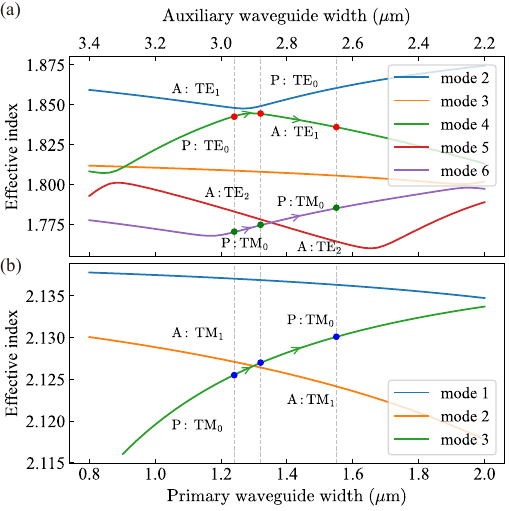}
    \caption{(a) Effective indices of supermodes in the APBS with a total width of 4.2 $\mu$m at a wavelength of 1550 nm, and (b) at a wavelength of 775 nm.
    }\label{fig:apbs_neff}
\end{figure}

The APBS device is designed to convert the TE0 signal mode of the primary waveguide into TE1 mode of a neighboring waveguide. Simultaneously, both TM0 idler and TM0 pump modes should have as little disruption as possible. To achieve this goal, we make use of the hybridization of TE0 supermode and TE1 supermode. Fig. \ref{fig:apbs_neff} shows the anti-crossing that we have utilized for the design. Meanwhile, the idler and pump should not couple with other supermodes, which can be confirmed by the fact that they do not have anticrossings.

\begin{figure*}
\includegraphics[width=\textwidth]{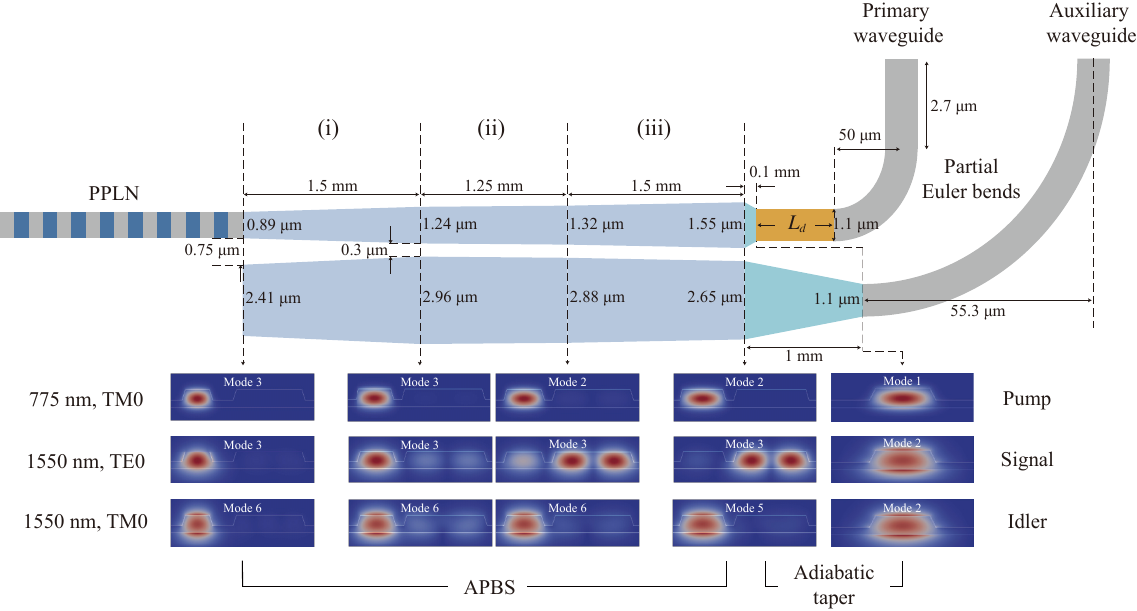}
    \caption{Schematic illustration of one half of the TWOC, featuring the APBS, an adiabatic taper, and a partial $\pi/2$-Euler bend. The APBS is segmented into three distinct phases: (i) the integration of an auxiliary waveguide; (ii) adiabatic coupling; and (iii) the escape stage. Subsequent to these stages, an adjustable delay is incorporated along the primary waveguide, whereas the adiabatic taper is aligned with the auxiliary waveguide. Both waveguides are then directed to the device's other half via $\pi/2$-Euler bends. The spatial mode profile involved in the TWOC is depicted at the bottom.
    } \label{fig:twoc_design}
\end{figure*}

The APBS is composed of three stages: In the region (i), the auxiliary waveguide is introduced. After that, both waveguides are tapered to prepare coupling. The widths of the primary waveguide and the auxiliary waveguide are linearly increased from 0.89 $\mu$m to 1.24 $\mu$m and from 2.41 $\mu$m to 2.95 $\mu$m, respectively. As the light propagates, the gap between the primary waveguide and the auxiliary waveguide, measured at the bottom of the waveguide rib, decreases from 0.75 $\mu$m to 0.3 $\mu$m. At the end of this region, the signal supermode is placed right before the hybridization region as depicted in Fig. \ref{fig:apbs_neff}(a); At the second stage, region (ii), the TE0 supermode is adiabatically converted into TE1 supermode. The total width, which is the sum of the two waveguides' top width, and bottom gap are maintained at 4.2 $\mu$m and 0.3 $\mu$m to reduce design complexity. The primary waveguide's width broadens to 1.32 $\mu$m while the coupler's width narrows down to 2.88 $\mu$m (Fig. \ref{fig:twoc_design}). The TM0 pump supermode crosses with TM1 supermode as shown in Fig. \ref{fig:apbs_neff}(b), but stays as TM0 since the two modes have different symmetries with respect to the vertical mirror plane placed in the middle of the waveguide. In other words, the power transfer between the two modes is forbidden by symmetry; In the third stage, region (iii), the modes are moved away from the coupling region such that the primary and auxiliary waveguides become independent each other. The widths of the primary and auxiliary waveguides are linearly tapered to 1.55 $\mu$m and 2.65 $\mu$m, respectively, maintaining the gap and the total width of the two waveguides. In this stage, the idler mode has a crossing with TE2 supermode, but it is retained in TM0 mode due to the different symmetries, as the case for the pump in the region (ii).

In summary, signal mode is coupled to TE1 mode of the auxiliary waveguide, while the other modes remain in their original spatial modes in the primary waveguide. The lengths of the three parts are 1.5 mm, 1.25 mm, and 1.5 mm, respectively, chosen to minimize the insertion loss. 

\begin{figure}
\centering
\includegraphics[width=\linewidth]{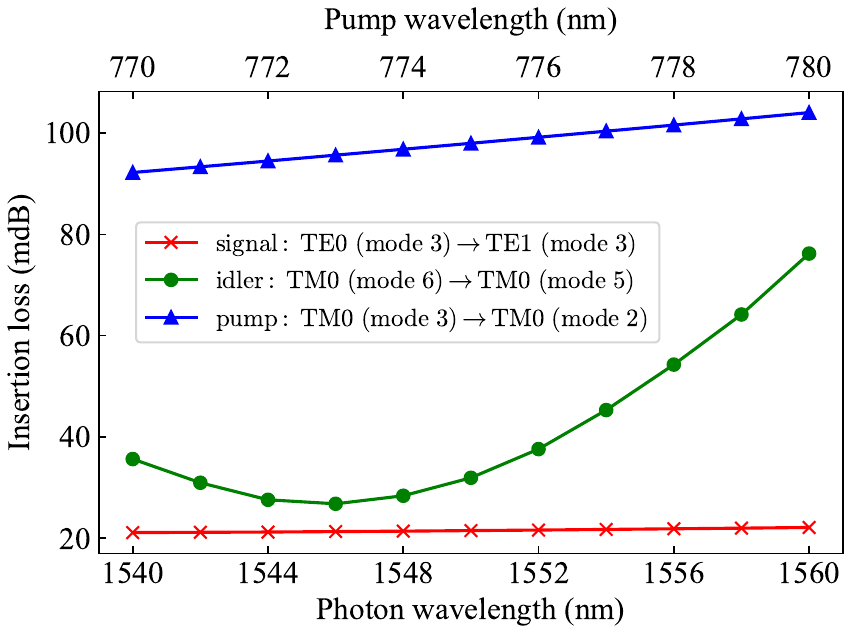}
    \caption{Insertion loss of each optical mode in the APBS, calculated using the EME method.
    }\label{fig:apbs_loss}
\end{figure}

The wavelength-dependent transmission is computed using the EME method. Fig. \ref{fig:apbs_loss} shows the insertion loss ($IL$) of the APBS, defined as follows:
\begin{equation}
\label{eq:insertion_loss}
IL\ \mathrm{(dB)} =  -10\log_{10}(T),
\end{equation}
where $T$ indicates the power transmission. The transmission is calculated for the final desired mode after three regions of APBS. The total insertion loss of signal, idler, and pump is less than 0.03 dB, 0.08 dB, and 0.11 dB each over the entire wavelength range shown in Fig. \ref{fig:apbs_loss}. 

The group delay through the APBS is calculated in two different ways. As the first method, we obtained it by differentiating the accumulated phase along the APBS obtained from the EME solver with respect to the angular frequency. The second method is to integrate time lapses of small segments calculated using the group indices $n_g$ obtained from the FDE simulation. The calculated group delays calculated by these methods are shown in Table \ref{tab:iapbs_gd}. Note that the values from two methods are very close to each other, where the difference is on the order of 0.1\%. 

\begin{table}[]
\centering
\caption{Group delay calculations of pump, signal, and idler ($\Delta\tau_p$, $\Delta\tau_s$, and $\Delta\tau_i$) for the APBS device using EME and FDE methods.
}
\begin{ruledtabular}
\begin{tabular}{cccc}
Simulation Method& $\Delta \tau_p  
\ \mathrm{(ps)}$& $\Delta \tau_s\ \mathrm{(ps)}$& $\Delta \tau_i\ \mathrm{(ps)}$\\
\hline
EME&   35.162&   32.422&   35.153\\
FDE&   35.145&   32.407&   35.141\\\end{tabular}
\end{ruledtabular}
\label{tab:iapbs_gd}
\end{table}

\subsection{Adiabatic taper}
\begin{figure}
\centering
\includegraphics[width=\linewidth]{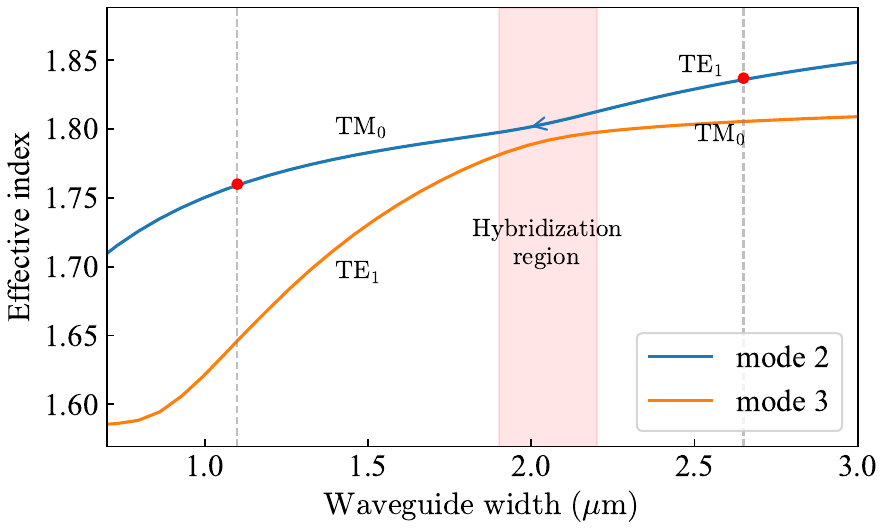}
    \caption{Effective indices of the auxiliary waveguide modes participating in the coupling within the adiabatic taper at a wavelength of 1550 nm.
    }\label{fig:taper_neff}
\end{figure}

Right after the APBS, the signal in TE1 mode is confined in the waveguide of 2.65 $\mu$m width. However, the higher-order spatial mode in a curved wide waveguide has a risk of significant losses due to inter-modal couplings and potential radiation to the free space. To mitigate the losses, we insert an adiabatic linear taper to convert TE1 mode into TM0 mode by narrowing the waveguide width to 1.1 $\mu$m utilizing the anti-crossing between TE1 and TM0 mode; we use the adiabatic coupling between TE1 and TM0 modes shown in Fig. \ref{fig:taper_neff}. A 1-mm-long linear taper allows the transmission approaching unity. Likewise, the primary waveguide confining idler and pump also narrows down from 1.55 $\mu$m to 1.1 $\mu$m. Since both TM0 pump and TM0 idler modes do not have a coupling with other modes at the given width range, 100-$\mu m$-long linear taper length is sufficient for transmission near unity.

\begin{figure}
\centering
\includegraphics[width=\linewidth]{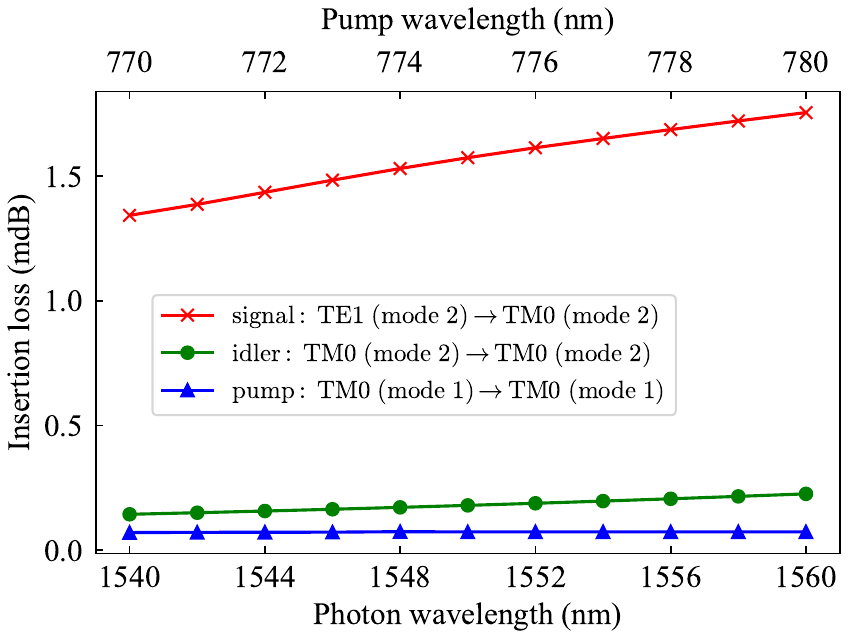}
    \caption{Insertion loss of each optical mode in the adiabatic taper, calculated using the EME method.
    }\label{fig:taper_loss}
\end{figure}

The wavelength dependent transmission and group delay are calculated as we did for APBS. The insertion loss and group delay are shown in Fig. \ref{fig:taper_loss} and Table \ref{tab:taper_gd}, respectively. The group delays calculated using EME and FDE show excellent agreement.

\begin{table}[]
\centering
\caption{Group delay calculations of pump, signal, and idler ($\Delta\tau_p$, $\Delta\tau_s$, and $\Delta\tau_i$) for the adiabatic taper using EME and FDE methods. 
}
\begin{ruledtabular}
\begin{tabular}{cccc}
                                 Simulation 
Method& $\Delta \tau_p  
\ \mathrm{(ps)}$& $\Delta \tau_s\ \mathrm{(ps)}$& $\Delta \tau_i\ \mathrm{(ps)}$\\
\hline
EME&   0.827&   8.023&   0.828\\
FDE&   0.827&   8.015&   0.827\\\end{tabular}
\end{ruledtabular}
\label{tab:taper_gd}
\end{table}

\subsection{Partial Euler bends}
To route the modes into the next nonlinear interaction stage and to reduce a total device footprint, we designed $\pi$-bend which consists of two $\pi/2$-partial-Euler bends \cite{ji2022,bahadori2019,vogelbacher2019}. The pump and idler modes are guided through the inner bend, and the signal mode is guided through the outer bend. For the inner bend, the effective radius $R_{\mathrm{eff}}$ is 50 $\mu $m and the $p$-value is optimized as $0.2$ to minimize the loss of the idler mode.

Meanwhile, due to the anisotropy of x-cut lithium niobate, eigenmode field profiles and corresponding $n_{\mathrm{eff}}$ continuously vary along the bend; accordingly, $n_{\mathrm{eff}}$s of TE0 and TM0 modes meet around the propagation angle $52^{\circ}$ at 775 nm wavelength. Due to the finite bend radius, the horizontal mirror symmetry of the waveguide is broken, allowing two modes to be coupled; it results in power transfer from TM0 mode into TE0 mode. Consequently, at the propagation angle of $90^{\circ}$, the power is found to be split into TE0 mode and TM0 mode, although it started as TM0 mode in the beginning. This phenomenon is called fundamental mode hybridization, investigated both in the simulation and the experiment \cite{wang2020, pan2019}.

\begin{figure}
\centering
\includegraphics[width=\linewidth]{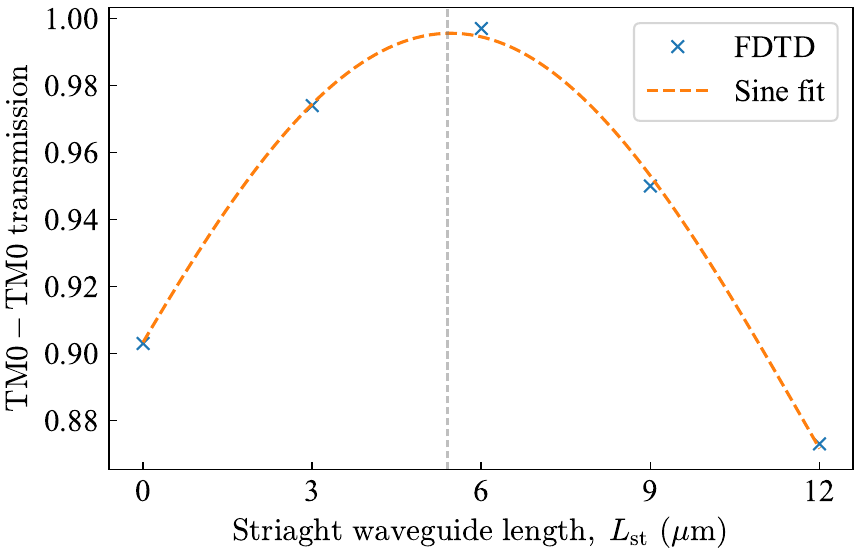}
    \caption{The power transmission of the pump mode from TM0 mode into TM0 mode after two $\pi/2$-partial-Euler bends is plotted against various lengths of straight waveguide segments between the two bends.
}\label{fig:euler_sinefit}
\end{figure}

We circumvented this issue based on the interferometric approach previously investigated for tapered waveguides \cite{guo2015}. Applying the same principles to our particular Euler bends, the straight waveguide of length $L_\mathrm{st}$ is inserted between the two $\pi/2$-partial Euler bends and controls the phase of TM0 and TE0 modes. It changes the transmission of TM0-TM0 and TM0-TE0 at the output of the second partial Euler bend as a function of its length, analogous to a conventional asymmetric Mach-Zehnder interferometer. Carrying out parameter sweeps in the FDTD solver, the transmission is calculated at five different lengths of the straight waveguide, ranging from 0 $\mu$m to 12 $\mu$m. Then, we fit the TM0-TM0 transmission to the sine function. The maximum transmission is obtained at 5.4 $\mu$m length of the straight waveguide as shown in Fig. \ref{fig:euler_sinefit}.

As a result, the inner bend of the primary waveguide consists of two $\pi/2$-partial Euler bends with $R_{1}=50 \ \mu$m, $p=0.2$, and the straight waveguide of length 5.4 $\mu m$ in the middle. The bend guides the pump and idler modes which are both TM0 modes. The outer bend is made of the two $\pi/2$-partial Euler bends with $R_{2}=55.3 \ \mu $m, $p=0.2$. $R_2$ is selected for the symmetric TWOC structure so that it satisfies $R_2 = R_1 + g + L_\mathrm{st}/2$, where $g$ is the gap between the primary and the auxiliary waveguides.

\begin{figure}
\centering
\includegraphics[width=\linewidth]{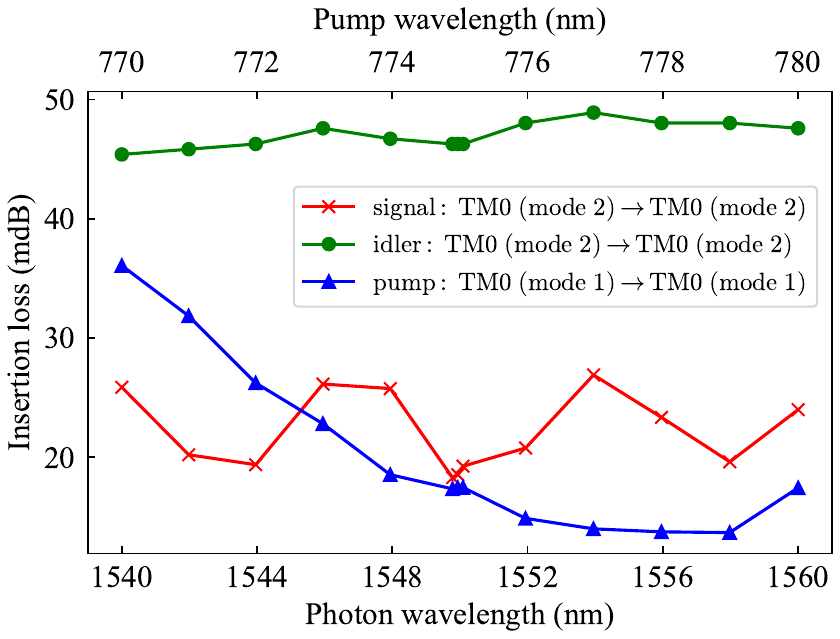}
    \caption{Insertion loss of each optical mode in the $\pi$-bend, calculated using the FDTD method.
    }\label{fig:euler_loss}
\end{figure}

The wavelength dependent insertion loss of both the inner and outer $\pi$-bends simulated by FDTD are shown in Fig. \ref{fig:euler_loss}. The loss of the signal, idler, and pump modes is less than 0.03 dB, 0.05 dB, and 0.04 dB, respectively. Also, we provide the group delays in Table \ref{tab:bend_gd}, comparing the results from the FDTD and the numerical integration values, which were calculated using group indices $n_g$ from the FDE.

\begin{table}[]
\centering
\caption{Group delay calculations of pump, signal, and idler ($\Delta\tau_p$, $\Delta\tau_s$, and $\Delta\tau_i$) through $\pi$-bend using FDTD and FDE methods. 
}
\begin{ruledtabular}
\begin{tabular}{cccc}
Simulation Method& 
$\Delta \tau_p  \ \mathrm{(ps)}$& $\Delta \tau_s\ \mathrm{(ps)}$& $\Delta \tau_i\ \mathrm{(ps)}$\\
\hline
FDTD&   1.374&   1.465&   1.366\\
FDE&   1.389&   1.484&   1.386\\
\end{tabular}
\end{ruledtabular}
\label{tab:bend_gd}
\end{table}

\bibliography{qunos}

\begin{thebibliography}{83}%
\makeatletter
\providecommand \@ifxundefined [1]{%
 \@ifx{#1\undefined}
}%
\providecommand \@ifnum [1]{%
 \ifnum #1\expandafter \@firstoftwo
 \else \expandafter \@secondoftwo
 \fi
}%
\providecommand \@ifx [1]{%
 \ifx #1\expandafter \@firstoftwo
 \else \expandafter \@secondoftwo
 \fi
}%
\providecommand \natexlab [1]{#1}%
\providecommand \enquote  [1]{``#1''}%
\providecommand \bibnamefont  [1]{#1}%
\providecommand \bibfnamefont [1]{#1}%
\providecommand \citenamefont [1]{#1}%
\providecommand \href@noop [0]{\@secondoftwo}%
\providecommand \href [0]{\begingroup \@sanitize@url \@href}%
\providecommand \@href[1]{\@@startlink{#1}\@@href}%
\providecommand \@@href[1]{\endgroup#1\@@endlink}%
\providecommand \@sanitize@url [0]{\catcode `\\12\catcode `\$12\catcode `\&12\catcode `\#12\catcode `\^12\catcode `\_12\catcode `\%12\relax}%
\providecommand \@@startlink[1]{}%
\providecommand \@@endlink[0]{}%
\providecommand \url  [0]{\begingroup\@sanitize@url \@url }%
\providecommand \@url [1]{\endgroup\@href {#1}{\urlprefix }}%
\providecommand \urlprefix  [0]{URL }%
\providecommand \Eprint [0]{\href }%
\providecommand \doibase [0]{https://doi.org/}%
\providecommand \selectlanguage [0]{\@gobble}%
\providecommand \bibinfo  [0]{\@secondoftwo}%
\providecommand \bibfield  [0]{\@secondoftwo}%
\providecommand \translation [1]{[#1]}%
\providecommand \BibitemOpen [0]{}%
\providecommand \bibitemStop [0]{}%
\providecommand \bibitemNoStop [0]{.\EOS\space}%
\providecommand \EOS [0]{\spacefactor3000\relax}%
\providecommand \BibitemShut  [1]{\csname bibitem#1\endcsname}%
\let\auto@bib@innerbib\@empty
\bibitem [{\citenamefont {Raussendorf}\ and\ \citenamefont {Briegel}(2001)}]{raussendorf2001}%
  \BibitemOpen
  \bibfield  {author} {\bibinfo {author} {\bibfnamefont {R.}~\bibnamefont {Raussendorf}}\ and\ \bibinfo {author} {\bibfnamefont {H.~J.}\ \bibnamefont {Briegel}},\ }\bibfield  {title} {\bibinfo {title} {A {{One-Way Quantum Computer}}},\ }\href {https://doi.org/10.1103/PhysRevLett.86.5188} {\bibfield  {journal} {\bibinfo  {journal} {Physical Review Letters}\ }\textbf {\bibinfo {volume} {86}},\ \bibinfo {pages} {5188} (\bibinfo {year} {2001})}\BibitemShut {NoStop}%
\bibitem [{\citenamefont {{Gimeno-Segovia}}\ \emph {et~al.}(2015)\citenamefont {{Gimeno-Segovia}}, \citenamefont {Shadbolt}, \citenamefont {Browne},\ and\ \citenamefont {Rudolph}}]{gimeno-segovia2015b}%
  \BibitemOpen
  \bibfield  {author} {\bibinfo {author} {\bibfnamefont {M.}~\bibnamefont {{Gimeno-Segovia}}}, \bibinfo {author} {\bibfnamefont {P.}~\bibnamefont {Shadbolt}}, \bibinfo {author} {\bibfnamefont {D.~E.}\ \bibnamefont {Browne}},\ and\ \bibinfo {author} {\bibfnamefont {T.}~\bibnamefont {Rudolph}},\ }\bibfield  {title} {\bibinfo {title} {From {{Three-Photon Greenberger-Horne-Zeilinger States}} to {{Ballistic Universal Quantum Computation}}},\ }\href {https://doi.org/10.1103/PhysRevLett.115.020502} {\bibfield  {journal} {\bibinfo  {journal} {Physical Review Letters}\ }\textbf {\bibinfo {volume} {115}},\ \bibinfo {pages} {020502} (\bibinfo {year} {2015})}\BibitemShut {NoStop}%
\bibitem [{\citenamefont {Bartolucci}\ \emph {et~al.}(2023)\citenamefont {Bartolucci}, \citenamefont {Birchall}, \citenamefont {Bomb{\'i}n}, \citenamefont {Cable}, \citenamefont {Dawson}, \citenamefont {{Gimeno-Segovia}}, \citenamefont {Johnston}, \citenamefont {Kieling}, \citenamefont {Nickerson}, \citenamefont {Pant}, \citenamefont {Pastawski}, \citenamefont {Rudolph},\ and\ \citenamefont {Sparrow}}]{bartolucci2023}%
  \BibitemOpen
  \bibfield  {author} {\bibinfo {author} {\bibfnamefont {S.}~\bibnamefont {Bartolucci}}, \bibinfo {author} {\bibfnamefont {P.}~\bibnamefont {Birchall}}, \bibinfo {author} {\bibfnamefont {H.}~\bibnamefont {Bomb{\'i}n}}, \bibinfo {author} {\bibfnamefont {H.}~\bibnamefont {Cable}}, \bibinfo {author} {\bibfnamefont {C.}~\bibnamefont {Dawson}}, \bibinfo {author} {\bibfnamefont {M.}~\bibnamefont {{Gimeno-Segovia}}}, \bibinfo {author} {\bibfnamefont {E.}~\bibnamefont {Johnston}}, \bibinfo {author} {\bibfnamefont {K.}~\bibnamefont {Kieling}}, \bibinfo {author} {\bibfnamefont {N.}~\bibnamefont {Nickerson}}, \bibinfo {author} {\bibfnamefont {M.}~\bibnamefont {Pant}}, \bibinfo {author} {\bibfnamefont {F.}~\bibnamefont {Pastawski}}, \bibinfo {author} {\bibfnamefont {T.}~\bibnamefont {Rudolph}},\ and\ \bibinfo {author} {\bibfnamefont {C.}~\bibnamefont {Sparrow}},\ }\bibfield  {title} {\bibinfo {title} {Fusion-based quantum computation},\ }\href {https://doi.org/10.1038/s41467-023-36493-1} {\bibfield  {journal} {\bibinfo
  {journal} {Nature Communications}\ }\textbf {\bibinfo {volume} {14}},\ \bibinfo {pages} {912} (\bibinfo {year} {2023})}\BibitemShut {NoStop}%
\bibitem [{\citenamefont {Silverstone}\ \emph {et~al.}(2016)\citenamefont {Silverstone}, \citenamefont {Bonneau}, \citenamefont {O'Brien},\ and\ \citenamefont {Thompson}}]{silverstone2016}%
  \BibitemOpen
  \bibfield  {author} {\bibinfo {author} {\bibfnamefont {J.~W.}\ \bibnamefont {Silverstone}}, \bibinfo {author} {\bibfnamefont {D.}~\bibnamefont {Bonneau}}, \bibinfo {author} {\bibfnamefont {J.~L.}\ \bibnamefont {O'Brien}},\ and\ \bibinfo {author} {\bibfnamefont {M.~G.}\ \bibnamefont {Thompson}},\ }\bibfield  {title} {\bibinfo {title} {Silicon {{Quantum Photonics}}},\ }\href {https://doi.org/10.1109/JSTQE.2016.2573218} {\bibfield  {journal} {\bibinfo  {journal} {IEEE Journal of Selected Topics in Quantum Electronics}\ }\textbf {\bibinfo {volume} {22}},\ \bibinfo {pages} {390} (\bibinfo {year} {2016})}\BibitemShut {NoStop}%
\bibitem [{\citenamefont {Christ}\ \emph {et~al.}(2013{\natexlab{a}})\citenamefont {Christ}, \citenamefont {Fedrizzi}, \citenamefont {H{\"u}bel}, \citenamefont {Jennewein},\ and\ \citenamefont {Silberhorn}}]{christ2013a}%
  \BibitemOpen
  \bibfield  {author} {\bibinfo {author} {\bibfnamefont {A.}~\bibnamefont {Christ}}, \bibinfo {author} {\bibfnamefont {A.}~\bibnamefont {Fedrizzi}}, \bibinfo {author} {\bibfnamefont {H.}~\bibnamefont {H{\"u}bel}}, \bibinfo {author} {\bibfnamefont {T.}~\bibnamefont {Jennewein}},\ and\ \bibinfo {author} {\bibfnamefont {C.}~\bibnamefont {Silberhorn}},\ }\bibfield  {title} {\bibinfo {title} {Parametric {{Down-Conversion}}},\ }in\ \href {https://doi.org/10.1016/B978-0-12-387695-9.00011-1} {\emph {\bibinfo {booktitle} {Experimental {{Methods}} in the {{Physical Sciences}}}}},\ Vol.~\bibinfo {volume} {45}\ (\bibinfo  {publisher} {Elsevier},\ \bibinfo {year} {2013})\ pp.\ \bibinfo {pages} {351--410}\BibitemShut {NoStop}%
\bibitem [{\citenamefont {Browne}\ and\ \citenamefont {Rudolph}(2005)}]{browne2005}%
  \BibitemOpen
  \bibfield  {author} {\bibinfo {author} {\bibfnamefont {D.~E.}\ \bibnamefont {Browne}}\ and\ \bibinfo {author} {\bibfnamefont {T.}~\bibnamefont {Rudolph}},\ }\bibfield  {title} {\bibinfo {title} {Resource-{{Efficient Linear Optical Quantum Computation}}},\ }\href {https://doi.org/10.1103/PhysRevLett.95.010501} {\bibfield  {journal} {\bibinfo  {journal} {Physical Review Letters}\ }\textbf {\bibinfo {volume} {95}},\ \bibinfo {pages} {010501} (\bibinfo {year} {2005})}\BibitemShut {NoStop}%
\bibitem [{\citenamefont {{Gimeno-Segovia}}(2015)}]{gimeno-segovia2015}%
  \BibitemOpen
  \bibfield  {author} {\bibinfo {author} {\bibfnamefont {M.}~\bibnamefont {{Gimeno-Segovia}}},\ }\emph {\bibinfo {title} {Towards Practical Linear Optical Quantum Computing}},\ \href@noop {} {Ph.D. thesis},\ \bibinfo  {school} {Imperial College London} (\bibinfo {year} {2015})\BibitemShut {NoStop}%
\bibitem [{\citenamefont {{Meyer-Scott}}\ \emph {et~al.}(2020)\citenamefont {{Meyer-Scott}}, \citenamefont {Silberhorn},\ and\ \citenamefont {Migdall}}]{meyer-scott2020}%
  \BibitemOpen
  \bibfield  {author} {\bibinfo {author} {\bibfnamefont {E.}~\bibnamefont {{Meyer-Scott}}}, \bibinfo {author} {\bibfnamefont {C.}~\bibnamefont {Silberhorn}},\ and\ \bibinfo {author} {\bibfnamefont {A.}~\bibnamefont {Migdall}},\ }\bibfield  {title} {\bibinfo {title} {Single-photon sources: {{Approaching}} the ideal through multiplexing},\ }\href {https://doi.org/10.1063/5.0003320} {\bibfield  {journal} {\bibinfo  {journal} {Review of Scientific Instruments}\ }\textbf {\bibinfo {volume} {91}},\ \bibinfo {pages} {041101} (\bibinfo {year} {2020})}\BibitemShut {NoStop}%
\bibitem [{\citenamefont {Wang}\ \emph {et~al.}(2018{\natexlab{a}})\citenamefont {Wang}, \citenamefont {Zhang}, \citenamefont {Chen}, \citenamefont {Bertrand}, \citenamefont {{Shams-Ansari}}, \citenamefont {Chandrasekhar}, \citenamefont {Winzer},\ and\ \citenamefont {Lon{\v c}ar}}]{wang2018a}%
  \BibitemOpen
  \bibfield  {author} {\bibinfo {author} {\bibfnamefont {C.}~\bibnamefont {Wang}}, \bibinfo {author} {\bibfnamefont {M.}~\bibnamefont {Zhang}}, \bibinfo {author} {\bibfnamefont {X.}~\bibnamefont {Chen}}, \bibinfo {author} {\bibfnamefont {M.}~\bibnamefont {Bertrand}}, \bibinfo {author} {\bibfnamefont {A.}~\bibnamefont {{Shams-Ansari}}}, \bibinfo {author} {\bibfnamefont {S.}~\bibnamefont {Chandrasekhar}}, \bibinfo {author} {\bibfnamefont {P.}~\bibnamefont {Winzer}},\ and\ \bibinfo {author} {\bibfnamefont {M.}~\bibnamefont {Lon{\v c}ar}},\ }\bibfield  {title} {\bibinfo {title} {Integrated lithium niobate electro-optic modulators operating at {{CMOS-compatible}} voltages},\ }\href {https://doi.org/10.1038/s41586-018-0551-y} {\bibfield  {journal} {\bibinfo  {journal} {Nature}\ }\textbf {\bibinfo {volume} {562}},\ \bibinfo {pages} {101} (\bibinfo {year} {2018}{\natexlab{a}})}\BibitemShut {NoStop}%
\bibitem [{\citenamefont {Newnham}(2005)}]{newnham2005}%
  \BibitemOpen
  \bibfield  {author} {\bibinfo {author} {\bibfnamefont {R.~E.}\ \bibnamefont {Newnham}},\ }\href {https://doi.org/10.1093/oso/9780198520757.001.0001} {\emph {\bibinfo {title} {Properties of Materials: Anisotropy, Symmetry, Structure}}}\ (\bibinfo  {publisher} {Oxford University Press},\ \bibinfo {address} {Oxford ; New York},\ \bibinfo {year} {2005})\BibitemShut {NoStop}%
\bibitem [{\citenamefont {Alexander}\ \emph {et~al.}(2018)\citenamefont {Alexander}, \citenamefont {George}, \citenamefont {Verbist}, \citenamefont {Neyts}, \citenamefont {Kuyken}, \citenamefont {Van~Thourhout},\ and\ \citenamefont {Beeckman}}]{alexander2018}%
  \BibitemOpen
  \bibfield  {author} {\bibinfo {author} {\bibfnamefont {K.}~\bibnamefont {Alexander}}, \bibinfo {author} {\bibfnamefont {J.~P.}\ \bibnamefont {George}}, \bibinfo {author} {\bibfnamefont {J.}~\bibnamefont {Verbist}}, \bibinfo {author} {\bibfnamefont {K.}~\bibnamefont {Neyts}}, \bibinfo {author} {\bibfnamefont {B.}~\bibnamefont {Kuyken}}, \bibinfo {author} {\bibfnamefont {D.}~\bibnamefont {Van~Thourhout}},\ and\ \bibinfo {author} {\bibfnamefont {J.}~\bibnamefont {Beeckman}},\ }\bibfield  {title} {\bibinfo {title} {Nanophotonic {{Pockels}} modulators on a silicon nitride platform},\ }\href {https://doi.org/10.1038/s41467-018-05846-6} {\bibfield  {journal} {\bibinfo  {journal} {Nature Communications}\ }\textbf {\bibinfo {volume} {9}},\ \bibinfo {pages} {3444} (\bibinfo {year} {2018})}\BibitemShut {NoStop}%
\bibitem [{\citenamefont {Abel}\ \emph {et~al.}(2019)\citenamefont {Abel}, \citenamefont {Eltes}, \citenamefont {Ortmann}, \citenamefont {Messner}, \citenamefont {Castera}, \citenamefont {Wagner}, \citenamefont {Urbonas}, \citenamefont {Rosa}, \citenamefont {Gutierrez}, \citenamefont {Tulli}, \citenamefont {Ma}, \citenamefont {Baeuerle}, \citenamefont {Josten}, \citenamefont {Heni}, \citenamefont {Caimi}, \citenamefont {Czornomaz}, \citenamefont {Demkov}, \citenamefont {Leuthold}, \citenamefont {Sanchis},\ and\ \citenamefont {Fompeyrine}}]{abel2019}%
  \BibitemOpen
  \bibfield  {author} {\bibinfo {author} {\bibfnamefont {S.}~\bibnamefont {Abel}}, \bibinfo {author} {\bibfnamefont {F.}~\bibnamefont {Eltes}}, \bibinfo {author} {\bibfnamefont {J.~E.}\ \bibnamefont {Ortmann}}, \bibinfo {author} {\bibfnamefont {A.}~\bibnamefont {Messner}}, \bibinfo {author} {\bibfnamefont {P.}~\bibnamefont {Castera}}, \bibinfo {author} {\bibfnamefont {T.}~\bibnamefont {Wagner}}, \bibinfo {author} {\bibfnamefont {D.}~\bibnamefont {Urbonas}}, \bibinfo {author} {\bibfnamefont {A.}~\bibnamefont {Rosa}}, \bibinfo {author} {\bibfnamefont {A.~M.}\ \bibnamefont {Gutierrez}}, \bibinfo {author} {\bibfnamefont {D.}~\bibnamefont {Tulli}}, \bibinfo {author} {\bibfnamefont {P.}~\bibnamefont {Ma}}, \bibinfo {author} {\bibfnamefont {B.}~\bibnamefont {Baeuerle}}, \bibinfo {author} {\bibfnamefont {A.}~\bibnamefont {Josten}}, \bibinfo {author} {\bibfnamefont {W.}~\bibnamefont {Heni}}, \bibinfo {author} {\bibfnamefont {D.}~\bibnamefont {Caimi}}, \bibinfo {author} {\bibfnamefont {L.}~\bibnamefont {Czornomaz}},
  \bibinfo {author} {\bibfnamefont {A.~A.}\ \bibnamefont {Demkov}}, \bibinfo {author} {\bibfnamefont {J.}~\bibnamefont {Leuthold}}, \bibinfo {author} {\bibfnamefont {P.}~\bibnamefont {Sanchis}},\ and\ \bibinfo {author} {\bibfnamefont {J.}~\bibnamefont {Fompeyrine}},\ }\bibfield  {title} {\bibinfo {title} {Large {{Pockels}} effect in micro- and nanostructured barium titanate integrated on silicon},\ }\href {https://doi.org/10.1038/s41563-018-0208-0} {\bibfield  {journal} {\bibinfo  {journal} {Nature Materials}\ }\textbf {\bibinfo {volume} {18}},\ \bibinfo {pages} {42} (\bibinfo {year} {2019})}\BibitemShut {NoStop}%
\bibitem [{\citenamefont {Zhang}\ \emph {et~al.}(2021)\citenamefont {Zhang}, \citenamefont {Wang}, \citenamefont {Kharel}, \citenamefont {Zhu},\ and\ \citenamefont {Lon{\v c}ar}}]{zhang2021}%
  \BibitemOpen
  \bibfield  {author} {\bibinfo {author} {\bibfnamefont {M.}~\bibnamefont {Zhang}}, \bibinfo {author} {\bibfnamefont {C.}~\bibnamefont {Wang}}, \bibinfo {author} {\bibfnamefont {P.}~\bibnamefont {Kharel}}, \bibinfo {author} {\bibfnamefont {D.}~\bibnamefont {Zhu}},\ and\ \bibinfo {author} {\bibfnamefont {M.}~\bibnamefont {Lon{\v c}ar}},\ }\bibfield  {title} {\bibinfo {title} {Integrated lithium niobate electro-optic modulators: When performance meets scalability},\ }\href {https://doi.org/10.1364/OPTICA.415762} {\bibfield  {journal} {\bibinfo  {journal} {Optica}\ }\textbf {\bibinfo {volume} {8}},\ \bibinfo {pages} {652} (\bibinfo {year} {2021})}\BibitemShut {NoStop}%
\bibitem [{\citenamefont {Jankowski}\ \emph {et~al.}(2021)\citenamefont {Jankowski}, \citenamefont {Mishra},\ and\ \citenamefont {Fejer}}]{jankowski2021}%
  \BibitemOpen
  \bibfield  {author} {\bibinfo {author} {\bibfnamefont {M.}~\bibnamefont {Jankowski}}, \bibinfo {author} {\bibfnamefont {J.}~\bibnamefont {Mishra}},\ and\ \bibinfo {author} {\bibfnamefont {M.~M.}\ \bibnamefont {Fejer}},\ }\bibfield  {title} {\bibinfo {title} {Dispersion-engineered {$\chi$}(2) nanophotonics: A flexible tool for nonclassical light},\ }\href {https://doi.org/10.1088/2515-7647/ac1729} {\bibfield  {journal} {\bibinfo  {journal} {Journal of Physics: Photonics}\ }\textbf {\bibinfo {volume} {3}},\ \bibinfo {pages} {042005} (\bibinfo {year} {2021})}\BibitemShut {NoStop}%
\bibitem [{\citenamefont {Wang}\ \emph {et~al.}(2018{\natexlab{b}})\citenamefont {Wang}, \citenamefont {Langrock}, \citenamefont {Marandi}, \citenamefont {Jankowski}, \citenamefont {Zhang}, \citenamefont {Desiatov}, \citenamefont {Fejer},\ and\ \citenamefont {Lon{\v c}ar}}]{wang2018}%
  \BibitemOpen
  \bibfield  {author} {\bibinfo {author} {\bibfnamefont {C.}~\bibnamefont {Wang}}, \bibinfo {author} {\bibfnamefont {C.}~\bibnamefont {Langrock}}, \bibinfo {author} {\bibfnamefont {A.}~\bibnamefont {Marandi}}, \bibinfo {author} {\bibfnamefont {M.}~\bibnamefont {Jankowski}}, \bibinfo {author} {\bibfnamefont {M.}~\bibnamefont {Zhang}}, \bibinfo {author} {\bibfnamefont {B.}~\bibnamefont {Desiatov}}, \bibinfo {author} {\bibfnamefont {M.~M.}\ \bibnamefont {Fejer}},\ and\ \bibinfo {author} {\bibfnamefont {M.}~\bibnamefont {Lon{\v c}ar}},\ }\bibfield  {title} {\bibinfo {title} {Ultrahigh-efficiency wavelength conversion in nanophotonic periodically poled lithium niobate waveguides},\ }\href {https://doi.org/10.1364/OPTICA.5.001438} {\bibfield  {journal} {\bibinfo  {journal} {Optica}\ }\textbf {\bibinfo {volume} {5}},\ \bibinfo {pages} {1438} (\bibinfo {year} {2018}{\natexlab{b}})}\BibitemShut {NoStop}%
\bibitem [{\citenamefont {Quesada}\ and\ \citenamefont {Sipe}(2014)}]{quesada2014}%
  \BibitemOpen
  \bibfield  {author} {\bibinfo {author} {\bibfnamefont {N.}~\bibnamefont {Quesada}}\ and\ \bibinfo {author} {\bibfnamefont {J.~E.}\ \bibnamefont {Sipe}},\ }\bibfield  {title} {\bibinfo {title} {Effects of time ordering in quantum nonlinear optics},\ }\href {https://doi.org/10.1103/PhysRevA.90.063840} {\bibfield  {journal} {\bibinfo  {journal} {Physical Review A}\ }\textbf {\bibinfo {volume} {90}},\ \bibinfo {pages} {063840} (\bibinfo {year} {2014})}\BibitemShut {NoStop}%
\bibitem [{\citenamefont {Triginer}\ \emph {et~al.}(2020)\citenamefont {Triginer}, \citenamefont {Vidrighin}, \citenamefont {Quesada}, \citenamefont {Eckstein}, \citenamefont {Moore}, \citenamefont {Kolthammer}, \citenamefont {Sipe},\ and\ \citenamefont {Walmsley}}]{triginer2020}%
  \BibitemOpen
  \bibfield  {author} {\bibinfo {author} {\bibfnamefont {G.}~\bibnamefont {Triginer}}, \bibinfo {author} {\bibfnamefont {M.~D.}\ \bibnamefont {Vidrighin}}, \bibinfo {author} {\bibfnamefont {N.}~\bibnamefont {Quesada}}, \bibinfo {author} {\bibfnamefont {A.}~\bibnamefont {Eckstein}}, \bibinfo {author} {\bibfnamefont {M.}~\bibnamefont {Moore}}, \bibinfo {author} {\bibfnamefont {W.~S.}\ \bibnamefont {Kolthammer}}, \bibinfo {author} {\bibfnamefont {J.~E.}\ \bibnamefont {Sipe}},\ and\ \bibinfo {author} {\bibfnamefont {I.~A.}\ \bibnamefont {Walmsley}},\ }\bibfield  {title} {\bibinfo {title} {Understanding {{High-Gain Twin-Beam Sources Using Cascaded Stimulated Emission}}},\ }\href {https://doi.org/10.1103/PhysRevX.10.031063} {\bibfield  {journal} {\bibinfo  {journal} {Physical Review X}\ }\textbf {\bibinfo {volume} {10}},\ \bibinfo {pages} {031063} (\bibinfo {year} {2020})}\BibitemShut {NoStop}%
\bibitem [{\citenamefont {Quesada}\ and\ \citenamefont {Bra{\'n}czyk}(2018)}]{quesada2018}%
  \BibitemOpen
  \bibfield  {author} {\bibinfo {author} {\bibfnamefont {N.}~\bibnamefont {Quesada}}\ and\ \bibinfo {author} {\bibfnamefont {A.~M.}\ \bibnamefont {Bra{\'n}czyk}},\ }\bibfield  {title} {\bibinfo {title} {Gaussian functions are optimal for waveguided nonlinear-quantum-optical processes},\ }\href {https://doi.org/10.1103/PhysRevA.98.043813} {\bibfield  {journal} {\bibinfo  {journal} {Physical Review A}\ }\textbf {\bibinfo {volume} {98}},\ \bibinfo {pages} {043813} (\bibinfo {year} {2018})}\BibitemShut {NoStop}%
\bibitem [{\citenamefont {Quesada}\ \emph {et~al.}(2020)\citenamefont {Quesada}, \citenamefont {Triginer}, \citenamefont {Vidrighin},\ and\ \citenamefont {Sipe}}]{quesada2020}%
  \BibitemOpen
  \bibfield  {author} {\bibinfo {author} {\bibfnamefont {N.}~\bibnamefont {Quesada}}, \bibinfo {author} {\bibfnamefont {G.}~\bibnamefont {Triginer}}, \bibinfo {author} {\bibfnamefont {M.~D.}\ \bibnamefont {Vidrighin}},\ and\ \bibinfo {author} {\bibfnamefont {J.~E.}\ \bibnamefont {Sipe}},\ }\bibfield  {title} {\bibinfo {title} {Theory of high-gain twin-beam generation in waveguides: {{From Maxwell}}'s equations to efficient simulation},\ }\href {https://doi.org/10.1103/PhysRevA.102.033519} {\bibfield  {journal} {\bibinfo  {journal} {Physical Review A}\ }\textbf {\bibinfo {volume} {102}},\ \bibinfo {pages} {033519} (\bibinfo {year} {2020})}\BibitemShut {NoStop}%
\bibitem [{\citenamefont {Smith}\ \emph {et~al.}(1998)\citenamefont {Smith}, \citenamefont {Armstrong},\ and\ \citenamefont {Alford}}]{smith1998}%
  \BibitemOpen
  \bibfield  {author} {\bibinfo {author} {\bibfnamefont {A.~V.}\ \bibnamefont {Smith}}, \bibinfo {author} {\bibfnamefont {D.~J.}\ \bibnamefont {Armstrong}},\ and\ \bibinfo {author} {\bibfnamefont {W.~J.}\ \bibnamefont {Alford}},\ }\bibfield  {title} {\bibinfo {title} {Increased acceptance bandwidths in optical frequency conversion by use of multiple walk-off-compensating nonlinear crystals},\ }\href {https://doi.org/10.1364/JOSAB.15.000122} {\bibfield  {journal} {\bibinfo  {journal} {Journal of the Optical Society of America B}\ }\textbf {\bibinfo {volume} {15}},\ \bibinfo {pages} {122} (\bibinfo {year} {1998})}\BibitemShut {NoStop}%
\bibitem [{\citenamefont {Huang}\ \emph {et~al.}(2004)\citenamefont {Huang}, \citenamefont {Kurz}, \citenamefont {Langrock}, \citenamefont {Schober},\ and\ \citenamefont {Fejer}}]{huang2004}%
  \BibitemOpen
  \bibfield  {author} {\bibinfo {author} {\bibfnamefont {J.}~\bibnamefont {Huang}}, \bibinfo {author} {\bibfnamefont {J.~R.}\ \bibnamefont {Kurz}}, \bibinfo {author} {\bibfnamefont {C.}~\bibnamefont {Langrock}}, \bibinfo {author} {\bibfnamefont {A.~M.}\ \bibnamefont {Schober}},\ and\ \bibinfo {author} {\bibfnamefont {M.~M.}\ \bibnamefont {Fejer}},\ }\bibfield  {title} {\bibinfo {title} {Quasi-group-velocity matching using integrated-optic structures},\ }\href {https://doi.org/10.1364/OL.29.002482} {\bibfield  {journal} {\bibinfo  {journal} {Optics Letters}\ }\textbf {\bibinfo {volume} {29}},\ \bibinfo {pages} {2482} (\bibinfo {year} {2004})}\BibitemShut {NoStop}%
\bibitem [{\citenamefont {Vidrighin}(2016)}]{vidrighin2016}%
  \BibitemOpen
  \bibfield  {author} {\bibinfo {author} {\bibfnamefont {M.~D.}\ \bibnamefont {Vidrighin}},\ }\emph {\bibinfo {title} {Quantum {{Optical Measurements}} for {{Practical Estimation}} and {{Information Thermodynamics}}}},\ \href@noop {} {Ph.D. thesis},\ \bibinfo  {school} {Imperial College London} (\bibinfo {year} {2016})\BibitemShut {NoStop}%
\bibitem [{\citenamefont {L{\ae}gsgaard}(2012)}]{laegsgaard2012}%
  \BibitemOpen
  \bibfield  {author} {\bibinfo {author} {\bibfnamefont {J.}~\bibnamefont {L{\ae}gsgaard}},\ }\bibfield  {title} {\bibinfo {title} {Modeling of nonlinear propagation in fiber tapers},\ }\href {https://doi.org/10.1364/JOSAB.29.003183} {\bibfield  {journal} {\bibinfo  {journal} {Journal of the Optical Society of America B}\ }\textbf {\bibinfo {volume} {29}},\ \bibinfo {pages} {3183} (\bibinfo {year} {2012})}\BibitemShut {NoStop}%
\bibitem [{\citenamefont {Thomas}\ \emph {et~al.}(2021)\citenamefont {Thomas}, \citenamefont {McCutcheon},\ and\ \citenamefont {McCutcheon}}]{thomas2021}%
  \BibitemOpen
  \bibfield  {author} {\bibinfo {author} {\bibfnamefont {O.~F.}\ \bibnamefont {Thomas}}, \bibinfo {author} {\bibfnamefont {W.}~\bibnamefont {McCutcheon}},\ and\ \bibinfo {author} {\bibfnamefont {D.~P.~S.}\ \bibnamefont {McCutcheon}},\ }\bibfield  {title} {\bibinfo {title} {A general framework for multimode {{Gaussian}} quantum optics and photo-detection: {{Application}} to {{Hong}}--{{Ou}}--{{Mandel}} interference with filtered heralded single photon sources},\ }\href {https://doi.org/10.1063/5.0044036} {\bibfield  {journal} {\bibinfo  {journal} {APL Photonics}\ }\textbf {\bibinfo {volume} {6}},\ \bibinfo {pages} {040801} (\bibinfo {year} {2021})}\BibitemShut {NoStop}%
\bibitem [{\citenamefont {Bures}(2009)}]{bures2009}%
  \BibitemOpen
  \bibfield  {author} {\bibinfo {author} {\bibfnamefont {J.}~\bibnamefont {Bures}},\ }\href@noop {} {\emph {\bibinfo {title} {Guided Optics: Optical Fibers and All-Fiber Components}}},\ Physics Textbook\ (\bibinfo  {publisher} {Wiley-VCH},\ \bibinfo {address} {Weinheim},\ \bibinfo {year} {2009})\BibitemShut {NoStop}%
\bibitem [{\citenamefont {{J E Sipe}}\ and\ \citenamefont {Steel}(2016)}]{jesipe2016}%
  \BibitemOpen
  \bibfield  {author} {\bibinfo {author} {\bibnamefont {{J E Sipe}}}\ and\ \bibinfo {author} {\bibfnamefont {M.~J.}\ \bibnamefont {Steel}},\ }\bibfield  {title} {\bibinfo {title} {A {{Hamiltonian}} treatment of stimulated {{Brillouin}} scattering in nanoscale integrated waveguides},\ }\href {https://doi.org/10.1088/1367-2630/18/4/045004} {\bibfield  {journal} {\bibinfo  {journal} {New Journal of Physics}\ }\textbf {\bibinfo {volume} {18}},\ \bibinfo {pages} {045004} (\bibinfo {year} {2016})}\BibitemShut {NoStop}%
\bibitem [{\citenamefont {Boyd}(2019)}]{boyd2019}%
  \BibitemOpen
  \bibfield  {author} {\bibinfo {author} {\bibfnamefont {R.~W.}\ \bibnamefont {Boyd}},\ }\href {https://doi.org/10.1016/C2015-0-05510-1} {\emph {\bibinfo {title} {Nonlinear Optics}}},\ \bibinfo {edition} {4th}\ ed.\ (\bibinfo  {publisher} {Academic Press},\ \bibinfo {address} {San Diego},\ \bibinfo {year} {2019})\BibitemShut {NoStop}%
\bibitem [{\citenamefont {Eisert}\ and\ \citenamefont {Wolf}(2007)}]{eisert2007}%
  \BibitemOpen
  \bibfield  {author} {\bibinfo {author} {\bibfnamefont {J.}~\bibnamefont {Eisert}}\ and\ \bibinfo {author} {\bibfnamefont {M.~M.}\ \bibnamefont {Wolf}},\ }\bibfield  {title} {\bibinfo {title} {Gaussian {{Quantum Channels}}},\ }in\ \href {https://doi.org/10.1142/9781860948169_0002} {\emph {\bibinfo {booktitle} {Quantum {{Information}} with {{Continuous Variables}} of {{Atoms}} and {{Light}}}}}\ (\bibinfo  {publisher} {Imperial College Press},\ \bibinfo {address} {London},\ \bibinfo {year} {2007})\ pp.\ \bibinfo {pages} {23--42}\BibitemShut {NoStop}%
\bibitem [{\citenamefont {Takeoka}\ \emph {et~al.}(2015)\citenamefont {Takeoka}, \citenamefont {Jin},\ and\ \citenamefont {Sasaki}}]{takeoka2015}%
  \BibitemOpen
  \bibfield  {author} {\bibinfo {author} {\bibfnamefont {M.}~\bibnamefont {Takeoka}}, \bibinfo {author} {\bibfnamefont {R.-B.}\ \bibnamefont {Jin}},\ and\ \bibinfo {author} {\bibfnamefont {M.}~\bibnamefont {Sasaki}},\ }\bibfield  {title} {\bibinfo {title} {Full analysis of multi-photon pair effects in spontaneous parametric down conversion based photonic quantum information processing},\ }\href {https://doi.org/10.1088/1367-2630/17/4/043030} {\bibfield  {journal} {\bibinfo  {journal} {New Journal of Physics}\ }\textbf {\bibinfo {volume} {17}},\ \bibinfo {pages} {043030} (\bibinfo {year} {2015})}\BibitemShut {NoStop}%
\bibitem [{\citenamefont {Xin}\ \emph {et~al.}(2022)\citenamefont {Xin}, \citenamefont {Mishra}, \citenamefont {Chen}, \citenamefont {Zhu}, \citenamefont {{Shams-Ansari}}, \citenamefont {Langrock}, \citenamefont {Sinclair}, \citenamefont {Wong}, \citenamefont {Fejer},\ and\ \citenamefont {Lon{\v c}ar}}]{xin2022}%
  \BibitemOpen
  \bibfield  {author} {\bibinfo {author} {\bibfnamefont {C.~J.}\ \bibnamefont {Xin}}, \bibinfo {author} {\bibfnamefont {J.}~\bibnamefont {Mishra}}, \bibinfo {author} {\bibfnamefont {C.}~\bibnamefont {Chen}}, \bibinfo {author} {\bibfnamefont {D.}~\bibnamefont {Zhu}}, \bibinfo {author} {\bibfnamefont {A.}~\bibnamefont {{Shams-Ansari}}}, \bibinfo {author} {\bibfnamefont {C.}~\bibnamefont {Langrock}}, \bibinfo {author} {\bibfnamefont {N.}~\bibnamefont {Sinclair}}, \bibinfo {author} {\bibfnamefont {F.~N.~C.}\ \bibnamefont {Wong}}, \bibinfo {author} {\bibfnamefont {M.~M.}\ \bibnamefont {Fejer}},\ and\ \bibinfo {author} {\bibfnamefont {M.}~\bibnamefont {Lon{\v c}ar}},\ }\bibfield  {title} {\bibinfo {title} {Spectrally separable photon-pair generation in dispersion engineered thin-film lithium niobate},\ }\href {https://doi.org/10.1364/OL.456873} {\bibfield  {journal} {\bibinfo  {journal} {Optics Letters}\ }\textbf {\bibinfo {volume} {47}},\ \bibinfo {pages} {2830} (\bibinfo {year} {2022})}\BibitemShut {NoStop}%
\bibitem [{\citenamefont {U'Ren}\ \emph {et~al.}(2005)\citenamefont {U'Ren}, \citenamefont {Silberhorn}, \citenamefont {Banaszek}, \citenamefont {Walmsley}, \citenamefont {Erdmann}, \citenamefont {Grice},\ and\ \citenamefont {Raymer}}]{uren2005}%
  \BibitemOpen
  \bibfield  {author} {\bibinfo {author} {\bibfnamefont {A.~B.}\ \bibnamefont {U'Ren}}, \bibinfo {author} {\bibfnamefont {C.}~\bibnamefont {Silberhorn}}, \bibinfo {author} {\bibfnamefont {K.}~\bibnamefont {Banaszek}}, \bibinfo {author} {\bibfnamefont {I.~A.}\ \bibnamefont {Walmsley}}, \bibinfo {author} {\bibfnamefont {R.}~\bibnamefont {Erdmann}}, \bibinfo {author} {\bibfnamefont {W.~P.}\ \bibnamefont {Grice}},\ and\ \bibinfo {author} {\bibfnamefont {M.~G.}\ \bibnamefont {Raymer}},\ }\bibfield  {title} {\bibinfo {title} {Generation of {{Pure-State Single-Photon Wavepackets}} by {{Conditional Preparation Based}} on {{Spontaneous Parametric Downconversion}}},\ }\href {https://doi.org/10.48550/arXiv.quant-ph/0611019} {\bibfield  {journal} {\bibinfo  {journal} {Laser Physics}\ }\textbf {\bibinfo {volume} {15}},\ \bibinfo {pages} {146} (\bibinfo {year} {2005})}\BibitemShut {NoStop}%
\bibitem [{\citenamefont {{Poveda-Hospital}}\ \emph {et~al.}(2023)\citenamefont {{Poveda-Hospital}}, \citenamefont {Peter},\ and\ \citenamefont {Quesada}}]{poveda-hospital2023}%
  \BibitemOpen
  \bibfield  {author} {\bibinfo {author} {\bibfnamefont {S.}~\bibnamefont {{Poveda-Hospital}}}, \bibinfo {author} {\bibfnamefont {Y.-A.}\ \bibnamefont {Peter}},\ and\ \bibinfo {author} {\bibfnamefont {N.}~\bibnamefont {Quesada}},\ }\bibfield  {title} {\bibinfo {title} {Custom {{Nonlinearity Profile}} for {{Integrated Quantum Light Sources}}},\ }\href {https://doi.org/10.1103/PhysRevApplied.19.054033} {\bibfield  {journal} {\bibinfo  {journal} {Physical Review Applied}\ }\textbf {\bibinfo {volume} {19}},\ \bibinfo {pages} {054033} (\bibinfo {year} {2023})}\BibitemShut {NoStop}%
\bibitem [{\citenamefont {Christ}\ \emph {et~al.}(2013{\natexlab{b}})\citenamefont {Christ}, \citenamefont {Brecht}, \citenamefont {Mauerer},\ and\ \citenamefont {Silberhorn}}]{christ2013}%
  \BibitemOpen
  \bibfield  {author} {\bibinfo {author} {\bibfnamefont {A.}~\bibnamefont {Christ}}, \bibinfo {author} {\bibfnamefont {B.}~\bibnamefont {Brecht}}, \bibinfo {author} {\bibfnamefont {W.}~\bibnamefont {Mauerer}},\ and\ \bibinfo {author} {\bibfnamefont {C.}~\bibnamefont {Silberhorn}},\ }\bibfield  {title} {\bibinfo {title} {Theory of quantum frequency conversion and type-{{II}} parametric down-conversion in the high-gain regime},\ }\href {https://doi.org/10.1088/1367-2630/15/5/053038} {\bibfield  {journal} {\bibinfo  {journal} {New Journal of Physics}\ }\textbf {\bibinfo {volume} {15}},\ \bibinfo {pages} {053038} (\bibinfo {year} {2013}{\natexlab{b}})}\BibitemShut {NoStop}%
\bibitem [{\citenamefont {Shin}\ \emph {et~al.}(2023)\citenamefont {Shin}, \citenamefont {Park}, \citenamefont {Kim}, \citenamefont {Lee}, \citenamefont {Kwon},\ and\ \citenamefont {Shin}}]{shin2023}%
  \BibitemOpen
  \bibfield  {author} {\bibinfo {author} {\bibfnamefont {W.}~\bibnamefont {Shin}}, \bibinfo {author} {\bibfnamefont {K.}~\bibnamefont {Park}}, \bibinfo {author} {\bibfnamefont {H.}~\bibnamefont {Kim}}, \bibinfo {author} {\bibfnamefont {D.}~\bibnamefont {Lee}}, \bibinfo {author} {\bibfnamefont {K.}~\bibnamefont {Kwon}},\ and\ \bibinfo {author} {\bibfnamefont {H.}~\bibnamefont {Shin}},\ }\bibfield  {title} {\bibinfo {title} {Photon-pair generation in a lossy waveguide},\ }\href {https://doi.org/10.1515/nanoph-2022-0582} {\bibfield  {journal} {\bibinfo  {journal} {Nanophotonics}\ }\textbf {\bibinfo {volume} {12}},\ \bibinfo {pages} {531} (\bibinfo {year} {2023})}\BibitemShut {NoStop}%
\bibitem [{\citenamefont {Fejer}\ \emph {et~al.}(1992)\citenamefont {Fejer}, \citenamefont {Magel}, \citenamefont {Jundt},\ and\ \citenamefont {Byer}}]{fejer1992}%
  \BibitemOpen
  \bibfield  {author} {\bibinfo {author} {\bibfnamefont {M.}~\bibnamefont {Fejer}}, \bibinfo {author} {\bibfnamefont {G.}~\bibnamefont {Magel}}, \bibinfo {author} {\bibfnamefont {D.}~\bibnamefont {Jundt}},\ and\ \bibinfo {author} {\bibfnamefont {R.}~\bibnamefont {Byer}},\ }\bibfield  {title} {\bibinfo {title} {Quasi-phase-matched second harmonic generation: Tuning and tolerances},\ }\href {https://doi.org/10.1109/3.161322} {\bibfield  {journal} {\bibinfo  {journal} {IEEE Journal of Quantum Electronics}\ }\textbf {\bibinfo {volume} {28}},\ \bibinfo {pages} {2631} (\bibinfo {year} {1992})}\BibitemShut {NoStop}%
\bibitem [{\citenamefont {Bra{\'n}czyk}\ \emph {et~al.}(2011)\citenamefont {Bra{\'n}czyk}, \citenamefont {Fedrizzi}, \citenamefont {Stace}, \citenamefont {Ralph},\ and\ \citenamefont {White}}]{branczyk2011}%
  \BibitemOpen
  \bibfield  {author} {\bibinfo {author} {\bibfnamefont {A.~M.}\ \bibnamefont {Bra{\'n}czyk}}, \bibinfo {author} {\bibfnamefont {A.}~\bibnamefont {Fedrizzi}}, \bibinfo {author} {\bibfnamefont {T.~M.}\ \bibnamefont {Stace}}, \bibinfo {author} {\bibfnamefont {T.~C.}\ \bibnamefont {Ralph}},\ and\ \bibinfo {author} {\bibfnamefont {A.~G.}\ \bibnamefont {White}},\ }\bibfield  {title} {\bibinfo {title} {Engineered optical nonlinearity for quantum light sources},\ }\href {https://doi.org/10.1364/OE.19.000055} {\bibfield  {journal} {\bibinfo  {journal} {Optics Express}\ }\textbf {\bibinfo {volume} {19}},\ \bibinfo {pages} {55} (\bibinfo {year} {2011})}\BibitemShut {NoStop}%
\bibitem [{\citenamefont {Ben~Dixon}\ \emph {et~al.}(2013)\citenamefont {Ben~Dixon}, \citenamefont {Shapiro},\ and\ \citenamefont {Wong}}]{bendixon2013}%
  \BibitemOpen
  \bibfield  {author} {\bibinfo {author} {\bibfnamefont {P.}~\bibnamefont {Ben~Dixon}}, \bibinfo {author} {\bibfnamefont {J.~H.}\ \bibnamefont {Shapiro}},\ and\ \bibinfo {author} {\bibfnamefont {F.~N.~C.}\ \bibnamefont {Wong}},\ }\bibfield  {title} {\bibinfo {title} {Spectral engineering by {{Gaussian}} phase-matching for quantum photonics},\ }\href {https://doi.org/10.1364/OE.21.005879} {\bibfield  {journal} {\bibinfo  {journal} {Optics Express}\ }\textbf {\bibinfo {volume} {21}},\ \bibinfo {pages} {5879} (\bibinfo {year} {2013})}\BibitemShut {NoStop}%
\bibitem [{\citenamefont {Dosseva}\ \emph {et~al.}(2016)\citenamefont {Dosseva}, \citenamefont {Cincio},\ and\ \citenamefont {Bra{\'n}czyk}}]{dosseva2016}%
  \BibitemOpen
  \bibfield  {author} {\bibinfo {author} {\bibfnamefont {A.}~\bibnamefont {Dosseva}}, \bibinfo {author} {\bibfnamefont {{\L}.}~\bibnamefont {Cincio}},\ and\ \bibinfo {author} {\bibfnamefont {A.~M.}\ \bibnamefont {Bra{\'n}czyk}},\ }\bibfield  {title} {\bibinfo {title} {Shaping the joint spectrum of down-converted photons through optimized custom poling},\ }\href {https://doi.org/10.1103/PhysRevA.93.013801} {\bibfield  {journal} {\bibinfo  {journal} {Physical Review A}\ }\textbf {\bibinfo {volume} {93}},\ \bibinfo {pages} {013801} (\bibinfo {year} {2016})}\BibitemShut {NoStop}%
\bibitem [{\citenamefont {Tambasco}\ \emph {et~al.}(2016)\citenamefont {Tambasco}, \citenamefont {Boes}, \citenamefont {Helt}, \citenamefont {Steel},\ and\ \citenamefont {Mitchell}}]{tambasco2016}%
  \BibitemOpen
  \bibfield  {author} {\bibinfo {author} {\bibfnamefont {J.-L.}\ \bibnamefont {Tambasco}}, \bibinfo {author} {\bibfnamefont {A.}~\bibnamefont {Boes}}, \bibinfo {author} {\bibfnamefont {L.~G.}\ \bibnamefont {Helt}}, \bibinfo {author} {\bibfnamefont {M.~J.}\ \bibnamefont {Steel}},\ and\ \bibinfo {author} {\bibfnamefont {A.}~\bibnamefont {Mitchell}},\ }\bibfield  {title} {\bibinfo {title} {Domain engineering algorithm for practical and effective photon sources},\ }\href {https://doi.org/10.1364/OE.24.019616} {\bibfield  {journal} {\bibinfo  {journal} {Optics Express}\ }\textbf {\bibinfo {volume} {24}},\ \bibinfo {pages} {19616} (\bibinfo {year} {2016})}\BibitemShut {NoStop}%
\bibitem [{\citenamefont {Graffitti}\ \emph {et~al.}(2018)\citenamefont {Graffitti}, \citenamefont {Barrow}, \citenamefont {Proietti}, \citenamefont {Kundys},\ and\ \citenamefont {Fedrizzi}}]{graffitti2018}%
  \BibitemOpen
  \bibfield  {author} {\bibinfo {author} {\bibfnamefont {F.}~\bibnamefont {Graffitti}}, \bibinfo {author} {\bibfnamefont {P.}~\bibnamefont {Barrow}}, \bibinfo {author} {\bibfnamefont {M.}~\bibnamefont {Proietti}}, \bibinfo {author} {\bibfnamefont {D.}~\bibnamefont {Kundys}},\ and\ \bibinfo {author} {\bibfnamefont {A.}~\bibnamefont {Fedrizzi}},\ }\bibfield  {title} {\bibinfo {title} {Independent high-purity photons created in domain-engineered crystals},\ }\href {https://doi.org/10.1364/OPTICA.5.000514} {\bibfield  {journal} {\bibinfo  {journal} {Optica}\ }\textbf {\bibinfo {volume} {5}},\ \bibinfo {pages} {514} (\bibinfo {year} {2018})}\BibitemShut {NoStop}%
\bibitem [{\citenamefont {Graffitti}\ \emph {et~al.}(2017)\citenamefont {Graffitti}, \citenamefont {Kundys}, \citenamefont {Reid}, \citenamefont {Bra{\'n}czyk},\ and\ \citenamefont {Fedrizzi}}]{graffitti2017}%
  \BibitemOpen
  \bibfield  {author} {\bibinfo {author} {\bibfnamefont {F.}~\bibnamefont {Graffitti}}, \bibinfo {author} {\bibfnamefont {D.}~\bibnamefont {Kundys}}, \bibinfo {author} {\bibfnamefont {D.~T.}\ \bibnamefont {Reid}}, \bibinfo {author} {\bibfnamefont {A.~M.}\ \bibnamefont {Bra{\'n}czyk}},\ and\ \bibinfo {author} {\bibfnamefont {A.}~\bibnamefont {Fedrizzi}},\ }\bibfield  {title} {\bibinfo {title} {Pure down-conversion photons through sub-coherence-length domain engineering},\ }\href {https://doi.org/10.1088/2058-9565/aa78d4} {\bibfield  {journal} {\bibinfo  {journal} {Quantum Science and Technology}\ }\textbf {\bibinfo {volume} {2}},\ \bibinfo {pages} {035001} (\bibinfo {year} {2017})}\BibitemShut {NoStop}%
\bibitem [{\citenamefont {Yang}\ \emph {et~al.}(2007)\citenamefont {Yang}, \citenamefont {Chak}, \citenamefont {Bristow}, \citenamefont {{van Driel}}, \citenamefont {Iyer}, \citenamefont {Aitchison}, \citenamefont {Smirl},\ and\ \citenamefont {Sipe}}]{yang2007}%
  \BibitemOpen
  \bibfield  {author} {\bibinfo {author} {\bibfnamefont {Z.}~\bibnamefont {Yang}}, \bibinfo {author} {\bibfnamefont {P.}~\bibnamefont {Chak}}, \bibinfo {author} {\bibfnamefont {A.~D.}\ \bibnamefont {Bristow}}, \bibinfo {author} {\bibfnamefont {H.~M.}\ \bibnamefont {{van Driel}}}, \bibinfo {author} {\bibfnamefont {R.}~\bibnamefont {Iyer}}, \bibinfo {author} {\bibfnamefont {J.~S.}\ \bibnamefont {Aitchison}}, \bibinfo {author} {\bibfnamefont {A.~L.}\ \bibnamefont {Smirl}},\ and\ \bibinfo {author} {\bibfnamefont {J.~E.}\ \bibnamefont {Sipe}},\ }\bibfield  {title} {\bibinfo {title} {Enhanced second-harmonic generation in {{AlGaAs}} microring resonators},\ }\href {https://doi.org/10.1364/OL.32.000826} {\bibfield  {journal} {\bibinfo  {journal} {Optics Letters}\ }\textbf {\bibinfo {volume} {32}},\ \bibinfo {pages} {826} (\bibinfo {year} {2007})}\BibitemShut {NoStop}%
\bibitem [{\citenamefont {Yang}\ and\ \citenamefont {Sipe}(2007)}]{yang2007a}%
  \BibitemOpen
  \bibfield  {author} {\bibinfo {author} {\bibfnamefont {Z.}~\bibnamefont {Yang}}\ and\ \bibinfo {author} {\bibfnamefont {J.~E.}\ \bibnamefont {Sipe}},\ }\bibfield  {title} {\bibinfo {title} {Generating entangled photons via enhanced spontaneous parametric downconversion in {{AlGaAs}} microring resonators},\ }\href {https://doi.org/10.1364/OL.32.003296} {\bibfield  {journal} {\bibinfo  {journal} {Optics Letters}\ }\textbf {\bibinfo {volume} {32}},\ \bibinfo {pages} {3296} (\bibinfo {year} {2007})}\BibitemShut {NoStop}%
\bibitem [{\citenamefont {Shoji}\ \emph {et~al.}(1997)\citenamefont {Shoji}, \citenamefont {Kondo}, \citenamefont {Kitamoto}, \citenamefont {Shirane},\ and\ \citenamefont {Ito}}]{shoji1997}%
  \BibitemOpen
  \bibfield  {author} {\bibinfo {author} {\bibfnamefont {I.}~\bibnamefont {Shoji}}, \bibinfo {author} {\bibfnamefont {T.}~\bibnamefont {Kondo}}, \bibinfo {author} {\bibfnamefont {A.}~\bibnamefont {Kitamoto}}, \bibinfo {author} {\bibfnamefont {M.}~\bibnamefont {Shirane}},\ and\ \bibinfo {author} {\bibfnamefont {R.}~\bibnamefont {Ito}},\ }\bibfield  {title} {\bibinfo {title} {Absolute scale of second-order nonlinear-optical coefficients},\ }\href {https://doi.org/10.1364/JOSAB.14.002268} {\bibfield  {journal} {\bibinfo  {journal} {Journal of the Optical Society of America B}\ }\textbf {\bibinfo {volume} {14}},\ \bibinfo {pages} {2268} (\bibinfo {year} {1997})}\BibitemShut {NoStop}%
\bibitem [{\citenamefont {Reddy}\ \emph {et~al.}(2014)\citenamefont {Reddy}, \citenamefont {Raymer},\ and\ \citenamefont {McKinstrie}}]{reddy2014}%
  \BibitemOpen
  \bibfield  {author} {\bibinfo {author} {\bibfnamefont {D.~V.}\ \bibnamefont {Reddy}}, \bibinfo {author} {\bibfnamefont {M.~G.}\ \bibnamefont {Raymer}},\ and\ \bibinfo {author} {\bibfnamefont {C.~J.}\ \bibnamefont {McKinstrie}},\ }\bibfield  {title} {\bibinfo {title} {Efficient sorting of quantum-optical wave packets by temporal-mode interferometry},\ }\href {https://doi.org/10.1364/OL.39.002924} {\bibfield  {journal} {\bibinfo  {journal} {Optics Letters}\ }\textbf {\bibinfo {volume} {39}},\ \bibinfo {pages} {2924} (\bibinfo {year} {2014})}\BibitemShut {NoStop}%
\bibitem [{\citenamefont {Reddy}\ \emph {et~al.}(2015)\citenamefont {Reddy}, \citenamefont {Raymer},\ and\ \citenamefont {McKinstrie}}]{reddy2015}%
  \BibitemOpen
  \bibfield  {author} {\bibinfo {author} {\bibfnamefont {D.~V.}\ \bibnamefont {Reddy}}, \bibinfo {author} {\bibfnamefont {M.~G.}\ \bibnamefont {Raymer}},\ and\ \bibinfo {author} {\bibfnamefont {C.~J.}\ \bibnamefont {McKinstrie}},\ }\bibfield  {title} {\bibinfo {title} {Sorting photon wave packets using temporal-mode interferometry based on multiple-stage quantum frequency conversion},\ }\href {https://doi.org/10.1103/PhysRevA.91.012323} {\bibfield  {journal} {\bibinfo  {journal} {Physical Review A}\ }\textbf {\bibinfo {volume} {91}},\ \bibinfo {pages} {012323} (\bibinfo {year} {2015})}\BibitemShut {NoStop}%
\bibitem [{\citenamefont {Ansari}\ \emph {et~al.}(2018{\natexlab{a}})\citenamefont {Ansari}, \citenamefont {Donohue}, \citenamefont {Brecht},\ and\ \citenamefont {Silberhorn}}]{ansari2018b}%
  \BibitemOpen
  \bibfield  {author} {\bibinfo {author} {\bibfnamefont {V.}~\bibnamefont {Ansari}}, \bibinfo {author} {\bibfnamefont {J.~M.}\ \bibnamefont {Donohue}}, \bibinfo {author} {\bibfnamefont {B.}~\bibnamefont {Brecht}},\ and\ \bibinfo {author} {\bibfnamefont {C.}~\bibnamefont {Silberhorn}},\ }\bibfield  {title} {\bibinfo {title} {Tailoring nonlinear processes for quantum optics with pulsed temporal-mode encodings},\ }\href {https://doi.org/10.1364/OPTICA.5.000534} {\bibfield  {journal} {\bibinfo  {journal} {Optica}\ }\textbf {\bibinfo {volume} {5}},\ \bibinfo {pages} {534} (\bibinfo {year} {2018}{\natexlab{a}})}\BibitemShut {NoStop}%
\bibitem [{\citenamefont {Houde}\ and\ \citenamefont {Quesada}(2023)}]{houde2023}%
  \BibitemOpen
  \bibfield  {author} {\bibinfo {author} {\bibfnamefont {M.}~\bibnamefont {Houde}}\ and\ \bibinfo {author} {\bibfnamefont {N.}~\bibnamefont {Quesada}},\ }\bibfield  {title} {\bibinfo {title} {Waveguided sources of consistent, single-temporal-mode squeezed light: The good, the bad, and the ugly},\ }\href {https://doi.org/10.1116/5.0133009} {\bibfield  {journal} {\bibinfo  {journal} {AVS Quantum Science}\ }\textbf {\bibinfo {volume} {5}},\ \bibinfo {pages} {011404} (\bibinfo {year} {2023})}\BibitemShut {NoStop}%
\bibitem [{\citenamefont {Eisenberg}\ \emph {et~al.}(2004)\citenamefont {Eisenberg}, \citenamefont {Khoury}, \citenamefont {Durkin}, \citenamefont {Simon},\ and\ \citenamefont {Bouwmeester}}]{eisenberg2004}%
  \BibitemOpen
  \bibfield  {author} {\bibinfo {author} {\bibfnamefont {H.~S.}\ \bibnamefont {Eisenberg}}, \bibinfo {author} {\bibfnamefont {G.}~\bibnamefont {Khoury}}, \bibinfo {author} {\bibfnamefont {G.~A.}\ \bibnamefont {Durkin}}, \bibinfo {author} {\bibfnamefont {C.}~\bibnamefont {Simon}},\ and\ \bibinfo {author} {\bibfnamefont {D.}~\bibnamefont {Bouwmeester}},\ }\bibfield  {title} {\bibinfo {title} {Quantum {{Entanglement}} of a {{Large Number}} of {{Photons}}},\ }\href {https://doi.org/10.1103/PhysRevLett.93.193901} {\bibfield  {journal} {\bibinfo  {journal} {Physical Review Letters}\ }\textbf {\bibinfo {volume} {93}},\ \bibinfo {pages} {193901} (\bibinfo {year} {2004})}\BibitemShut {NoStop}%
\bibitem [{\citenamefont {Zhong}\ \emph {et~al.}(2021)\citenamefont {Zhong}, \citenamefont {Deng}, \citenamefont {Qin}, \citenamefont {Wang}, \citenamefont {Chen}, \citenamefont {Peng}, \citenamefont {Luo}, \citenamefont {Wu}, \citenamefont {Gong}, \citenamefont {Su}, \citenamefont {Hu}, \citenamefont {Hu}, \citenamefont {Yang}, \citenamefont {Zhang}, \citenamefont {Li}, \citenamefont {Li}, \citenamefont {Jiang}, \citenamefont {Gan}, \citenamefont {Yang}, \citenamefont {You}, \citenamefont {Wang}, \citenamefont {Li}, \citenamefont {Liu}, \citenamefont {Renema}, \citenamefont {Lu},\ and\ \citenamefont {Pan}}]{zhong2021}%
  \BibitemOpen
  \bibfield  {author} {\bibinfo {author} {\bibfnamefont {H.-S.}\ \bibnamefont {Zhong}}, \bibinfo {author} {\bibfnamefont {Y.-H.}\ \bibnamefont {Deng}}, \bibinfo {author} {\bibfnamefont {J.}~\bibnamefont {Qin}}, \bibinfo {author} {\bibfnamefont {H.}~\bibnamefont {Wang}}, \bibinfo {author} {\bibfnamefont {M.-C.}\ \bibnamefont {Chen}}, \bibinfo {author} {\bibfnamefont {L.-C.}\ \bibnamefont {Peng}}, \bibinfo {author} {\bibfnamefont {Y.-H.}\ \bibnamefont {Luo}}, \bibinfo {author} {\bibfnamefont {D.}~\bibnamefont {Wu}}, \bibinfo {author} {\bibfnamefont {S.-Q.}\ \bibnamefont {Gong}}, \bibinfo {author} {\bibfnamefont {H.}~\bibnamefont {Su}}, \bibinfo {author} {\bibfnamefont {Y.}~\bibnamefont {Hu}}, \bibinfo {author} {\bibfnamefont {P.}~\bibnamefont {Hu}}, \bibinfo {author} {\bibfnamefont {X.-Y.}\ \bibnamefont {Yang}}, \bibinfo {author} {\bibfnamefont {W.-J.}\ \bibnamefont {Zhang}}, \bibinfo {author} {\bibfnamefont {H.}~\bibnamefont {Li}}, \bibinfo {author} {\bibfnamefont {Y.}~\bibnamefont {Li}}, \bibinfo {author}
  {\bibfnamefont {X.}~\bibnamefont {Jiang}}, \bibinfo {author} {\bibfnamefont {L.}~\bibnamefont {Gan}}, \bibinfo {author} {\bibfnamefont {G.}~\bibnamefont {Yang}}, \bibinfo {author} {\bibfnamefont {L.}~\bibnamefont {You}}, \bibinfo {author} {\bibfnamefont {Z.}~\bibnamefont {Wang}}, \bibinfo {author} {\bibfnamefont {L.}~\bibnamefont {Li}}, \bibinfo {author} {\bibfnamefont {N.-L.}\ \bibnamefont {Liu}}, \bibinfo {author} {\bibfnamefont {J.~J.}\ \bibnamefont {Renema}}, \bibinfo {author} {\bibfnamefont {C.-Y.}\ \bibnamefont {Lu}},\ and\ \bibinfo {author} {\bibfnamefont {J.-W.}\ \bibnamefont {Pan}},\ }\bibfield  {title} {\bibinfo {title} {Phase-{{Programmable Gaussian Boson Sampling Using Stimulated Squeezed Light}}},\ }\href {https://doi.org/10.1103/PhysRevLett.127.180502} {\bibfield  {journal} {\bibinfo  {journal} {Physical Review Letters}\ }\textbf {\bibinfo {volume} {127}},\ \bibinfo {pages} {180502} (\bibinfo {year} {2021})}\BibitemShut {NoStop}%
\bibitem [{\citenamefont {Reddy}\ and\ \citenamefont {Raymer}(2018)}]{reddy2018}%
  \BibitemOpen
  \bibfield  {author} {\bibinfo {author} {\bibfnamefont {D.~V.}\ \bibnamefont {Reddy}}\ and\ \bibinfo {author} {\bibfnamefont {M.~G.}\ \bibnamefont {Raymer}},\ }\bibfield  {title} {\bibinfo {title} {High-selectivity quantum pulse gating of photonic temporal modes using all-optical {{Ramsey}} interferometry},\ }\href {https://doi.org/10.1364/OPTICA.5.000423} {\bibfield  {journal} {\bibinfo  {journal} {Optica}\ }\textbf {\bibinfo {volume} {5}},\ \bibinfo {pages} {423} (\bibinfo {year} {2018})}\BibitemShut {NoStop}%
\bibitem [{\citenamefont {Kim}\ \emph {et~al.}(2000{\natexlab{a}})\citenamefont {Kim}, \citenamefont {Chekhova}, \citenamefont {Kulik}, \citenamefont {Shih},\ and\ \citenamefont {Rubin}}]{kim2000}%
  \BibitemOpen
  \bibfield  {author} {\bibinfo {author} {\bibfnamefont {Y.-H.}\ \bibnamefont {Kim}}, \bibinfo {author} {\bibfnamefont {M.~V.}\ \bibnamefont {Chekhova}}, \bibinfo {author} {\bibfnamefont {S.~P.}\ \bibnamefont {Kulik}}, \bibinfo {author} {\bibfnamefont {Y.}~\bibnamefont {Shih}},\ and\ \bibinfo {author} {\bibfnamefont {M.~H.}\ \bibnamefont {Rubin}},\ }\bibfield  {title} {\bibinfo {title} {First-order interference of nonclassical light emitted spontaneously at different times},\ }\href {https://doi.org/10.1103/PhysRevA.61.051803} {\bibfield  {journal} {\bibinfo  {journal} {Physical Review A}\ }\textbf {\bibinfo {volume} {61}},\ \bibinfo {pages} {051803} (\bibinfo {year} {2000}{\natexlab{a}})}\BibitemShut {NoStop}%
\bibitem [{\citenamefont {Kim}\ \emph {et~al.}(2000{\natexlab{b}})\citenamefont {Kim}, \citenamefont {Berardi}, \citenamefont {Chekhova}, \citenamefont {Garuccio},\ and\ \citenamefont {Shih}}]{kim2000c}%
  \BibitemOpen
  \bibfield  {author} {\bibinfo {author} {\bibfnamefont {Y.-H.}\ \bibnamefont {Kim}}, \bibinfo {author} {\bibfnamefont {V.}~\bibnamefont {Berardi}}, \bibinfo {author} {\bibfnamefont {M.~V.}\ \bibnamefont {Chekhova}}, \bibinfo {author} {\bibfnamefont {A.}~\bibnamefont {Garuccio}},\ and\ \bibinfo {author} {\bibfnamefont {Y.}~\bibnamefont {Shih}},\ }\bibfield  {title} {\bibinfo {title} {Temporal indistinguishability and quantum interference},\ }\href {https://doi.org/10.1103/PhysRevA.62.043820} {\bibfield  {journal} {\bibinfo  {journal} {Physical Review A}\ }\textbf {\bibinfo {volume} {62}},\ \bibinfo {pages} {043820} (\bibinfo {year} {2000}{\natexlab{b}})}\BibitemShut {NoStop}%
\bibitem [{\citenamefont {Zou}\ \emph {et~al.}(1991)\citenamefont {Zou}, \citenamefont {Wang},\ and\ \citenamefont {Mandel}}]{zou1991}%
  \BibitemOpen
  \bibfield  {author} {\bibinfo {author} {\bibfnamefont {X.~Y.}\ \bibnamefont {Zou}}, \bibinfo {author} {\bibfnamefont {L.~J.}\ \bibnamefont {Wang}},\ and\ \bibinfo {author} {\bibfnamefont {L.}~\bibnamefont {Mandel}},\ }\bibfield  {title} {\bibinfo {title} {Induced coherence and indistinguishability in optical interference},\ }\href {https://doi.org/10.1103/PhysRevLett.67.318} {\bibfield  {journal} {\bibinfo  {journal} {Physical Review Letters}\ }\textbf {\bibinfo {volume} {67}},\ \bibinfo {pages} {318} (\bibinfo {year} {1991})}\BibitemShut {NoStop}%
\bibitem [{\citenamefont {Chekhova}\ and\ \citenamefont {Ou}(2016)}]{chekhova2016}%
  \BibitemOpen
  \bibfield  {author} {\bibinfo {author} {\bibfnamefont {M.~V.}\ \bibnamefont {Chekhova}}\ and\ \bibinfo {author} {\bibfnamefont {Z.~Y.}\ \bibnamefont {Ou}},\ }\bibfield  {title} {\bibinfo {title} {Nonlinear interferometers in quantum optics},\ }\href {https://doi.org/10.1364/AOP.8.000104} {\bibfield  {journal} {\bibinfo  {journal} {Advances in Optics and Photonics}\ }\textbf {\bibinfo {volume} {8}},\ \bibinfo {pages} {104} (\bibinfo {year} {2016})}\BibitemShut {NoStop}%
\bibitem [{\citenamefont {Reddy}\ \emph {et~al.}(2013)\citenamefont {Reddy}, \citenamefont {Raymer}, \citenamefont {McKinstrie}, \citenamefont {Mejling},\ and\ \citenamefont {Rottwitt}}]{reddy2013}%
  \BibitemOpen
  \bibfield  {author} {\bibinfo {author} {\bibfnamefont {D.~V.}\ \bibnamefont {Reddy}}, \bibinfo {author} {\bibfnamefont {M.~G.}\ \bibnamefont {Raymer}}, \bibinfo {author} {\bibfnamefont {C.~J.}\ \bibnamefont {McKinstrie}}, \bibinfo {author} {\bibfnamefont {L.}~\bibnamefont {Mejling}},\ and\ \bibinfo {author} {\bibfnamefont {K.}~\bibnamefont {Rottwitt}},\ }\bibfield  {title} {\bibinfo {title} {Temporal mode selectivity by frequency conversion in second-order nonlinear optical waveguides},\ }\href {https://doi.org/10.1364/OE.21.013840} {\bibfield  {journal} {\bibinfo  {journal} {Optics Express}\ }\textbf {\bibinfo {volume} {21}},\ \bibinfo {pages} {13840} (\bibinfo {year} {2013})}\BibitemShut {NoStop}%
\bibitem [{\citenamefont {Onodera}\ \emph {et~al.}(2016)\citenamefont {Onodera}, \citenamefont {Liscidini}, \citenamefont {Sipe},\ and\ \citenamefont {Helt}}]{onodera2016}%
  \BibitemOpen
  \bibfield  {author} {\bibinfo {author} {\bibfnamefont {T.}~\bibnamefont {Onodera}}, \bibinfo {author} {\bibfnamefont {M.}~\bibnamefont {Liscidini}}, \bibinfo {author} {\bibfnamefont {J.~E.}\ \bibnamefont {Sipe}},\ and\ \bibinfo {author} {\bibfnamefont {L.~G.}\ \bibnamefont {Helt}},\ }\bibfield  {title} {\bibinfo {title} {Parametric fluorescence in a sequence of resonators: {{An}} analogy with {{Dicke}} superradiance},\ }\href {https://doi.org/10.1103/PhysRevA.93.043837} {\bibfield  {journal} {\bibinfo  {journal} {Physical Review A}\ }\textbf {\bibinfo {volume} {93}},\ \bibinfo {pages} {043837} (\bibinfo {year} {2016})}\BibitemShut {NoStop}%
\bibitem [{\citenamefont {Zhang}\ \emph {et~al.}(2017)\citenamefont {Zhang}, \citenamefont {Wang}, \citenamefont {Cheng}, \citenamefont {{Shams-Ansari}},\ and\ \citenamefont {Lon{\v c}ar}}]{zhang2017}%
  \BibitemOpen
  \bibfield  {author} {\bibinfo {author} {\bibfnamefont {M.}~\bibnamefont {Zhang}}, \bibinfo {author} {\bibfnamefont {C.}~\bibnamefont {Wang}}, \bibinfo {author} {\bibfnamefont {R.}~\bibnamefont {Cheng}}, \bibinfo {author} {\bibfnamefont {A.}~\bibnamefont {{Shams-Ansari}}},\ and\ \bibinfo {author} {\bibfnamefont {M.}~\bibnamefont {Lon{\v c}ar}},\ }\bibfield  {title} {\bibinfo {title} {Monolithic ultra-high-{{Q}} lithium niobate microring resonator},\ }\href {https://doi.org/10.1364/OPTICA.4.001536} {\bibfield  {journal} {\bibinfo  {journal} {Optica}\ }\textbf {\bibinfo {volume} {4}},\ \bibinfo {pages} {1536} (\bibinfo {year} {2017})}\BibitemShut {NoStop}%
\bibitem [{\citenamefont {Roslund}\ \emph {et~al.}(2014)\citenamefont {Roslund}, \citenamefont {De~Ara{\'u}jo}, \citenamefont {Jiang}, \citenamefont {Fabre},\ and\ \citenamefont {Treps}}]{roslund2014}%
  \BibitemOpen
  \bibfield  {author} {\bibinfo {author} {\bibfnamefont {J.}~\bibnamefont {Roslund}}, \bibinfo {author} {\bibfnamefont {R.~M.}\ \bibnamefont {De~Ara{\'u}jo}}, \bibinfo {author} {\bibfnamefont {S.}~\bibnamefont {Jiang}}, \bibinfo {author} {\bibfnamefont {C.}~\bibnamefont {Fabre}},\ and\ \bibinfo {author} {\bibfnamefont {N.}~\bibnamefont {Treps}},\ }\bibfield  {title} {\bibinfo {title} {Wavelength-multiplexed quantum networks with ultrafast frequency combs},\ }\href {https://doi.org/10.1038/nphoton.2013.340} {\bibfield  {journal} {\bibinfo  {journal} {Nature Photonics}\ }\textbf {\bibinfo {volume} {8}},\ \bibinfo {pages} {109} (\bibinfo {year} {2014})}\BibitemShut {NoStop}%
\bibitem [{\citenamefont {Zhong}\ \emph {et~al.}(2020)\citenamefont {Zhong}, \citenamefont {Wang}, \citenamefont {Deng}, \citenamefont {Chen}, \citenamefont {Peng}, \citenamefont {Luo}, \citenamefont {Qin}, \citenamefont {Wu}, \citenamefont {Ding}, \citenamefont {Hu}, \citenamefont {Hu}, \citenamefont {Yang}, \citenamefont {Zhang}, \citenamefont {Li}, \citenamefont {Li}, \citenamefont {Jiang}, \citenamefont {Gan}, \citenamefont {Yang}, \citenamefont {You}, \citenamefont {Wang}, \citenamefont {Li}, \citenamefont {Liu}, \citenamefont {Lu},\ and\ \citenamefont {Pan}}]{zhong2020}%
  \BibitemOpen
  \bibfield  {author} {\bibinfo {author} {\bibfnamefont {H.-S.}\ \bibnamefont {Zhong}}, \bibinfo {author} {\bibfnamefont {H.}~\bibnamefont {Wang}}, \bibinfo {author} {\bibfnamefont {Y.-H.}\ \bibnamefont {Deng}}, \bibinfo {author} {\bibfnamefont {M.-C.}\ \bibnamefont {Chen}}, \bibinfo {author} {\bibfnamefont {L.-C.}\ \bibnamefont {Peng}}, \bibinfo {author} {\bibfnamefont {Y.-H.}\ \bibnamefont {Luo}}, \bibinfo {author} {\bibfnamefont {J.}~\bibnamefont {Qin}}, \bibinfo {author} {\bibfnamefont {D.}~\bibnamefont {Wu}}, \bibinfo {author} {\bibfnamefont {X.}~\bibnamefont {Ding}}, \bibinfo {author} {\bibfnamefont {Y.}~\bibnamefont {Hu}}, \bibinfo {author} {\bibfnamefont {P.}~\bibnamefont {Hu}}, \bibinfo {author} {\bibfnamefont {X.-Y.}\ \bibnamefont {Yang}}, \bibinfo {author} {\bibfnamefont {W.-J.}\ \bibnamefont {Zhang}}, \bibinfo {author} {\bibfnamefont {H.}~\bibnamefont {Li}}, \bibinfo {author} {\bibfnamefont {Y.}~\bibnamefont {Li}}, \bibinfo {author} {\bibfnamefont {X.}~\bibnamefont {Jiang}}, \bibinfo {author}
  {\bibfnamefont {L.}~\bibnamefont {Gan}}, \bibinfo {author} {\bibfnamefont {G.}~\bibnamefont {Yang}}, \bibinfo {author} {\bibfnamefont {L.}~\bibnamefont {You}}, \bibinfo {author} {\bibfnamefont {Z.}~\bibnamefont {Wang}}, \bibinfo {author} {\bibfnamefont {L.}~\bibnamefont {Li}}, \bibinfo {author} {\bibfnamefont {N.-L.}\ \bibnamefont {Liu}}, \bibinfo {author} {\bibfnamefont {C.-Y.}\ \bibnamefont {Lu}},\ and\ \bibinfo {author} {\bibfnamefont {J.-W.}\ \bibnamefont {Pan}},\ }\bibfield  {title} {\bibinfo {title} {Quantum computational advantage using photons},\ }\href {https://doi.org/10.1126/science.abe8770} {\bibfield  {journal} {\bibinfo  {journal} {Science}\ }\textbf {\bibinfo {volume} {370}},\ \bibinfo {pages} {1460} (\bibinfo {year} {2020})}\BibitemShut {NoStop}%
\bibitem [{\citenamefont {Arrazola}\ \emph {et~al.}(2021)\citenamefont {Arrazola}, \citenamefont {Bergholm}, \citenamefont {Br{\'a}dler}, \citenamefont {Bromley}, \citenamefont {Collins}, \citenamefont {Dhand}, \citenamefont {Fumagalli}, \citenamefont {Gerrits}, \citenamefont {Goussev}, \citenamefont {Helt}, \citenamefont {Hundal}, \citenamefont {Isacsson}, \citenamefont {Israel}, \citenamefont {Izaac}, \citenamefont {Jahangiri}, \citenamefont {Janik}, \citenamefont {Killoran}, \citenamefont {Kumar}, \citenamefont {Lavoie}, \citenamefont {Lita}, \citenamefont {Mahler}, \citenamefont {Menotti}, \citenamefont {Morrison}, \citenamefont {Nam}, \citenamefont {Neuhaus}, \citenamefont {Qi}, \citenamefont {Quesada}, \citenamefont {Repingon}, \citenamefont {Sabapathy}, \citenamefont {Schuld}, \citenamefont {Su}, \citenamefont {Swinarton}, \citenamefont {Sz{\'a}va}, \citenamefont {Tan}, \citenamefont {Tan}, \citenamefont {Vaidya}, \citenamefont {Vernon}, \citenamefont {Zabaneh},\ and\ \citenamefont
  {Zhang}}]{arrazola2021}%
  \BibitemOpen
  \bibfield  {author} {\bibinfo {author} {\bibfnamefont {J.~M.}\ \bibnamefont {Arrazola}}, \bibinfo {author} {\bibfnamefont {V.}~\bibnamefont {Bergholm}}, \bibinfo {author} {\bibfnamefont {K.}~\bibnamefont {Br{\'a}dler}}, \bibinfo {author} {\bibfnamefont {T.~R.}\ \bibnamefont {Bromley}}, \bibinfo {author} {\bibfnamefont {M.~J.}\ \bibnamefont {Collins}}, \bibinfo {author} {\bibfnamefont {I.}~\bibnamefont {Dhand}}, \bibinfo {author} {\bibfnamefont {A.}~\bibnamefont {Fumagalli}}, \bibinfo {author} {\bibfnamefont {T.}~\bibnamefont {Gerrits}}, \bibinfo {author} {\bibfnamefont {A.}~\bibnamefont {Goussev}}, \bibinfo {author} {\bibfnamefont {L.~G.}\ \bibnamefont {Helt}}, \bibinfo {author} {\bibfnamefont {J.}~\bibnamefont {Hundal}}, \bibinfo {author} {\bibfnamefont {T.}~\bibnamefont {Isacsson}}, \bibinfo {author} {\bibfnamefont {R.~B.}\ \bibnamefont {Israel}}, \bibinfo {author} {\bibfnamefont {J.}~\bibnamefont {Izaac}}, \bibinfo {author} {\bibfnamefont {S.}~\bibnamefont {Jahangiri}}, \bibinfo {author} {\bibfnamefont
  {R.}~\bibnamefont {Janik}}, \bibinfo {author} {\bibfnamefont {N.}~\bibnamefont {Killoran}}, \bibinfo {author} {\bibfnamefont {S.~P.}\ \bibnamefont {Kumar}}, \bibinfo {author} {\bibfnamefont {J.}~\bibnamefont {Lavoie}}, \bibinfo {author} {\bibfnamefont {A.~E.}\ \bibnamefont {Lita}}, \bibinfo {author} {\bibfnamefont {D.~H.}\ \bibnamefont {Mahler}}, \bibinfo {author} {\bibfnamefont {M.}~\bibnamefont {Menotti}}, \bibinfo {author} {\bibfnamefont {B.}~\bibnamefont {Morrison}}, \bibinfo {author} {\bibfnamefont {S.~W.}\ \bibnamefont {Nam}}, \bibinfo {author} {\bibfnamefont {L.}~\bibnamefont {Neuhaus}}, \bibinfo {author} {\bibfnamefont {H.~Y.}\ \bibnamefont {Qi}}, \bibinfo {author} {\bibfnamefont {N.}~\bibnamefont {Quesada}}, \bibinfo {author} {\bibfnamefont {A.}~\bibnamefont {Repingon}}, \bibinfo {author} {\bibfnamefont {K.~K.}\ \bibnamefont {Sabapathy}}, \bibinfo {author} {\bibfnamefont {M.}~\bibnamefont {Schuld}}, \bibinfo {author} {\bibfnamefont {D.}~\bibnamefont {Su}}, \bibinfo {author} {\bibfnamefont
  {J.}~\bibnamefont {Swinarton}}, \bibinfo {author} {\bibfnamefont {A.}~\bibnamefont {Sz{\'a}va}}, \bibinfo {author} {\bibfnamefont {K.}~\bibnamefont {Tan}}, \bibinfo {author} {\bibfnamefont {P.}~\bibnamefont {Tan}}, \bibinfo {author} {\bibfnamefont {V.~D.}\ \bibnamefont {Vaidya}}, \bibinfo {author} {\bibfnamefont {Z.}~\bibnamefont {Vernon}}, \bibinfo {author} {\bibfnamefont {Z.}~\bibnamefont {Zabaneh}},\ and\ \bibinfo {author} {\bibfnamefont {Y.}~\bibnamefont {Zhang}},\ }\bibfield  {title} {\bibinfo {title} {Quantum circuits with many photons on a programmable nanophotonic chip},\ }\href {https://doi.org/10.1038/s41586-021-03202-1} {\bibfield  {journal} {\bibinfo  {journal} {Nature}\ }\textbf {\bibinfo {volume} {591}},\ \bibinfo {pages} {54} (\bibinfo {year} {2021})}\BibitemShut {NoStop}%
\bibitem [{\citenamefont {Larsen}\ \emph {et~al.}(2019)\citenamefont {Larsen}, \citenamefont {Guo}, \citenamefont {Breum}, \citenamefont {{Neergaard-Nielsen}},\ and\ \citenamefont {Andersen}}]{larsen2019}%
  \BibitemOpen
  \bibfield  {author} {\bibinfo {author} {\bibfnamefont {M.~V.}\ \bibnamefont {Larsen}}, \bibinfo {author} {\bibfnamefont {X.}~\bibnamefont {Guo}}, \bibinfo {author} {\bibfnamefont {C.~R.}\ \bibnamefont {Breum}}, \bibinfo {author} {\bibfnamefont {J.~S.}\ \bibnamefont {{Neergaard-Nielsen}}},\ and\ \bibinfo {author} {\bibfnamefont {U.~L.}\ \bibnamefont {Andersen}},\ }\bibfield  {title} {\bibinfo {title} {Deterministic generation of a two-dimensional cluster state},\ }\href {https://doi.org/10.1126/science.aay4354} {\bibfield  {journal} {\bibinfo  {journal} {Science}\ }\textbf {\bibinfo {volume} {366}},\ \bibinfo {pages} {369} (\bibinfo {year} {2019})}\BibitemShut {NoStop}%
\bibitem [{\citenamefont {Asavanant}\ \emph {et~al.}(2019)\citenamefont {Asavanant}, \citenamefont {Shiozawa}, \citenamefont {Yokoyama}, \citenamefont {Charoensombutamon}, \citenamefont {Emura}, \citenamefont {Alexander}, \citenamefont {Takeda}, \citenamefont {Yoshikawa}, \citenamefont {Menicucci}, \citenamefont {Yonezawa},\ and\ \citenamefont {Furusawa}}]{asavanant2019}%
  \BibitemOpen
  \bibfield  {author} {\bibinfo {author} {\bibfnamefont {W.}~\bibnamefont {Asavanant}}, \bibinfo {author} {\bibfnamefont {Y.}~\bibnamefont {Shiozawa}}, \bibinfo {author} {\bibfnamefont {S.}~\bibnamefont {Yokoyama}}, \bibinfo {author} {\bibfnamefont {B.}~\bibnamefont {Charoensombutamon}}, \bibinfo {author} {\bibfnamefont {H.}~\bibnamefont {Emura}}, \bibinfo {author} {\bibfnamefont {R.~N.}\ \bibnamefont {Alexander}}, \bibinfo {author} {\bibfnamefont {S.}~\bibnamefont {Takeda}}, \bibinfo {author} {\bibfnamefont {J.-i.}\ \bibnamefont {Yoshikawa}}, \bibinfo {author} {\bibfnamefont {N.~C.}\ \bibnamefont {Menicucci}}, \bibinfo {author} {\bibfnamefont {H.}~\bibnamefont {Yonezawa}},\ and\ \bibinfo {author} {\bibfnamefont {A.}~\bibnamefont {Furusawa}},\ }\bibfield  {title} {\bibinfo {title} {Generation of time-domain-multiplexed two-dimensional cluster state},\ }\href {https://doi.org/10.1126/science.aay2645} {\bibfield  {journal} {\bibinfo  {journal} {Science}\ }\textbf {\bibinfo {volume} {366}},\ \bibinfo {pages}
  {373} (\bibinfo {year} {2019})}\BibitemShut {NoStop}%
\bibitem [{\citenamefont {Takeda}\ \emph {et~al.}(2013)\citenamefont {Takeda}, \citenamefont {Mizuta}, \citenamefont {Fuwa}, \citenamefont {Van~Loock},\ and\ \citenamefont {Furusawa}}]{takeda2013}%
  \BibitemOpen
  \bibfield  {author} {\bibinfo {author} {\bibfnamefont {S.}~\bibnamefont {Takeda}}, \bibinfo {author} {\bibfnamefont {T.}~\bibnamefont {Mizuta}}, \bibinfo {author} {\bibfnamefont {M.}~\bibnamefont {Fuwa}}, \bibinfo {author} {\bibfnamefont {P.}~\bibnamefont {Van~Loock}},\ and\ \bibinfo {author} {\bibfnamefont {A.}~\bibnamefont {Furusawa}},\ }\bibfield  {title} {\bibinfo {title} {Deterministic quantum teleportation of photonic quantum bits by a hybrid technique},\ }\href {https://doi.org/10.1038/nature12366} {\bibfield  {journal} {\bibinfo  {journal} {Nature}\ }\textbf {\bibinfo {volume} {500}},\ \bibinfo {pages} {315} (\bibinfo {year} {2013})}\BibitemShut {NoStop}%
\bibitem [{\citenamefont {He}\ and\ \citenamefont {Malaney}(2022)}]{he2022}%
  \BibitemOpen
  \bibfield  {author} {\bibinfo {author} {\bibfnamefont {M.}~\bibnamefont {He}}\ and\ \bibinfo {author} {\bibfnamefont {R.}~\bibnamefont {Malaney}},\ }\bibfield  {title} {\bibinfo {title} {Teleportation of hybrid entangled states with continuous-variable entanglement},\ }\href {https://doi.org/10.1038/s41598-022-21283-4} {\bibfield  {journal} {\bibinfo  {journal} {Scientific Reports}\ }\textbf {\bibinfo {volume} {12}},\ \bibinfo {pages} {17169} (\bibinfo {year} {2022})}\BibitemShut {NoStop}%
\bibitem [{\citenamefont {Ansari}\ \emph {et~al.}(2018{\natexlab{b}})\citenamefont {Ansari}, \citenamefont {Roccia}, \citenamefont {Santandrea}, \citenamefont {Doostdar}, \citenamefont {Eigner}, \citenamefont {Padberg}, \citenamefont {Gianani}, \citenamefont {Sbroscia}, \citenamefont {Donohue}, \citenamefont {Mancino}, \citenamefont {Barbieri},\ and\ \citenamefont {Silberhorn}}]{ansari2018a}%
  \BibitemOpen
  \bibfield  {author} {\bibinfo {author} {\bibfnamefont {V.}~\bibnamefont {Ansari}}, \bibinfo {author} {\bibfnamefont {E.}~\bibnamefont {Roccia}}, \bibinfo {author} {\bibfnamefont {M.}~\bibnamefont {Santandrea}}, \bibinfo {author} {\bibfnamefont {M.}~\bibnamefont {Doostdar}}, \bibinfo {author} {\bibfnamefont {C.}~\bibnamefont {Eigner}}, \bibinfo {author} {\bibfnamefont {L.}~\bibnamefont {Padberg}}, \bibinfo {author} {\bibfnamefont {I.}~\bibnamefont {Gianani}}, \bibinfo {author} {\bibfnamefont {M.}~\bibnamefont {Sbroscia}}, \bibinfo {author} {\bibfnamefont {J.~M.}\ \bibnamefont {Donohue}}, \bibinfo {author} {\bibfnamefont {L.}~\bibnamefont {Mancino}}, \bibinfo {author} {\bibfnamefont {M.}~\bibnamefont {Barbieri}},\ and\ \bibinfo {author} {\bibfnamefont {C.}~\bibnamefont {Silberhorn}},\ }\bibfield  {title} {\bibinfo {title} {Heralded generation of high-purity ultrashort single photons in programmable temporal shapes},\ }\href {https://doi.org/10.1364/OE.26.002764} {\bibfield  {journal} {\bibinfo  {journal}
  {Optics Express}\ }\textbf {\bibinfo {volume} {26}},\ \bibinfo {pages} {2764} (\bibinfo {year} {2018}{\natexlab{b}})}\BibitemShut {NoStop}%
\bibitem [{\citenamefont {Brecht}\ \emph {et~al.}(2015)\citenamefont {Brecht}, \citenamefont {Reddy}, \citenamefont {Silberhorn},\ and\ \citenamefont {Raymer}}]{brecht2015}%
  \BibitemOpen
  \bibfield  {author} {\bibinfo {author} {\bibfnamefont {B.}~\bibnamefont {Brecht}}, \bibinfo {author} {\bibfnamefont {D.~V.}\ \bibnamefont {Reddy}}, \bibinfo {author} {\bibfnamefont {C.}~\bibnamefont {Silberhorn}},\ and\ \bibinfo {author} {\bibfnamefont {M.~G.}\ \bibnamefont {Raymer}},\ }\bibfield  {title} {\bibinfo {title} {Photon {{Temporal Modes}}: {{A Complete Framework}} for {{Quantum Information Science}}},\ }\href {https://doi.org/10.1103/PhysRevX.5.041017} {\bibfield  {journal} {\bibinfo  {journal} {Physical Review X}\ }\textbf {\bibinfo {volume} {5}},\ \bibinfo {pages} {041017} (\bibinfo {year} {2015})}\BibitemShut {NoStop}%
\bibitem [{\citenamefont {Mejling}\ \emph {et~al.}(2012)\citenamefont {Mejling}, \citenamefont {McKinstrie}, \citenamefont {Raymer},\ and\ \citenamefont {Rottwitt}}]{mejling2012a}%
  \BibitemOpen
  \bibfield  {author} {\bibinfo {author} {\bibfnamefont {L.}~\bibnamefont {Mejling}}, \bibinfo {author} {\bibfnamefont {C.~J.}\ \bibnamefont {McKinstrie}}, \bibinfo {author} {\bibfnamefont {M.~G.}\ \bibnamefont {Raymer}},\ and\ \bibinfo {author} {\bibfnamefont {K.}~\bibnamefont {Rottwitt}},\ }\bibfield  {title} {\bibinfo {title} {Quantum frequency translation by four-wave mixing in a fiber: Low-conversion regime},\ }\href {https://doi.org/10.1364/OE.20.008367} {\bibfield  {journal} {\bibinfo  {journal} {Optics Express}\ }\textbf {\bibinfo {volume} {20}},\ \bibinfo {pages} {8367} (\bibinfo {year} {2012})}\BibitemShut {NoStop}%
\bibitem [{\citenamefont {Reddy}\ and\ \citenamefont {Raymer}(2017)}]{reddy2017a}%
  \BibitemOpen
  \bibfield  {author} {\bibinfo {author} {\bibfnamefont {D.~V.}\ \bibnamefont {Reddy}}\ and\ \bibinfo {author} {\bibfnamefont {M.~G.}\ \bibnamefont {Raymer}},\ }\bibfield  {title} {\bibinfo {title} {Engineering temporal-mode-selective frequency conversion in nonlinear optical waveguides: From theory to experiment},\ }\href {https://doi.org/10.1364/OE.25.012952} {\bibfield  {journal} {\bibinfo  {journal} {Optics Express}\ }\textbf {\bibinfo {volume} {25}},\ \bibinfo {pages} {12952} (\bibinfo {year} {2017})}\BibitemShut {NoStop}%
\bibitem [{\citenamefont {Brecht}\ \emph {et~al.}(2014)\citenamefont {Brecht}, \citenamefont {Eckstein}, \citenamefont {Ricken}, \citenamefont {Quiring}, \citenamefont {Suche}, \citenamefont {Sansoni},\ and\ \citenamefont {Silberhorn}}]{brecht2014}%
  \BibitemOpen
  \bibfield  {author} {\bibinfo {author} {\bibfnamefont {B.}~\bibnamefont {Brecht}}, \bibinfo {author} {\bibfnamefont {A.}~\bibnamefont {Eckstein}}, \bibinfo {author} {\bibfnamefont {R.}~\bibnamefont {Ricken}}, \bibinfo {author} {\bibfnamefont {V.}~\bibnamefont {Quiring}}, \bibinfo {author} {\bibfnamefont {H.}~\bibnamefont {Suche}}, \bibinfo {author} {\bibfnamefont {L.}~\bibnamefont {Sansoni}},\ and\ \bibinfo {author} {\bibfnamefont {C.}~\bibnamefont {Silberhorn}},\ }\bibfield  {title} {\bibinfo {title} {Demonstration of coherent time-frequency {{Schmidt}} mode selection using dispersion-engineered frequency conversion},\ }\href {https://doi.org/10.1103/PhysRevA.90.030302} {\bibfield  {journal} {\bibinfo  {journal} {Physical Review A}\ }\textbf {\bibinfo {volume} {90}},\ \bibinfo {pages} {030302} (\bibinfo {year} {2014})}\BibitemShut {NoStop}%
\bibitem [{\citenamefont {Chen}\ \emph {et~al.}(2024)\citenamefont {Chen}, \citenamefont {Briggs}, \citenamefont {Cui}, \citenamefont {Zhang}, \citenamefont {Shah},\ and\ \citenamefont {Fan}}]{chen2024}%
  \BibitemOpen
  \bibfield  {author} {\bibinfo {author} {\bibfnamefont {P.-K.}\ \bibnamefont {Chen}}, \bibinfo {author} {\bibfnamefont {I.}~\bibnamefont {Briggs}}, \bibinfo {author} {\bibfnamefont {C.}~\bibnamefont {Cui}}, \bibinfo {author} {\bibfnamefont {L.}~\bibnamefont {Zhang}}, \bibinfo {author} {\bibfnamefont {M.}~\bibnamefont {Shah}},\ and\ \bibinfo {author} {\bibfnamefont {L.}~\bibnamefont {Fan}},\ }\bibfield  {title} {\bibinfo {title} {Adapted poling to break the nonlinear efficiency limit in nanophotonic lithium niobate waveguides},\ }\href {https://doi.org/10.1038/s41565-023-01525-w} {\bibfield  {journal} {\bibinfo  {journal} {Nature Nanotechnology}\ }\textbf {\bibinfo {volume} {19}},\ \bibinfo {pages} {44} (\bibinfo {year} {2024})}\BibitemShut {NoStop}%
\bibitem [{\citenamefont {Stegeman}(1997)}]{stegeman1997}%
  \BibitemOpen
  \bibfield  {author} {\bibinfo {author} {\bibfnamefont {G.~I.}\ \bibnamefont {Stegeman}},\ }\bibfield  {title} {\bibinfo {title} {\$chi\^{}\{(2)\}\$ cascading: Nonlinear phase shifts},\ }\href {https://doi.org/10.1088/1355-5111/9/2/003} {\bibfield  {journal} {\bibinfo  {journal} {Quantum and Semiclassical Optics: Journal of the European Optical Society Part B}\ }\textbf {\bibinfo {volume} {9}},\ \bibinfo {pages} {139} (\bibinfo {year} {1997})}\BibitemShut {NoStop}%
\bibitem [{\citenamefont {Kozub}\ \emph {et~al.}(2023)\citenamefont {Kozub}, \citenamefont {Gerstmann},\ and\ \citenamefont {Schmidt}}]{kozub2023}%
  \BibitemOpen
  \bibfield  {author} {\bibinfo {author} {\bibfnamefont {A.~L.}\ \bibnamefont {Kozub}}, \bibinfo {author} {\bibfnamefont {U.}~\bibnamefont {Gerstmann}},\ and\ \bibinfo {author} {\bibfnamefont {W.~G.}\ \bibnamefont {Schmidt}},\ }\bibfield  {title} {\bibinfo {title} {Third-{{Order Susceptibility}} of {{Lithium Niobate}}: {{Influence}} of {{Polarons}} and {{Bipolarons}}},\ }\href {https://doi.org/10.1002/pssb.202200453} {\bibfield  {journal} {\bibinfo  {journal} {physica status solidi (b)}\ }\textbf {\bibinfo {volume} {260}},\ \bibinfo {pages} {2200453} (\bibinfo {year} {2023})}\BibitemShut {NoStop}%
\bibitem [{\citenamefont {{Shams-Ansari}}\ \emph {et~al.}(2022)\citenamefont {{Shams-Ansari}}, \citenamefont {Huang}, \citenamefont {He}, \citenamefont {Li}, \citenamefont {Holzgrafe}, \citenamefont {Jankowski}, \citenamefont {Churaev}, \citenamefont {Kharel}, \citenamefont {Cheng}, \citenamefont {Zhu}, \citenamefont {Sinclair}, \citenamefont {Desiatov}, \citenamefont {Zhang}, \citenamefont {Kippenberg},\ and\ \citenamefont {Lon{\v c}ar}}]{shams-ansari2022}%
  \BibitemOpen
  \bibfield  {author} {\bibinfo {author} {\bibfnamefont {A.}~\bibnamefont {{Shams-Ansari}}}, \bibinfo {author} {\bibfnamefont {G.}~\bibnamefont {Huang}}, \bibinfo {author} {\bibfnamefont {L.}~\bibnamefont {He}}, \bibinfo {author} {\bibfnamefont {Z.}~\bibnamefont {Li}}, \bibinfo {author} {\bibfnamefont {J.}~\bibnamefont {Holzgrafe}}, \bibinfo {author} {\bibfnamefont {M.}~\bibnamefont {Jankowski}}, \bibinfo {author} {\bibfnamefont {M.}~\bibnamefont {Churaev}}, \bibinfo {author} {\bibfnamefont {P.}~\bibnamefont {Kharel}}, \bibinfo {author} {\bibfnamefont {R.}~\bibnamefont {Cheng}}, \bibinfo {author} {\bibfnamefont {D.}~\bibnamefont {Zhu}}, \bibinfo {author} {\bibfnamefont {N.}~\bibnamefont {Sinclair}}, \bibinfo {author} {\bibfnamefont {B.}~\bibnamefont {Desiatov}}, \bibinfo {author} {\bibfnamefont {M.}~\bibnamefont {Zhang}}, \bibinfo {author} {\bibfnamefont {T.~J.}\ \bibnamefont {Kippenberg}},\ and\ \bibinfo {author} {\bibfnamefont {M.}~\bibnamefont {Lon{\v c}ar}},\ }\bibfield  {title} {\bibinfo {title} {Reduced
  material loss in thin-film lithium niobate waveguides},\ }\href {https://doi.org/10.1063/5.0095146} {\bibfield  {journal} {\bibinfo  {journal} {APL Photonics}\ }\textbf {\bibinfo {volume} {7}},\ \bibinfo {pages} {081301} (\bibinfo {year} {2022})}\BibitemShut {NoStop}%
\bibitem [{\citenamefont {Quesada}\ \emph {et~al.}(2022)\citenamefont {Quesada}, \citenamefont {Helt}, \citenamefont {Menotti}, \citenamefont {Liscidini},\ and\ \citenamefont {Sipe}}]{quesada2022}%
  \BibitemOpen
  \bibfield  {author} {\bibinfo {author} {\bibfnamefont {N.}~\bibnamefont {Quesada}}, \bibinfo {author} {\bibfnamefont {L.~G.}\ \bibnamefont {Helt}}, \bibinfo {author} {\bibfnamefont {M.}~\bibnamefont {Menotti}}, \bibinfo {author} {\bibfnamefont {M.}~\bibnamefont {Liscidini}},\ and\ \bibinfo {author} {\bibfnamefont {J.~E.}\ \bibnamefont {Sipe}},\ }\bibfield  {title} {\bibinfo {title} {Beyond photon pairs: {{Nonlinear}} quantum photonics in the high-gain regime},\ }\href {https://doi.org/10.1364/AOP.445496} {\bibfield  {journal} {\bibinfo  {journal} {Advances in Optics and Photonics}\ }\textbf {\bibinfo {volume} {14}},\ \bibinfo {pages} {291} (\bibinfo {year} {2022})}\BibitemShut {NoStop}%
\bibitem [{\citenamefont {Chung}\ \emph {et~al.}(2017)\citenamefont {Chung}, \citenamefont {Huang}, \citenamefont {Wang}, \citenamefont {Yang}, \citenamefont {Yang}, \citenamefont {Sung}, \citenamefont {Solntsev}, \citenamefont {Sukhorukov}, \citenamefont {Neshev},\ and\ \citenamefont {Chen}}]{chung2017}%
  \BibitemOpen
  \bibfield  {author} {\bibinfo {author} {\bibfnamefont {H.-P.}\ \bibnamefont {Chung}}, \bibinfo {author} {\bibfnamefont {K.-H.}\ \bibnamefont {Huang}}, \bibinfo {author} {\bibfnamefont {K.}~\bibnamefont {Wang}}, \bibinfo {author} {\bibfnamefont {S.-L.}\ \bibnamefont {Yang}}, \bibinfo {author} {\bibfnamefont {S.-Y.}\ \bibnamefont {Yang}}, \bibinfo {author} {\bibfnamefont {C.-I.}\ \bibnamefont {Sung}}, \bibinfo {author} {\bibfnamefont {A.~S.}\ \bibnamefont {Solntsev}}, \bibinfo {author} {\bibfnamefont {A.~A.}\ \bibnamefont {Sukhorukov}}, \bibinfo {author} {\bibfnamefont {D.~N.}\ \bibnamefont {Neshev}},\ and\ \bibinfo {author} {\bibfnamefont {Y.-H.}\ \bibnamefont {Chen}},\ }\bibfield  {title} {\bibinfo {title} {Asymmetric adiabatic couplers for fully-integrated broadband quantum-polarization state preparation},\ }\href {https://doi.org/10.1038/s41598-017-17094-7} {\bibfield  {journal} {\bibinfo  {journal} {Scientific Reports}\ }\textbf {\bibinfo {volume} {7}},\ \bibinfo {pages} {16841} (\bibinfo {year}
  {2017})}\BibitemShut {NoStop}%
\bibitem [{\citenamefont {Chen}\ \emph {et~al.}(2021)\citenamefont {Chen}, \citenamefont {Yang}, \citenamefont {Wong}, \citenamefont {Pun},\ and\ \citenamefont {Wang}}]{chen2021}%
  \BibitemOpen
  \bibfield  {author} {\bibinfo {author} {\bibfnamefont {Z.}~\bibnamefont {Chen}}, \bibinfo {author} {\bibfnamefont {J.}~\bibnamefont {Yang}}, \bibinfo {author} {\bibfnamefont {W.-H.}\ \bibnamefont {Wong}}, \bibinfo {author} {\bibfnamefont {E.~Y.-B.}\ \bibnamefont {Pun}},\ and\ \bibinfo {author} {\bibfnamefont {C.}~\bibnamefont {Wang}},\ }\bibfield  {title} {\bibinfo {title} {Broadband adiabatic polarization rotator-splitter based on a lithium niobate on insulator platform},\ }\href {https://doi.org/10.1364/PRJ.432906} {\bibfield  {journal} {\bibinfo  {journal} {Photonics Research}\ }\textbf {\bibinfo {volume} {9}},\ \bibinfo {pages} {2319} (\bibinfo {year} {2021})}\BibitemShut {NoStop}%
\bibitem [{\citenamefont {Ji}\ \emph {et~al.}(2022)\citenamefont {Ji}, \citenamefont {Liu}, \citenamefont {He}, \citenamefont {Wang}, \citenamefont {Qiu}, \citenamefont {Riemensberger},\ and\ \citenamefont {Kippenberg}}]{ji2022}%
  \BibitemOpen
  \bibfield  {author} {\bibinfo {author} {\bibfnamefont {X.}~\bibnamefont {Ji}}, \bibinfo {author} {\bibfnamefont {J.}~\bibnamefont {Liu}}, \bibinfo {author} {\bibfnamefont {J.}~\bibnamefont {He}}, \bibinfo {author} {\bibfnamefont {R.~N.}\ \bibnamefont {Wang}}, \bibinfo {author} {\bibfnamefont {Z.}~\bibnamefont {Qiu}}, \bibinfo {author} {\bibfnamefont {J.}~\bibnamefont {Riemensberger}},\ and\ \bibinfo {author} {\bibfnamefont {T.~J.}\ \bibnamefont {Kippenberg}},\ }\bibfield  {title} {\bibinfo {title} {Compact, spatial-mode-interaction-free, ultralow-loss, nonlinear photonic integrated circuits},\ }\href {https://doi.org/10.1038/s42005-022-00851-0} {\bibfield  {journal} {\bibinfo  {journal} {Communications Physics}\ }\textbf {\bibinfo {volume} {5}},\ \bibinfo {pages} {84} (\bibinfo {year} {2022})}\BibitemShut {NoStop}%
\bibitem [{\citenamefont {Bahadori}\ \emph {et~al.}(2019)\citenamefont {Bahadori}, \citenamefont {Nikdast}, \citenamefont {Cheng},\ and\ \citenamefont {Bergman}}]{bahadori2019}%
  \BibitemOpen
  \bibfield  {author} {\bibinfo {author} {\bibfnamefont {M.}~\bibnamefont {Bahadori}}, \bibinfo {author} {\bibfnamefont {M.}~\bibnamefont {Nikdast}}, \bibinfo {author} {\bibfnamefont {Q.}~\bibnamefont {Cheng}},\ and\ \bibinfo {author} {\bibfnamefont {K.}~\bibnamefont {Bergman}},\ }\bibfield  {title} {\bibinfo {title} {Universal {{Design}} of {{Waveguide Bends}} in {{Silicon-on-Insulator Photonics Platform}}},\ }\href {https://doi.org/10.1109/JLT.2019.2909983} {\bibfield  {journal} {\bibinfo  {journal} {Journal of Lightwave Technology}\ }\textbf {\bibinfo {volume} {37}},\ \bibinfo {pages} {3044} (\bibinfo {year} {2019})}\BibitemShut {NoStop}%
\bibitem [{\citenamefont {Vogelbacher}\ \emph {et~al.}(2019)\citenamefont {Vogelbacher}, \citenamefont {Nevlacsil}, \citenamefont {Sagmeister}, \citenamefont {Kraft}, \citenamefont {Unterrainer},\ and\ \citenamefont {Hainberger}}]{vogelbacher2019}%
  \BibitemOpen
  \bibfield  {author} {\bibinfo {author} {\bibfnamefont {F.}~\bibnamefont {Vogelbacher}}, \bibinfo {author} {\bibfnamefont {S.}~\bibnamefont {Nevlacsil}}, \bibinfo {author} {\bibfnamefont {M.}~\bibnamefont {Sagmeister}}, \bibinfo {author} {\bibfnamefont {J.}~\bibnamefont {Kraft}}, \bibinfo {author} {\bibfnamefont {K.}~\bibnamefont {Unterrainer}},\ and\ \bibinfo {author} {\bibfnamefont {R.}~\bibnamefont {Hainberger}},\ }\bibfield  {title} {\bibinfo {title} {Analysis of silicon nitride partial {{Euler}} waveguide bends},\ }\href {https://doi.org/10.1364/OE.27.031394} {\bibfield  {journal} {\bibinfo  {journal} {Optics Express}\ }\textbf {\bibinfo {volume} {27}},\ \bibinfo {pages} {31394} (\bibinfo {year} {2019})}\BibitemShut {NoStop}%
\bibitem [{\citenamefont {Wang}\ \emph {et~al.}(2020)\citenamefont {Wang}, \citenamefont {Chen}, \citenamefont {Dai},\ and\ \citenamefont {Liu}}]{wang2020}%
  \BibitemOpen
  \bibfield  {author} {\bibinfo {author} {\bibfnamefont {J.}~\bibnamefont {Wang}}, \bibinfo {author} {\bibfnamefont {P.}~\bibnamefont {Chen}}, \bibinfo {author} {\bibfnamefont {D.}~\bibnamefont {Dai}},\ and\ \bibinfo {author} {\bibfnamefont {L.}~\bibnamefont {Liu}},\ }\bibfield  {title} {\bibinfo {title} {Polarization {{Coupling}} of \${{X}}\$-{{Cut Thin Film Lithium Niobate Based Waveguides}}},\ }\href {https://doi.org/10.1109/JPHOT.2020.2995317} {\bibfield  {journal} {\bibinfo  {journal} {IEEE Photonics Journal}\ }\textbf {\bibinfo {volume} {12}},\ \bibinfo {pages} {1} (\bibinfo {year} {2020})}\BibitemShut {NoStop}%
\bibitem [{\citenamefont {Pan}\ \emph {et~al.}(2019)\citenamefont {Pan}, \citenamefont {Hu}, \citenamefont {Zeng},\ and\ \citenamefont {Xia}}]{pan2019}%
  \BibitemOpen
  \bibfield  {author} {\bibinfo {author} {\bibfnamefont {A.}~\bibnamefont {Pan}}, \bibinfo {author} {\bibfnamefont {C.}~\bibnamefont {Hu}}, \bibinfo {author} {\bibfnamefont {C.}~\bibnamefont {Zeng}},\ and\ \bibinfo {author} {\bibfnamefont {J.}~\bibnamefont {Xia}},\ }\bibfield  {title} {\bibinfo {title} {Fundamental mode hybridization in a thin film lithium niobate ridge waveguide},\ }\href {https://doi.org/10.1364/OE.27.035659} {\bibfield  {journal} {\bibinfo  {journal} {Optics Express}\ }\textbf {\bibinfo {volume} {27}},\ \bibinfo {pages} {35659} (\bibinfo {year} {2019})}\BibitemShut {NoStop}%
\bibitem [{\citenamefont {Guo}\ and\ \citenamefont {Zhao}(2015)}]{guo2015}%
  \BibitemOpen
  \bibfield  {author} {\bibinfo {author} {\bibfnamefont {J.}~\bibnamefont {Guo}}\ and\ \bibinfo {author} {\bibfnamefont {Y.}~\bibnamefont {Zhao}},\ }\bibfield  {title} {\bibinfo {title} {Analysis of {{Mode Hybridization}} in {{Tapered Waveguides}}},\ }\href {https://doi.org/10.1109/LPT.2015.2468059} {\bibfield  {journal} {\bibinfo  {journal} {IEEE Photonics Technology Letters}\ }\textbf {\bibinfo {volume} {27}},\ \bibinfo {pages} {2441} (\bibinfo {year} {2015})}\BibitemShut {NoStop}%
\end{thebibliography}%
\end{document}